\documentclass[11pt,twoside]{report}

\usepackage[headheight=14pt,bottom=2.5cm,inner=4cm,outer=2.5cm]{geometry} % The minimal margins as specified by Code of Practice.
\usepackage{setspace}
\usepackage{datetime}
\usepackage{fancyhdr}
\usepackage{tikz}
\usepackage{rotating}

\usepackage{booktabs}
\usepackage{algorithm}
\usepackage{algpseudocode}

\usepackage{amssymb,amsmath,amsthm}
\usepackage{hyperref} % It is often advices to load hyperref last (there are exceptions).
\hypersetup{
colorlinks=true,
linkcolor=black,
citecolor=black
}

\graphicspath{{./figures/}{../figures/}} % Location of the figures and graphics files so that you can insert them without specifying a path to the figures folder.

\makeatletter
\title{Towards the Blockchain Massive Adoption with Permissionless Storage} \let\Title\@title
\author{Jia Kan} \let\Author\@author
\makeatother

\begin{document}
%% The title page uses the title and author name specified above. No need to edit title.tex. Note that the current month is at the bottom of the title page. Edit it manually, if you want to have some other month there.
\pagestyle{empty}
\begin{titlepage}
	\centering
	\vspace*{1cm}
	\includegraphics[width=65mm]{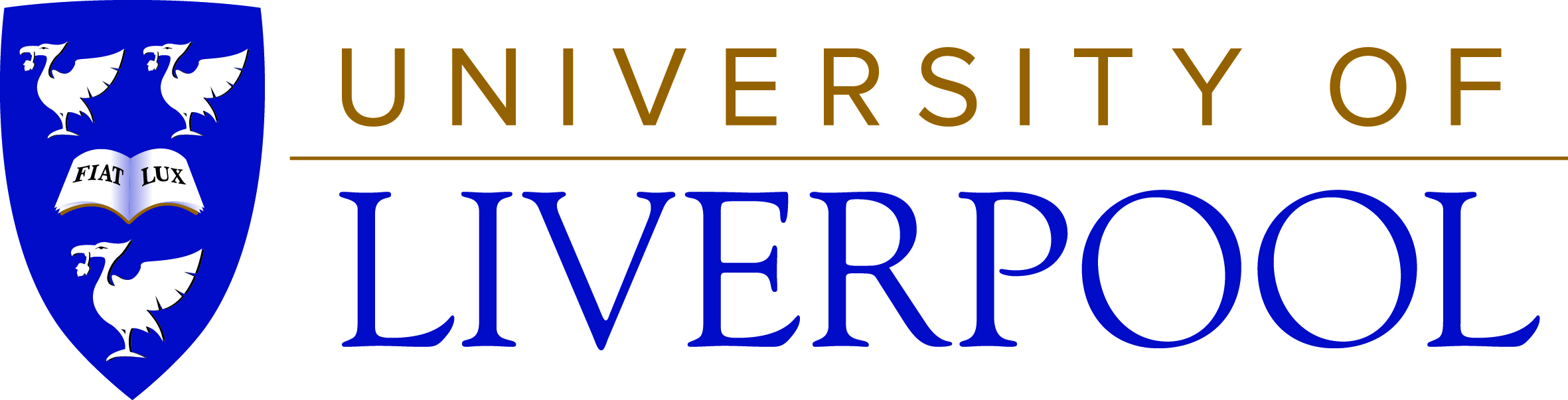}\par\vspace{1cm}
		\vspace{2cm}
	{\huge \Title\par}
	\vspace{5cm}
	{Thesis submitted in accordance with the requirements of the University of Liverpool for the  degree  of  Doctor  in  Philosophy by  \par}
	\vspace{1cm}
	{\textbf{\Author}\par}
	\vfill

% Bottom of the page
	{\large \monthname\, \the\year\par}
\end{titlepage}
\cleardoublepage

%% In the preface, pages are numbered by roman numerals (i,ii,...) and in the header Author and Title.
\pagenumbering{roman}
\pagestyle{fancy}
\fancyhead{} % clear all header fields
\fancyhead[LO,RE]{PhD Thesis}
\fancyhead[LE,RO]{\Author}
%\fancyfoot{} % clear all footer fields
%\fancyfoot[C]{\thepage}

\begin{doublespace}

\addcontentsline{toc}{chapter}{Abstract} % Add the abstract to the table of contents.
\chapter*{Abstract}

Blockchain technology emerged with the advent of Bitcoin and rapidly developed over the past few decades, becoming widely accepted and known by the public.
% This technology is highly anticipated by the public, as before its invention, many services had to rely on a trusted third party, such as banks.
% People expect blockchain to be widely used.
However, in the past decades, the massive adoption of blockchain technology has yet to come.
% Blockchain application scenarios are still limited in finance and restricted by technological limitations.
% Bitcoin has already shown its outstanding features of permissionless transactions and value stores.
% In the Ethereum ecosystem, DeFi was evolved and accepted by early adopters.
Rather than the scalability issue, the blockchain application is challenged by its expensive usage cost.
% The end-users are discouraged from using blockchain with such a high transaction/gas fee.
However, the high cost of blockchain usage is deeply connected with the blockchain consensus and security mechanism.
The permissionless blockchain must maintain its high cost for security against the 51\% Attack.
Chain users indirectly cover the cost as coins are appointed for blockchain usage fees.
This conflict prevents the massive adoption of blockchain.
Thus, blockchain must be improved to solve those problems:
1. The cost of blockchain usage should be low enough.
2. The blockchain should remain decentralized. % Otherwise, users perfer to stay in the current web2 system.
3. The scalability of blockchain must meet the demand.

In my thesis, new approaches are applied to solve the issues above.
The key contribution is the discovery of the useful PoW.
It extends the Nakamoto PoW with another usage of file data encoding during the same Nakamoto Consensus computation to prove honest data preservation.
Based on this theory, a permissionless storage network is proposed as the new security engine for the blockchain.
It bridges the high blockchain security cost to the storage users with real demands who are willing to pay for the storage resource.
On the other hand, the chain users can benefit from the low transaction fee.
Meanwhile, we also provide a scalability solution to shard the blockchain.
It enables high TPS and keeps decentralization.
The solutions in this thesis provide the answers to all the dependencies of the massive adoption.

\cleardoublepage

\addcontentsline{toc}{chapter}{Acknowledgements} % Add the acknowledgements to the table of contents.
\chapter*{Acknowledgements}

First of all, I would like to express my sincere gratitude to my supervisors Professor Jie Zhang and Professor Xin Huang for their support and guidance throughout my time of Ph.D. research. They have made many contributions to this work. They have given me a lot of valuable guidance and suggestions for conducting the Ph.D. research and completing the Ph.D. thesis.

Secondly, I would like to thank to my third supervisor Professor Junqing Zhang for their valuable suggestion on my Ph.D. research. I would also like to thank Dr. Qi Chen, Dr. Jiaping Wang, Dr. Ming Wu for their help.

This work is supported by Department of Computer Science, University of Liverpool and School of Advanced Technology, Xi'an Jiaotong-Liverpool University. Special thanks to all the colleagues from these institutes.

Finally, I would like to thank my family and friends for their love and friendship. I would especially like to thank my wife Zhen Liu for her support and love, and my little daughter Luyu Kan.

\cleardoublepage

\tableofcontents
\addcontentsline{toc}{chapter}{Contents} % Add the the table of contents to the table of contents.
\clearpage

\listoffigures
\addcontentsline{toc}{chapter}{List of Figures} % Add the list of figures to the table of contents.
\clearpage

\listoftables
\addcontentsline{toc}{chapter}{List of Tables} % Add the list of tables to the table of contents.
\clearpage

%% You might need a similar construction to include the list of algorithms.

%\end{singlespace}

\cleardoublepage

%% In the thesis, pages are numbered with arabic numerals in the top outer corners (left top corner on even numbered pages, and right top corner on odd numbered pages) and alternatively Author and Chapter: Chapter name in the other top corner.
\pagenumbering{arabic}
\pagestyle{fancy}
\fancyhead{} % clear all header fields
\fancyhead[RO,LE]{\thepage}
\fancyhead[LO]{\nouppercase\leftmark}
\fancyhead[RE]{\Author}
\fancyfoot{} % clear all footer fields

%\begin{doublespace}

\chapter{Introduction} % Write in your own chapter title
\label{ch:intro} % Add a label in case you want to refer to the chapter.
%
%An introduction to the thesis. An example reference \cite{Ref17}.
%
%\newpage
%Even pages look like this.
%
%\newpage
%Odd pages look like this.
% Blockchain technology has born with Bitcoin for more than a decade.Blockchain technology emerged with the advent of Bitcoin and has rapidly developed over the past years, becoming widely accepted and well-known to the public.
Bitcoin\cite{2008Bitcoin} was created with the good intention of using technology to counter the social turmoil caused by capital greed and proposed a potential solution to avoid the huge harm caused by economic crises to society.
The public highly anticipates this technology because many services had to rely on a trusted third party, such as banks, before its invention.
The characteristics of decentralization and avoidance of trusting institutions make us see more significant potential for this technology in the future.
For the first time, blockchain technology allows services to operate independently without relying on any trusted entity, which is considered one of the epoch-making technologies.

The public expects more blockchain applications to be implemented for massive adoption, but various limitations restrict the development of blockchain applications.
There are several prerequisites for large-scale use: first, the blockchain end-users cost of use should be reduced to an acceptable level; second, the blockchain infrastructure must handle enough requests; third, the blockchain must remain decentralized as the bottom line.
However, achieving those prerequisites is challenging. 

First, using a public chain requires paying transaction/gas fees with the chain's native coins.
With the drastic fluctuations of Bitcoin and Ethereum\cite{buterin2014next} coin prices, the cost of single-use has become very high, making it difficult for ordinary people to accept.
This mechanism behind this is that a public chain is required to issue coins in exchange for more mining power to increase its security protection\cite{cbeci}.
Even in Ethereum, blockchain cannot suppress the skyrocketing usage cost, as the chain must keep the high coin price to ensure security with staking.
Thus, the public chains need to seek new mechanisms to obtain security protection without putting a high burden on users to enable blockchain large-scale use.
In our thesis, the useful Proof of Work solution proposed is an extended Nakamoto Proof of Work that bridges the cost of security to useful tasks.
Since the PoW computation is a by-product of computing useful tasks, the electricity cost of PoW is paid by users who request the tasks.
Blockchain users can enjoy a low fee for chain usage.

To realize this, there are few scenarios that the Nakamoto PoW algorithm can apply because of the hash algorithm.
Permissionless storage network is a scenario with fundamental and wide demands.
Different from traditional cloud-computing storage, the permissionless storage provider is required to prove the preservation of the file data honestly.
Proof of Replication algorithm is indispensable in the permissionless network against Outsourcing Attack\cite{2017PoRep}.
The Nakamoto PoW matches the requirements of the encoding algorithm for Proof of Replication exactly.
This discovery makes the Nakamoto PoW useful in the tasks first time.
The consensus and the encoding can be calculated within the same hash computation.

To bridge the high security cost to the useful tasks, it is required to build a storage market.
The more files and data are sent to the storage system, the higher the security blockchain would reach.
With the mechanism change, the blockchain security cost is no longer shared by the end-users gas fee payment, except that users still pay some tips for the miners to decide which transaction should be processed in priority.
This brings down the chain users' cost to almost zero, which makes blockchain massive usage possible.
The miners could get more income from the storage users during the PoW computation.
Unlike solo mining, most miners could get sustainable rewards during the recurring billing of the storage service.

% Due to the huge energy consumption of the PoW algorithm and the fact that many decentralized blockchains claim to be able to improve performance by sacrificing decentralization, most blockchains have begun to abandon PoW and choose alternative algorithms such as PoS and PoC.
% Prior to this, people tried to find other useful Proof of Work algorithms, such as using consensus computing for machine learning and other useful scenarios.
% After many years of effort, people realized that such algorithms may not exist.
% This is because the consensus algorithm requires multiple parties to calculate a deterministic and verifiable result based on the input, making it difficult to apply the hash algorithm used in the Satoshi Nakamoto consensus to meaningful computing.

% Although there has been little progress in modifying and finding meaningful alternatives to the Satoshi Nakamoto PoW algorithm, we have unexpectedly found a way to use hash power to perform meaningful work without modifying the Satoshi Nakamoto consensus algorithm itself.
% This method can be considered an upgrade to Bitcoin's consensus algorithm: PoW 2.0.
% Subsequent research has shown that making hash computing power useful may seem like a small change, but it has the potential for a huge impact on the design of blockchain and the entire industry.
% Rather than environment friendly, useful PoW.

Second, as the first blockchain, Bitcoin has a transaction capacity of only seven transactions per second (TPS).
Currently, the DAG-based blockchain\cite{pervez2018comparative} can achieve thousands of transactions per second.
However, the entire node that keeps all the history transactions will grow fast in its storage requirement until it cannot meet the storage capacity demand.
High TPS blockchain is still facing the challenge of centralization.
Sharding technology must be applied before the massive adoption.

The sharding solution proposed in this thesis improves the Monoxide\cite{wang2019monoxide}.
The history transactions can be split up to one user account per shard.
In the design, without a smart contract, the coins can be transferred between shards.
Meanwhile, by keeping the global state on each node, the sharding also supports Ethereum-like smart contracts.
This sharding can achieve TPS as high as a DAG-based blockchain.
The data structure is related simply compared to DAG.
It is possible to achieve higher performance and scalability without sacrificing decentralization.

Decentralization ensures the asset's security.
Layer 1 blockchain verifies the transactions by all the nodes and miners.
The network is permissionless and anyone can join as a miner and storage provider.
The end-users no longer share the security cost.
The on-chain transaction can be as cheap as almost free.
Those aspects make chain users willing to participate in blockchain transactions and smart contract interaction.

% Many blockchain projects are trying to improve performance, but most performance improvement solutions are built on the premise of sacrificing decentralization, such as using the DPoS or PBFT consensus algorithm, where consensus is reached by a committee that is elected in advance.
% In contrast to Bitcoin's consensus algorithm, Bitcoin achieves consensus among all miners rather than forming a committee.
% Although all nodes still verify the correctness of transaction history, committee members may collude to change the order of transactions entering the blockchain, and even making profit from it.

% Consensus algorithm is not the bottle neck of blockchain performance, DAG solution already showed that blockchain data structure changing could increase the TPS.
% Followed with Monoxide design, we proposed Wider solution for sharding and high performance.

% Both our consensus and performance improvements remain the decentralization.
% The initial blockchain project Bitcoin uses decentralization to keep the safty of the virtual currency.
% The suffiency Bitcoin full nodes verify the ledger for the correctness.
% Anyone could run the Bitcoin software as a node to join the permissionless blockchain network.
% Decentralization ensures the assets security and service continuation.
% Thus, all our approaches must keep the blockchain system decentralized, which is the core reason people moving from web2 towards web3.

\section{Motivation}

% 研究动机：通过研究和理论创新，推动区块链和web3应用的落地和大规模使用。
% 将区块链共识与区块链存储技术结合，降低安全成本和用户使用门槛, 增加已经消费的PoW算力资源的产出。
% 在不牺牲区块链的去中心化的特性的前提下, 提高区块链的性能和吞吐量。
% 使用代理重加密来保障区块链各种应用的保密性，安全性和隐私性。
% 数据可以被存储, 构建可持续运行的去中心化服务。

The thesis aims to achieve massive blockchain adoption by resolving the core problems.
How to make blockchain technology land on a large scale and enable society to enjoy the benefits of this new technology truly is the goal. 
After more than ten years of development, various new blockchain technologies have emerged, but from the current perspective, large-scale use is still a long way off.
The improvement should remain the decentralization, which is the most significant feature of blockchain.
However, many of the improvements to blockchain sacrifice decentralization for performance.
Thus, it is time to rethink the approaches to address existing problems.

The fundamental problems include:
a) Maintain the blockchain's high-security cost and reduce the blockchain usage cost.
b) improve the blockchain performance and capacity.
c) keep and improve the decentralization of blockchain.
In addition, we would like to see:
d) reuse consensus algorithm computation for useful tasks.
e) onboard more real-world services onto the blockchain.

Unlike other modular systems, the blockchain system was born with a unique design.
Simply changing one aspect would affect another.
Therefore, my research goes back to the origin of blockchain, starting from Satoshi Nakamoto's work, and seeks breakthroughs under the premise of decentralization and PoW.
At the same time, we also have some new reflections and improvements on the latest technologies that have emerged in blockchain development over the years.

%Of course, some radical designs, such as the total limit of 21 million Bitcoins, also make it impossible for Bitcoin to truly become a currency.
%After more than ten years of development, Bitcoin is mainly used for storage rather than payment.

%The main motivation of this article is to make blockchain technology more closely related to practical applications in real life through doctoral studies and research.
%Regarding the consensus part, the mainstream idea is to replace computing power with capital, and PoW is gradually replaced by PoS.
%Although the energy consumption problem has been solved, various new problems have emerged.

%The blockchain aims to build the decentralize system.
%Such new generation of system can avoid the users to trust a third party.
%The blockchain and web3 give the users back the control of their own data.
%The motivation lies in two aspects: (1) the importance of decentralize and (2) the importance of security.

\subsection{The conflict between security cost and user usage cost}

In permissionless blockchain networks, security must be guaranteed by the system's mechanisms.
To protect security against attacks such as 51\% Attack, blockchains need to maintain high-security costs.
However, current blockchains obtain security computing power protection through the mechanism of virtual currency (native coin).
This mechanism ultimately transfers the security cost to the end-users of the blockchain, and users need to pay expensive fees to cover the blockchain cost.
Reducing the final user cost while not sacrificing blockchain security is a contradiction of the current blockchain infrastructure, and it is also one of the important problems that need to be solved for the large-scale use of blockchain technology.

\subsection{Importance of Sharding to Scalability}

Blockchain scalability is a long-term goal.
Many works focused on this topic, and much progress has been made.
Currently, the DAG-based blockchain has already achieved thousands of TPS in performance.
Higher performance requires higher throughput and increases the hardware requirement.

With limited scalability, the hardware cannot support higher performance and throughput.
The disk space of a blockchain node would run out, and the CPU can hardly verify incoming transactions in time.
The scalability also impacts the decentralization.
Bitcoin limited its block size to 1M data to prevent the full chain disk usage from growing too fast.
It enables anyone to afford to host a full node, making the Bitcoin blockchain more decentralized.

Instead of all the transactions being kept and processed by all the nodes, sharding reduces the pressure on the full nodes.
The history transactions can be hosted by different nodes, and the verification can be adjusted to probabilistic.

\subsection{Importance of Decentralization}

Decentralization is the motivation and goal of blockchain invention.
As an infrastructure, blockchain can provide us with essential services that do not rely on third parties.
Such services are difficult to shut down actively.

Decentralization means that the nodes in the blockchain network are equal, and no single node controls the entire network.
It also encourages everyone to verify the transactions rather than trust in agents.
This makes blockchain more secure, transparent, and reliable.

\section{Objective}

The objective of this thesis is approching blockchain massive adoption.
The blockchain for assets must remain decentralized while keep low usage cost for the end users.
The chain infrastructure also demands the high throughput to support large amount of usage.
High TPS and sharding are required.
We propose solutions from security engine to sharding scheme to improve blockchain system.

\section{Solutions}

To achieve massive blockchain adoption and reduce the gap between the challenge problems and real-world demands, we have proposed several solutions.
Our methods remain and even improve the decentralization.

The key idea to solve the conflict between security and user usage costs is transferring the high computation cost during PoW consensus onto another useful scenario.
Currently, blockchain uses dedicated PoW computation for consensus.
The computation has no extra purpose except for blockchain security.
Thus, the high cost of security is paid by chain users.
The useful PoW as the new engine protects the chain security from a by-product of useful tasks.
With the change, blockchain can receive free security protection.
Another group of users with real demand for useful tasks would cover the cost.

In the direction of scalability and decentralization, we improve the chain data structure.
The DAG structure lacks the scalability.
It is hard to shard the history of transactions in different nodes.
In our solution, the main chain with subchains solves the issue.
Each subchain is a shard for a unique user, and all the transactions are cross-shard.
The main chain is used for the confirmation of all the subchains updated.
The consensus algorithm is only applied on the main chain.

\section{Contributions}

The thesis has the following contributions:

\begin{enumerate}
    \item The first useful PoW algorithm by extending the current Nakamoto PoW with no change to the consensus algorithm based on the hash.
    \item The protocol for storage integrity verification for a permissionless storage network to provide blockchain security with PoW computation.
    \item The Proxy Re-Encryption scheme for cryptographic file permission grant.
    \item The blockchain sharding solution to increase the scalability, performance, and throughput without sacrificing the decentralization.
\end{enumerate}

\section{Deliverable Outcomes}

This thesis is partly based on the following publication.
The contributions of the author are also listed.

\begin{enumerate}
    \item J Kan, J Zhang, D Liu, X Huang. Proxy Re-Encryption Scheme for Decentralized Storage Networks\cite{kan2022proxy}.
%   \item Jie Zhang, Nian Xue, Xin Huang. A Secure System for Pervasive Social Network-based Healthcare. IEEE Access, Vol. 4, Pages 9239-9250, 2016.

%       Contributions of the author: (1) design of the improved IEEE 802.15.6 display authenticated association protocol; (2) design of the protocol for blockchain consensus mechanism; (3) design of the healthcare blockchain; (4) security analysis for protocols; and (5) performance evaluation.

\end{enumerate}

This thesis is mainly based on the following pre-print papers.
The paper Economic Proof of Work is in submission to the ATC24 conference with the updated title Useful Nakamoto Proof of Work.

\begin{enumerate}
    \item J Kan. Economic Proof of Work\cite{kan2020economic}.
    \item J Kan, J Zhang, X Huang. Wider: Scale Out Blockchain With Sharding by Account\cite{kan2022wider}.
\end{enumerate}

The research direction is based on the works during the post-guaranteed studies in XJTLU.

\begin{enumerate}
    \item J Kan, S Chen, X Huang. Improve blockchain performance using graph data structure and parallel mining\cite{2018Parallel}.
    \item J Kan, L Zou, B Liu, X Huang. Boost blockchain broadcast propagation with tree routing\cite{kan2018boost}.
    \item J Kan, KS Kim. MTFS: Merkle-tree-based file system\cite{2019MTFS}.
\end{enumerate}

While the theories proposed in the thesis solve the critical problems in the blockchain field, they come with practice.
The complete implementation of a sharding blockchain is developed with a Python VM and bytecode-based smart contract.
We build the blockchain and the virtual machine from scratch since 2018.

\begin{enumerate}
    \item \url{https://github.com/kernel1983/BitPoW}.
\end{enumerate}

The chain code is implemented in Python 3.10 with the execution environment of Ubuntu 22.04 LTS, which is the most easy and popular programming language so far.
We use rocksdb for the database.
It works for both study and production purpose.
The chain code is develeoping during the study of consensus and sharding.
Wider sharding part is implemented.
However, the permissionless storage network is not a part of the chain code.

Once important thing we learn from the practice is that we should keep the complex applications off-chain, such as permissionless storage network, to remain the chain performance and avoid complex.

\section{Thesis Outline}

The remaining of this thesis is outlined as following:
\vspace{6mm}

\textbf{Chapter \ref{ch:preliminaries}: Preliminaries}

This chapter introduces the basic knowledge of blockchain, cryptography, and distributed systems.
Bitcoin and Ethereum basics: Hash chain, Proof of Work mining and longest chain rule are the core concepts of Bitcoin.
Ethereum has different components such as Proof of Stake, EVM and Account Model.
The state machine can be used to model the blockchain systems.
Consensus algorithms: traditional distributed consensus (consistency) and blockchain consensus, including Proof of Work, Proof of Stake, PoW/PoS hybrid consensus.
We also introduce fundamental cryptography: hash functions, digital signatures, symmetric encryption, asymmetric encryption in this chapter.

\vspace{6mm}
\textbf{Chapter \ref{ch:consensus}: Useful Proof of Work}

This chapter proposes the useful PoW as the key algorithm in the thesis to improve the blockchain system.
The useful PoW extends Nakamoto PoW without replacing the hash algorithm.
It brings new use scenario to the original Nakamoto PoW computation, which can be used as an encoding algorithm.
This enables PoW computation used for data integrity checking in a permissionless storage network.
It is the first useful PoW beyond the consensus purpose without modification to Nakamoto Consensus.

% \begin{itemize}
%     \item How to use meaningless computing power in consensus calculation for useful tasks without changing the consensus algorithm.
%     \item How to ensure the security of file storage.
% \end{itemize}

\vspace{6mm}
\textbf{Chapter \ref{ch:storage}: Permissionless Storage as Security Engine}

This chapter designs a permissionless storage network system based on the useful PoW in chapter \ref{ch:consensus}.
We learn from existing storage systems and design the permissionless storage system to provide free security to the PoW-based blockchain.
Moreover, a Proxy Re-Encryption scheme is proposed to solve the cryptographic permission grant in the permissionless storage system.

\vspace{10mm}
\textbf{Chapter \ref{ch:sharding}: Sharding for High Performance Blockchain}

This chapter introduces sharding scheme without sacrificing the decentralization.
We proposed the Wider sharding scheme with a multi-chain data structure.
It provides the infinite shards for the transactions with high TPS and throughput.

% \begin{itemize}
%     \item How to improve the throughput of blockchain.
%     \item How to prevent the infinite expansion of full nodes.
% \end{itemize}

\vspace{6mm}
\textbf{Chapter \ref{ch:conclusion}: Conclusion and Future Work}

This chapter discusses our theory improvements for blockchain massive adoption and the future mission to implement.

\chapter{Preliminaries} % Write in your own chapter title
\label{ch:preliminaries} % Add a label in case you want to refer to the chapter.

We review the blockchain and cryptography-related knowledge in this chapter.
It covers the fundamental concepts of the blockchain and permissionless storage network.
It also introduces cryptography and distributed system theories such as consensus and sharding.

\section{Bitcoin}

Bitcoin was the first blockchain, created by Satoshi Nakamoto in 2008.
The system was initially proposed in a whitepaper\cite{2008Bitcoin} and soon released along with its source code.
Several years later, people realized that the blockchain technology behind Bitcoin had huge potential for the future, attracting more research and study.
Bitcoin includes several essential elements, such as Proof of Work (PoW), the longest chain rule, a hash chain with blocks, block size limit, difficulty adjustment, digital signatures, Unspent Transaction Outputs (UTXO), and a gossip protocol.

\subsection{Hash Chain}

A hash chain is a data structure linking blocks with a cryptographic hash.
A block's data will be hashed into a fixed-size output(the hash value or message digest), which is placed in the next block.
The blocks can always find their previous block unless it is the first block in the chain.

Due to the hash algorithm attribute, any modification of the block data in bits will result in the hash output change.
This prevents the tamper of the previous blocks.
To modify the blockchain, the following blocks must be changed as well.
The chain can be forked.
Hash chain is used in blockchain technology to ensure the integrity and immutability of data.

% applied to data to produce a fixed-size output (the hash value or message digest), which is then used as input for the next hash function in the chain. Each hash function in the chain takes as input the output of the previous hash function, creating a "chain" of hash values that are all dependent on each other.

% In a blockchain, each block contains a hash of the previous block in the chain, creating a chain of blocks that are linked together by their hash values. The hash of the entire chain is also included in the block, ensuring that any attempt to tamper with the data in a single block would be detected and rejected by the network.

% Hash chains are also used in password-based authentication protocols, where a user's password is hashed and the resulting hash value is used as input to another hash function in the chain. This process is repeated a certain number of times, creating a long hash chain that is difficult to reverse-engineer or attack by brute force.

\begin{figure}[!t]
    \centering
    \includegraphics[width=1\textwidth]{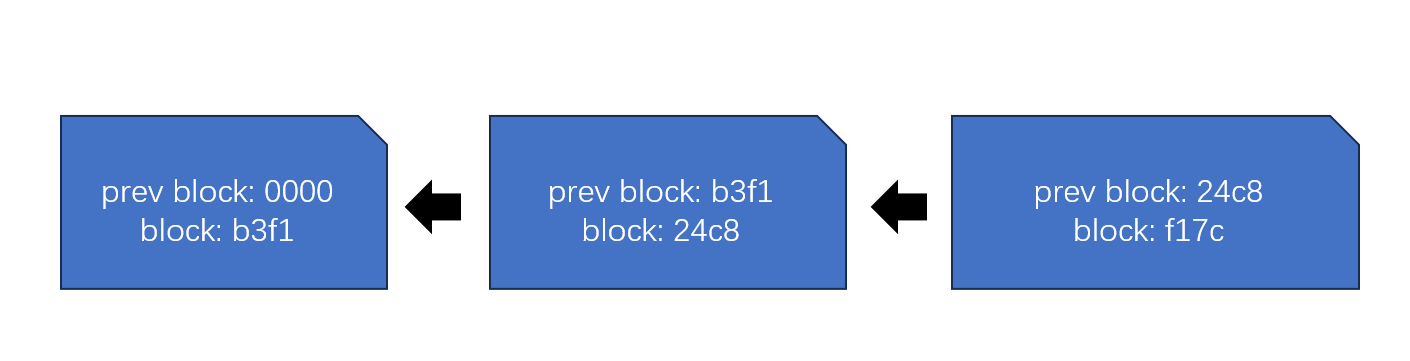}
    \caption{Hashchain: a block links to the previous block with hash value}
    \label{fig_hashchain}
\end{figure}

In Bitcoin, a block contains a block header and transactions.
The previous block hash is within the block header, shown in Figure~\ref{fig_hashchain}.
The header also contains the Merkle tree of the transactions and the mining information.
In the permissionless blockchain system, the miner generates the new block.
A blockchain has additional rules to prevent the miners from arbitrarily appending, such as the PoW difficulty and longest chain rule.

\subsection{Mining for Consensus}

The miners run the PoW algorithm to win the coins reward in Bitcoin.
However, this incentive is only the mechanism to attract miners to contribute their computation.
The goal of mining is to achieve an agreement for the next block among the miners.

In Bitcoin, a rule is set that a block contains new coins (coinbase).
Whoever mined the new block can get the coins as a reward.
The consensus is formed during the coin reward competition, just like voting.
The miners run the Proof of Work algorithm to propose new blocks.
Once a new block is found, the miner should announce the new block information to the world as soon as possible.
This announcement will stop the other miners from calculating the block at the same height and continue to look for the next block based on the announced block.

\begin{figure}[!t]
    \centering
    \includegraphics[width=0.8\textwidth]{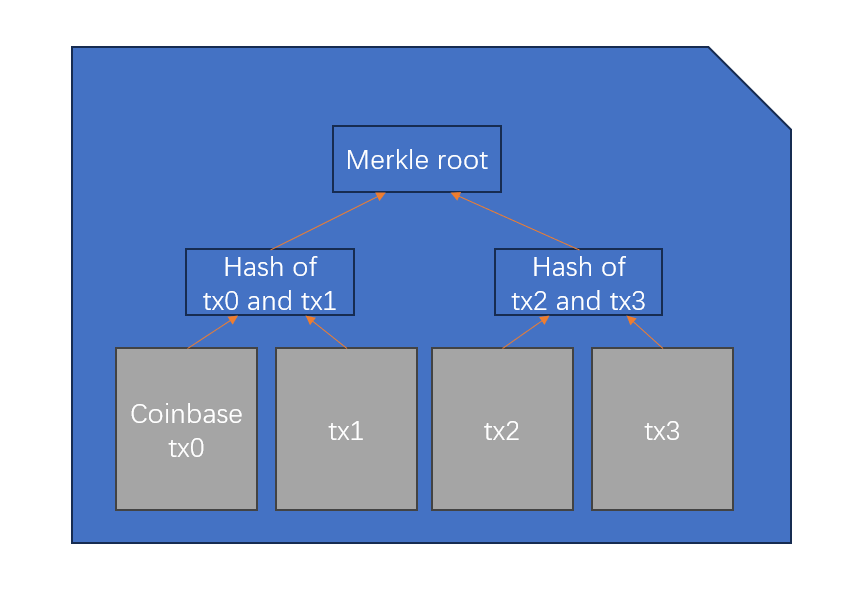}
    \caption{Merkle root of transactions in the block header}
    \label{fig_btc_transactions_merkle_root}
\end{figure}

Miners collect as many transactions as ordered by the transaction fee and calculate the Merkle root of the transactions, shown in Figure~\ref{fig_btc_transactions_merkle_root}.
They run the PoW to look for new blocks with difficulty criteria.
The mining difficulty prevents miners from creating large amounts of new blocks or forking the current blockchain.
Table~\ref{btc_block_header} shows the data fields to construct a block header.

\begin{table}[h!]
    \centering
    \caption{Bitcoin block header fields}
    \label{btc_block_header}
    \begin{tabular}{| c | c | c |} 
        \hline
        Field & Purpose & Size (Bytes)  \\ 
        \hline
        Version & Block version number & 4 \\
        \hline
        hashPrevBlock & 256-bit hash of the previous block header & 32  \\
        \hline
        hashMerkleRoot & 256-bit Merkle root of all of the transactions & 4 \\
        \hline
        Time & Current block timestamp as seconds since 1970-01-01 & 4 \\
        \hline
        Bits & Current target in compact format & 4 \\
        \hline
        Nonce & 32-bit number (starts at 0) & 4 \\
        \hline
    \end{tabular}
\end{table}

\subsection{Longest Chain Rule}

The miners can run the PoW algorithm in the hash chain to append new blocks.
If two miners find the new blocks almost simultaneously, the chain is forked.
The longest chain rule is used to detect which fork the miners should base on to generate the next new block.

The longest chain rule is simple: the miners should continue mining on the longest chain fork.
When there are more than one forks with the same height, the miners can choose any of them.
Depending on the choice of miners, different amounts of computation will be distributed to various forks.
One of the forks can randomly gain more computation and become the longest chain.

The longest chain rule causes the blockchain to have no finality.
Thus, the output of a coinbase transaction cannot be spent for 100 blocks to avoid attacking.
However, this attribute encourages a truly decentralized blockchain system, as the majority forms the consensus.

The longest chain rule ensures that the blockchain remains secure and tamper-proof, as any attempts to modify a previous block would require recalculating all subsequent blocks in the chain, making it computationally infeasible.
The Longest Chain Rule is a key feature of many blockchain networks, including Bitcoin and PoW Ethereum.

\subsection{Bitcoin Halving}

\begin{figure}[!t]
    \centering
    \includegraphics[width=0.8\textwidth]{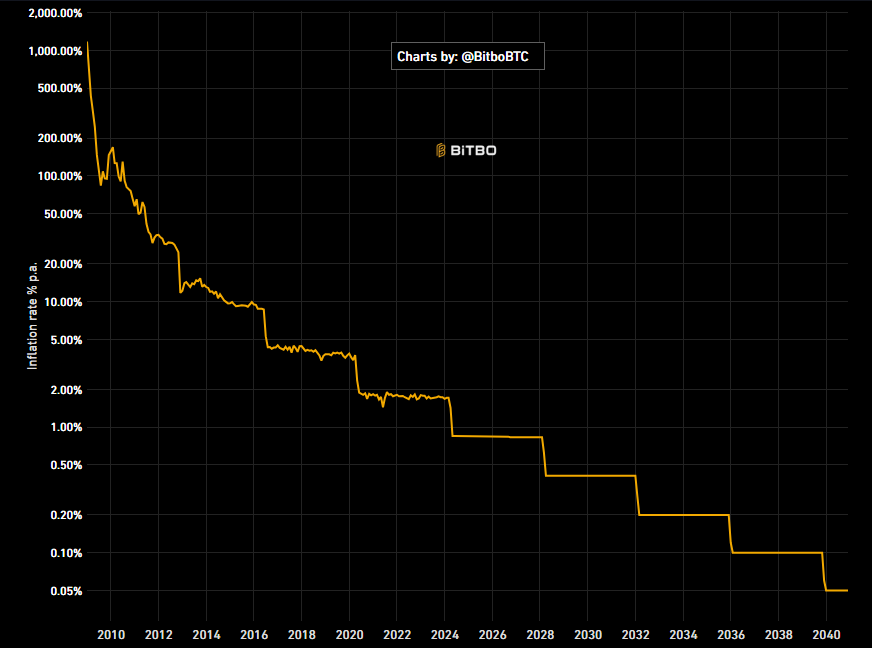}
    \caption{Bitcoin inflation since 2009}
    \label{fig_btc_inflation}
\end{figure}

The Bitcoin halving is an important mechanism in the Bitcoin blockchain system.
The Bitcoin halving refers to the fact that the block reward for Bitcoin is halved every 210,000 blocks.
Before the first halving, the reward for each block is 50 coins; after the first halving, the reward for each block is 25 Bitcoin.
The total supply of Bitcoin is 21 million.
Half of the total coins are rewarded to the miners at the first halving.

After the halving, the same computation would get less reward, which increases Bitcoin production cost.
However, due to market competition, the price of Bitcoin would not double at once.
Some of the miners with low energy efficiency mining equipment have to quit mining.

According to Figure~\ref{fig_btc_inflation} (data source \cite{bitbo_inflation}), the halving always comes with a rise in Bitcoin price.
The produced coins must be sold at a price to cover the production cost.
When the coin production rate halved, the price raising helped the Bitcoin system remain secure.
In theory, the higher price stands for higher market value.
This brings Bitcoin miners money to purchase electricity from the real world.

\subsection{Difficulty Adjust Algorithm}

The rise in the Bitcoin price has attracted more people to join in mining.
As more computation is put into mining, the interval of block generation narrows down.
Bitcoin has a mechanism to adjust the new block generation criteria.
The difficulty of the PoW algorithm is adjusted every 2016 block.

Bitcoin difficulty adjustment ensures the miners generate a block every 10 minutes.
The algorithm enables Bitcoin to adapt to the entire network's computing power automatically.

\subsection{Block Size Limition}

In Bitcoin, each block on the chain is data with a limited size.
The size is no more than 1MB.
This block size limitation with the average 10-minute block interval leads to around 7 TPS blockchain performance.
Although it is proposed to increase the block size and decrease the interval, Bitcoin still keeps the original settings.

The 1MB block size also has its strength, making the blockchain full node grow linearly.
Unlike the other recent blockchain systems, including Ethereum, a full node requires far more computation and storage capacity than the Bitcoin system.
The gentle growth of a Bitcoin full node makes it a better-decentralized system than the others, as many individuals can afford to run a full node.

\subsection{UTXO}

\begin{figure}[!t]
    \centering
    \includegraphics[width=0.8\textwidth]{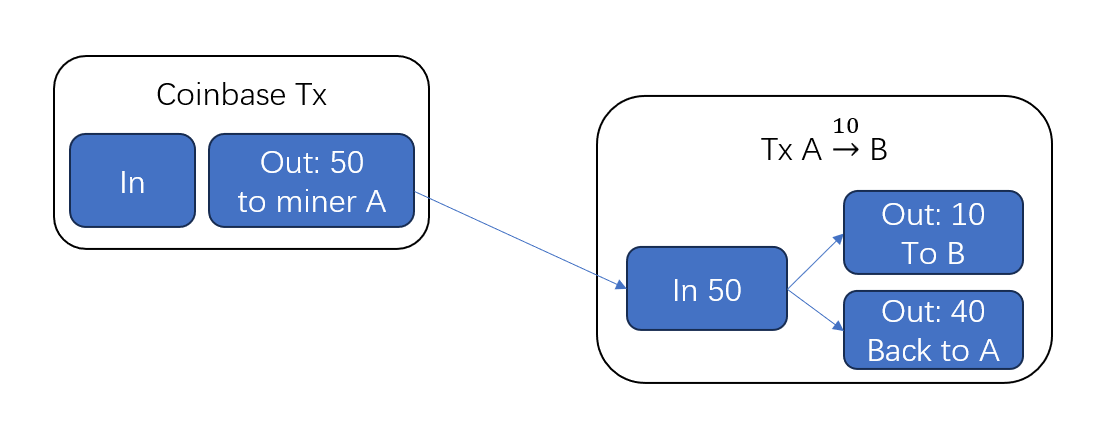}
    \caption{Unspent Transaction Output}
    \label{fig_btc_utxo}
\end{figure}

Unspent Transaction Output (UTXO) is the Bitcoin design to handle coin balance.
In Bitcoin, the coin assets are distributed to the miners when a new block is generated.
In the new block, the first transaction is called a coinbase transaction, which creates the coins from nothing and rewards the miner.

The miner who earned the coinbase can spend the coins after 100 blocks by creating the UTXO transaction and signing with his private key.
A UTXO contains the inputs and outputs.
The miner may use the coinbase as the input, pay someone with the amount, and pay himself for the change.
In this case, a transaction has one input and two outputs, like Figure~\ref{fig_btc_utxo}.

\subsection{Transaction and Signature}

Transactions in Bitcoin contain the UTXO data structure and the witness data.
Witness refers to the signature, a UTXO input requires the witness to spend the value from the previous output.
It is possible to create an output that anyone can claim without any authorization.
It is also possible to require that an input be signed by multiple keys or redeemable with a password instead of a key.

The cryptographic signature is used to sign the Bitcoin transaction to spend the output.
ECDSA is used for signatures in early Bitcoin.
Schnorr is introduced for better multi-sign after the taproot upgrade.

\subsection{Gossip Protocol}

Bitcoin uses peer-to-peer (P2P) communication to ensure people can set up the full node behind the firewall.
However, the blockchain is desired for broadcast communication rather than P2P for both new block propagation and transaction sending.
Gossip protocol is the robust broadcast protocol used by Bitcoin.

Both the 10-minute block interval and Gossip protocol ensure the new blocks and transactions arrive in the full nodes worldwide.
The spreading of information reduces the blockchain forks.

\section{Ethereum}

Based on the concept of Bitcoin, Ethereum\cite{buterin2014next} was proposed in 2014.
It extends the blockchain by introducing smart contracts\cite{wood2014ethereum}, incorporating a Turing-complete programming language into the blockchain.
In this section, we will introduce the concepts of the Ethereum Virtual Machine (EVM), Proof of Stake (PoS), the Account Model, and the Merkle Patricia Tree (MPT).

\subsection{EVM and Gas Limit}

Ethereum Virtual Machine (EVM) is the core component of Ethereum.
While Bitcoin allows the user to send coin assets over a permissionless blockchain, Ethereum allows users to create customized apps named smart contracts.

The smart contract is immutable.
The Turing complete EVM bytecode defines the behavior of a smart contract.
Due to the famous halting problem\cite{turing1936computable}, it is impossible to analyze the EVM bytecode to prevent a dead loop.
Thus, Ethereum introduces a gas fee to avoid smart contract execution consuming too much computation.
Meanwhile, the gas price is used to adjust the overall cost of blockchain usage.
The user may delay the on-chain operation if the current gas price is high.

\subsection{PoS and Staking}

Ethereum started with a PoW algorithm similar to Bitcoin, then switched to PoS as planned in 2022.
With PoS consensus, Ethereum no longer consumes large amounts of electricity for mining.
The finance resource replaces the electricity resource, as ETH coins can be staked for the reward and secure the blockchain.

Unlike electricity, which is used for computation, finance resources are not burned but require interest.
As the blockchain does not generate value, the end-users pay to afford the staking interest.
Similar to PoW blockchain, PoS also has a high security cost.

\subsection{Account Model and Global State}

Different from the UTXO model, the account model is used in Ethereum.
When transferring ETH coins from one account to another, the balance is deducted from the sender and increased for the receiver.
Blockchain does not alter the current state but creates a new state with changes.
Merkle Patricia Tree (MPT) works great with the state and provides minimal change to the data structure between block states.

Ethereum uses the state not only for coin balance but also for the smart contract.
The smart contract stores variables and K/V mapping in the global state.
In Ethereum, each block corresponds to a global state.
As the block is generated with more transaction input, more K/V data is put into the state.
The size of the global state keeps growing.
It is called an Ethereum state explosion.

\subsection{MPT Trees}

Merkle Patricia Tree (MPT) data structure is used to represent the Ethereum transactions, global state, and receipts.
Each kind of data is stored in a delegated MPT tree, such as the transaction tree and the state tree.
A block contains the roots of the three trees.

The MPT is effective in a single node, avoiding duplicate data.
However, it requires calculating the overall K/V pairs into a root hash.
It makes the MPT hard to split over the shards.

\section{State Machine}

By studying both Bitcoin and Ethereum systems, we can abstract different blockchains into a similar model.
The state machine, a computer science concept, is often used to model blockchains.
A current state transitions to a new state after applying an input.

\begin{figure}[!t]
    \centering
    \includegraphics[width=1\textwidth]{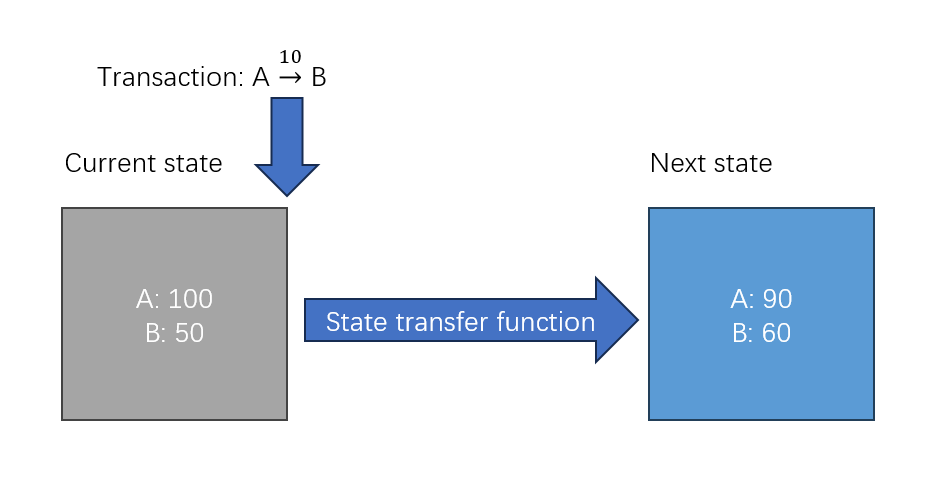}
    \caption{Transaction as the State Machine}
    \label{fig_state_machine_transaction}
\end{figure}

In blockchains, a block corresponds to a state, representing the collection of account balances or the global state of smart contracts at that block height.
All the transactions in the block are regarded as the input to the state machine.
The state transitions to a new state by applying a new block, as shown in Figure~\ref{fig_state_machine_transaction}.

% From computer program level, we learned that a blockchain is a distributed system and every node run the program locally.

% Program = data structure + algorithm

% The fundamental task of blockchain program is to generate new block on the current chain data structure.
% The consensus algorithm runs over the hash chain, keeping the world wide nodes sharing a chain of consistent blocks.

\subsection{State Transfer Function}

An important concept of state machines is the state transition function.
The state transition function is fixed in blockchains without smart contracts, such as Bitcoin.

When Alice transfers money to Bob, Alice's balance decreases and Bob's balance increases. At the same time, Alice's account balance cannot be less than zero, and the transfer amount must be positive (with a little transaction fee).

\begin{figure}[!t]
    \centering
    \includegraphics[width=1\textwidth]{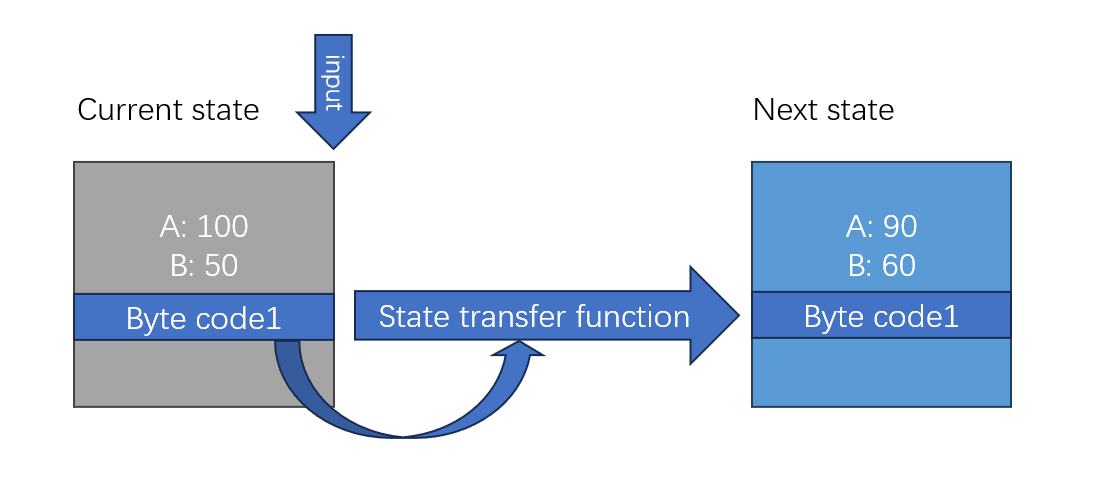}
    \caption{State Transfer Function on the State}
    \label{fig_state_machine_bytecode}
\end{figure}

In a blockchain like Ethereum, each smart contract is defined by the end-users with programming language.
This indicates that the state transfer function is defined on the state, shown in Figure~\ref{fig_state_machine_bytecode}.
A smart contract with a poor design may cause unpredictable results.
Due to the immutable of smart contracts, it is impossible to fix the bug in a deployed smart contract.
However, people use proxy technology to build the upgradable smart contract as a walkaround.

\section{Traditional Consensus}

The consensus algorithm was studied long before the blockchain system.
It is widely used in highly available systems, such as databases and cloud computing.
In this section, we review the traditional consensus algorithm overall.

\subsection{Byzantine Generals' Problem}

The Byzantine Generals' Problem\cite{lamport2019byzantine} is a theoretical problem in distributed computing and communication that was first introduced in a paper published in 1982 by Leslie Lamport, Robert Shostak, and Marshall Pease.
The problem describes a situation where a group of Byzantine generals needs to agree on a coordinated attack or retreat plan.
Still, they are separated by distance and can only communicate with each other through unreliable messengers.

In this scenario, some generals may be traitors who want to undermine the plan and confuse the loyal generals by sending false messages.
The loyal generals need to come up with a consensus on whether to attack or retreat while accounting for the possibility of traitors and faulty communication channels.

The Byzantine Generals' Problem is fundamental in distributed systems, as it highlights the challenge of achieving consensus in a decentralized network where nodes may be faulty or malicious.
The problem is often used as a benchmark for evaluating the resilience and effectiveness of different consensus algorithms, including those used in blockchain technology.

\subsection{Paxos and Raft}

Paxos\cite{lamport2001paxos} and Raft\cite{moad2005living} are two consensus algorithms used to reach agreement among distributed nodes in a network.
Both algorithms are designed to solve the consensus problem, which arises when a group of nodes needs to agree on a common value or state in the presence of faulty or malicious nodes.

Paxos\cite{lamport2001paxos} is an older consensus algorithm that was first introduced in the late 1980s by Leslie Lamport.
The algorithm is based on a sequence of rounds, during which nodes communicate messages and propose a shared value.
In each round, a node acts as a leader and proposes a value to be adopted by the group.
The other nodes can either accept or reject the proposal and if the proposal is rejected, the leader will need to retry with a new proposal in the next round.
Paxos is known for its complexity and difficulty to implement correctly, but it is still widely used in many distributed systems today.

Raft\cite{moad2005living} is a newer consensus algorithm that was introduced in 2014 by Diego Ongaro and John Ousterhout.
The algorithm is based on a leader-follower model, where one node acts as a leader and coordinates the consensus process among the other nodes.
In Raft, the leader manages the log of committed values, and the other nodes follow the leader's instructions to maintain consistency.
If the leader fails or becomes unavailable, a new leader is elected through a voting process.
Raft is designed to be simpler and easier to understand than Paxos, making it a popular choice for many distributed systems.

Both Paxos and Raft are widely used in practice, and each has its strengths and weaknesses.
While Paxos is more flexible and can tolerate a higher number of faulty nodes, it is also more complex and difficult to implement.
Conversely, Raft is easier to understand and implement but may not be as fault-tolerant as Paxos in some scenarios.
Ultimately, the choice of consensus algorithm will depend on the specific requirements and constraints of the distributed system being developed.

\subsection{PBFT}

PBFT\cite{castro1999practical} stands for Practical Byzantine Fault Tolerance.
It is a consensus algorithm that aims to solve the Byzantine Generals' Problem, which is a problem in distributed computing where multiple nodes must agree on a single value despite the presence of faulty nodes.

PBFT is designed to be used in a network of computers, where each computer is a node that participates in the consensus process.
In PBFT, each node has its own copy of the transaction log, which records all the transactions that have occurred in the network.

The consensus process in PBFT consists of three phases: the pre-prepare, prepare, and commit phases.
In the pre-prepare phase, the node that initiates the transaction sends a message to all other nodes, asking them to prepare for the transaction.
In the preparation phase, each node verifies that the transaction is valid and then sends a message to all other nodes to indicate that it is prepared to commit the transaction.
Finally, in the commit phase, each node sends a message to all other nodes to indicate that it is committing the transaction.

If a node fails to respond or provides an invalid response, it is considered faulty, and the algorithm continues without it.
PBFT requires that more than two-thirds of the nodes in the network be correct for the algorithm to work correctly.

Overall, PBFT is a well-known consensus algorithm that is used in many blockchain systems and distributed databases.
It is designed to provide a high degree of fault tolerance while ensuring that transactions are processed quickly and efficiently.

\section{Permissionless Consensus}

% 我们把区块链共识和传统的分布式系统共识区分开来, 主要还是因为使用的场景有很大的不同

% 传统的分布式系统共识大多数在permissioned网络中, 而区块链所需要的共识算法运行在无许可的网络环境中
% 这一改变使得大多数传统的共识算法无法适应区块链的运行环境
% 传统共识算法多基于多轮通信, 对于小组成员有限的情况下, 通信复杂度是可控的
% 区块链所需要的共识, 需要去中心化, 意味着需要在上万个节点, 甚至更大的规模下运行, 基于多轮通信的共识算法基本无法满足区块链的共识需求

Blockchain consensus refers to the mechanism or protocol used to achieve agreement among nodes in a distributed network about the validity and order of transactions or blocks.
In a blockchain network, all nodes maintain a copy of the ledger and work together to ensure its integrity and security.
The consensus algorithm is essential to ensuring that all nodes agree on the same network state.

The two most common consensus algorithms in blockchain networks are Proof of Work (PoW) and Proof of Stake (PoS).
PoW is the original consensus algorithm used by Bitcoin, and it requires nodes to perform mathematical calculations to validate transactions and create new blocks.
PoS, on the other hand, requires nodes to hold a certain amount of cryptocurrency as a stake in the network, which they can lose if they act maliciously.

Other consensus algorithms include Delegated Proof of Stake (DPoS), Practical Byzantine Fault Tolerance (PBFT), and Raft.
DPoS is used by networks such as EOS and requires stakeholders to elect delegates who validate transactions and create new blocks on their behalf.
PBFT is used by permission networks such as Hyperledger Fabric and relies on a set of nodes to validate transactions and reach consensus through a voting process.
Raft is a consensus algorithm used in permission blockchain networks that relies on a leader node to coordinate the network and validate transactions.

The choice of consensus algorithm depends on the requirements of the network, including security, scalability, energy efficiency, and decentralization.
While PoW is secure and decentralized, it consumes a lot of energy waste and has scalability issues.
PoS is more energy-efficient and scalable but may be less decentralized.
Other consensus algorithms offer different trade-offs between these factors and may be better suited for certain use cases.

\subsection{Proof of Work}

\begin{figure}[!t]
    \centering
    \includegraphics[width=1\textwidth]{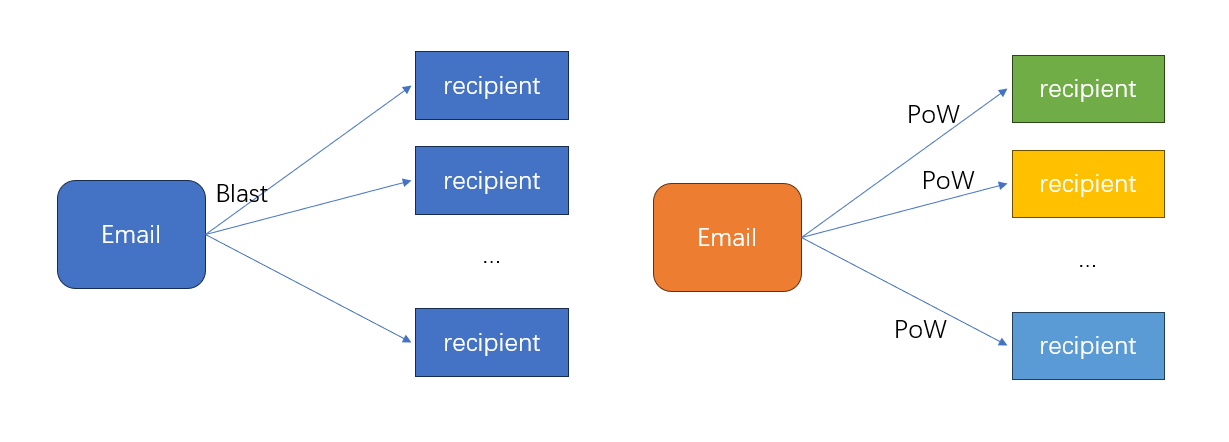}
    \caption{Email Blast Spam and PoW}
    \label{fig_email_spam_and_pow}
\end{figure}

Proof of Work was first proposed in 1992 last century.
It was proposed far before Bitcoin.
At that time, PoW is used for email anti-spam\cite{dwork1992pricing}.
Due to the simplicity, anyone can blast an email to arbitrary recipients at a very low cost quickly, as shown in Figure~\ref{fig_email_spam_and_pow}.
This caused an email spamming issue.
The email system delivers and stores huge amounts of emails, most of which are junk emails.
The PoW can be applied to prove that the sender applied a heavy computation on the email.
This could indicate the importance of the email in a certain way.
The email client software may use it as a factor to classify if it is junk mail.
However, due to the success of the machine learning approach, the Bayes method is applied in the email anti-spam in most client software.
Later, in 1997, Adem Back proposed to use PoW computation as the postage of the email\cite{Hashcash}.
This takes one more step to treat the PoW computation as a sort of money.

However, the power of the PoW algorithm has finally made a breakthrough.
In 2008, Bitcoin\cite{2008Bitcoin} used PoW as the consensus algorithm.
It majorly solves the Sybal Attack and Double Spending Attack in a permissionless network.
Simply, the consensus is voting.
In the cyber world, especially the pure cryptographic permissionless network, identities can be created at a very low cost.
A simple program can generate an arbitrary number of public/private key pairs.
Consensus with PoW uses CPU computation as the voting criteria, like one CPU, one vote.
More details about the Nakamoto Consensus will be described in section \ref{nakamoto_consensus}.

\subsection{Proof of Stake}

Proof of Stake (PoS) is the alternative consensus that was first proposed by Peercoin\cite{king2012ppcoin} in 2012.
Due to the heavy electricity consumed by Bitcoin and PoW, the idea of using financial resources to replace electricity resources has gained widespread attention.
The early PoS is bothered by the nothing at stake issue.
In Proof of Stake, if there is a fork in the chain, the optimal strategy for any validator is to validate on every chain so that the validator gets their reward regardless of the outcome of the fork.

After many years of research and development, PoS is generally mature enough to adopt.
Ethereum switched from PoW to PoS\cite{2017Casper}, making PoS the second blockchain consensus in wide usage.
However, successful switching is just the beginning, and the impact of PoS is still under observation.

In PoW, the electricity resource is consumed in exchange for coins.
In PoS, the staked capital is not burned; reversely, the capital gains interest.
Due to the liquidity of financial resources, capital will always look for a higher interest reward.
Thus, Ethereum must provide an APR higher or equal to the existing rate of interest, e.g., the traditional bank.
Currently, the more ETH coins are in staking; the fewer coins are used for transaction gas.
Will this cause a navigated impact on the Ethereum ecosystem?
We are still determining this.

Another issue we observed from the Ethereum PoW/PoS switching is the security cost.
Although lots of electricity was saved after the merge, the end-users did not benefit from the change.
The users are still required to pay ETH coins as a fee to use the Ethereum network, but the coin price and the amount remain the same.
Therefore, the user's cost for blockchain usage has not decreased after switching to PoS.
In other words, PoS still keeps the high blockchain security cost for the chain security.

\subsection{Proof of Space}

Besides PoW and PoS, another permissionless consensus was proposed with less energy cost.
Different from the PoS, Proof of Space (sometimes called Proof of Capacity, PoC) also uses non-finance resources for consensus.
Unlike PoW, PoSpace uses storage instead of computation.
The concept was formulated in 2013 by Dziembowski et al.\cite{Dziembowski2013PoSpace}\cite{Dziembowski2015PoSpace} and (with a different formulation) by Ateniese et al.\cite{2013Proofs}\cite{2014Proofs}

The Chia blockchain adopted Proof of Space in 2018 and was launched in 2021.
The PoC does not store useful data but only uses the storage resource for the consensus.
The miners pre-calculate(plot) the proofs and store them on disk to avoid exhaustive computation like PoW.
The proofs are sorted on disk so the miner can find the pre-calculate proof in a very cheap operation and submit it.
This avoids the energy cost in the long run, but during the plot, it still costs the energy.

\subsection{Consensus Algorithms Comparsion}

PoW has better decentralization characteristics but is more energy-consuming.
PoS almost does not consume energy, but users still incur interest costs for staking.
PoC (Proof of Space) is a scheme that uses space to exchange for time, consuming less energy but still occupying resources.

\begin{table}[h!]
    \centering
    \caption{Consensus algorithm comparsion}
    \label{consensus_comparsion}
    \begin{tabular}{| c | c | c |} 
        \hline
        Consensus algorithm & Energy comsumption & Description \\ 
        \hline
        Proof of Work & High & Hash computation based \\
        \hline
        Proof of Stake & Low & Asset staking based \\
        \hline
        Proof of Space & Medium & Disk storage space based \\
        \hline
        PBFT & Low & Voting based \\
        \hline
    \end{tabular}
\end{table}

The modern blockchain combines PBFT and PoS to reduce block generation interval time.
This is a kind of hybird consensus.
Table~\ref{consensus_comparsion} shows the comparsion for different consensus algorithms.

\section{Nakamoto Consensus}
\label{nakamoto_consensus}

% 从数据结构的角度来看，区块链就是哈希链
% 我们认为构成比特币的最重要三元素，是哈希链，最长链法则，以及工作量证明。
% 本section我们讨论哈希链的知识，以及最长链法则
% 工作量证明被中本聪创造性的应用于共识，我们会在2.2和2.5节中更深入的讨论。

% The Longest Chain Rule is a fundamental concept in blockchain technology that determines which block is considered the valid one in a blockchain network. When multiple miners or nodes create new blocks simultaneously, the network needs a way to reach consensus on which block should be added to the blockchain.

% According to the Longest Chain Rule, the blockchain with the longest valid chain of blocks is considered the true blockchain. A valid block in a blockchain network is one that meets certain criteria, such as having a valid Proof of Work or being signed by the correct private key.

% When two or more blocks are created at the same time, miners work to add them to the blockchain network. Each node will choose the block with the longest chain of valid blocks, adding it to their own copy of the blockchain. Once a majority of nodes agree on the longest chain, it becomes the accepted version of the blockchain for the entire network.

Nakamoto Consensus is the core technology behind Bitcoin.
It not only relies on PoW but also includes the following key concepts:
Hash Chain Structure: Blocks are linked together through hash values, forming an immutable chain. Each block contains the hash value of the previous block, ensuring the security and order of the blockchain.
Longest Chain Rule: All nodes in the network always recognize the longest valid chain as the correct ledger.
This means that when multiple chain forks appear, nodes will choose the chain with the most work (the longest chain) to ensure consensus consistency.
By combining PoW, the hash chain structure, and the longest chain rule, the Nakamoto consensus achieves a secure, reliable, and decentralized permissionless ledger system.
This not only solves the Double Spending problem but also ensures data consistency and reliability in distributed systems.

Nakamoto Consensus has the best decentralization characteristics.
Its disadvantage is that it consumes a large amount of energy, and due to the existence of mining pools, it does not achieve complete decentralization.
We believe that there is still room for improvement in Nakamoto Consensus. 
By improving the mechanism to encourage more people to mine independently,  Nakamoto Consensus can achieve a more complete decentralization. This is something other consensus algorithms do not possess.

\section{Cryptography}

\subsection{Hash Algorithm}

A hash algorithm is a mathematical function that takes input data of arbitrary length and produces a fixed-size output called a hash.
The process of computing a hash involves passing the input data through a set of algorithms that generates a unique digital fingerprint of the data.
This fingerprint is a string of characters typically represented in hexadecimal format and is used to verify the integrity of the original data.

Hash algorithms are commonly used in computer security, cryptography, and data management to ensure the authenticity and integrity of digital information.
They are used to verify that data has not been tampered with or corrupted in transit, to secure passwords and other sensitive information, and to enable efficient data retrieval and storage.

Hash algorithms are designed to be one-way functions, meaning it is practically impossible to reverse-engineer the original input data from the hash output.
They are also designed to produce unique hash outputs for different input data, with the probability of two different input data producing the same hash output (known as a collision) being very low.

Some commonly used hash algorithms include MD5, SHA1, SHA2, and SHA3, each with different security and cryptographic strength levels.

\subsection{Group}

In mathematics, a group is a set equipped with a binary operation that combines any two elements to form a third element, subject to certain axioms.
These axioms include closure, associativity, identity, and inverse.

Closure means that the operation must combine two elements from the group to produce another element in the group.
Associativity means that the order in which the operation is performed does not affect the result. Identity means that there is an element in the group that, when combined with any other element, produces the same element.
Inverse means that for every element in the group, there is another element that, when combined with the first element, produces the identity element.

Groups are used in a wide range of mathematical disciplines, including algebra, geometry, and number theory.
They are also used in many areas of computer science, including cryptography and coding theory.
The study of groups is known as group theory.

\subsection{Discrete Logarithm}

Discrete logarithm is a mathematical problem that arises in cryptography, particularly in public-key cryptography.
It is the opposite of the exponential problem, which asks for the result of raising a fixed base to an unknown exponent.
In the discrete logarithm problem, one is given a base, a modulus, and a result, and the goal is to find the exponent that yields that result.

More formally, given a prime number $p$, a base $g$, and an integer $y$, the discrete logarithm problem is to find an integer $x$ such that:

\begin{equation}
    g^x = y (mod\ p)
\end{equation}

The discrete logarithm problem is considered difficult to solve for large values of $p$, $g$, and $y$, and is believed to be intractable for sufficiently large values.
As a result, it is widely used in cryptographic systems, such as the Diffie-Hellman key exchange and the Digital Signature Algorithm (DSA), to provide security against eavesdropping and other attacks.

\subsection{Elliptic Curves}

Elliptic curves are a type of mathematical curve defined by an equation of the form, where a and b are constants.

\begin{equation}
    y^2 = x^3 + ax + b
\end{equation}

They have the property that any straight line drawn on the curve intersects it in exactly three points.
This makes them useful in cryptography, as they can be used to create public-private key pairs, which are widely used to secure data in modern digital communications.

In elliptic curve cryptography, the key pair consists of a private key and a public key, which are mathematically related to each other by a point on the curve.
The private key is kept secret by the user, while the public key can be freely shared with others.
The security of the system relies on the fact that it is very difficult to compute the private key given only the public key.

Elliptic curves are also used in other areas of mathematics, including number theory, algebraic geometry, and algebraic topology.
They have interesting geometric properties and can be used to study various mathematical phenomena.

\subsection{Digital Signature}

A digital signature is a cryptographic mechanism used to authenticate the sender and integrity of a digital message or document. It provides a way to ensure that the message or document has not been tampered with during transmission and that the sender is who they claim to be.

A digital signature is created using the sender's private key and the message or document being signed. The private key is used to create a unique digital signature for the message, which can be verified using the sender's public key. If the signature is valid, it proves that the message has not been altered since it was signed and that the sender indeed signed it.

The process of creating a digital signature involves several steps, including hashing the message or document to create a fixed-length message digest, encrypting the message digest using the sender's private key, and attaching the resulting signature to the original message. The message's recipient can then use the sender's public key to verify the signature and ensure the message has not been tampered with.

Digital signatures are widely used in electronic transactions, such as online banking and e-commerce, as well as in secure communications, such as email and messaging apps.
They provide a way to ensure the authenticity and integrity of digital communications and are an essential component of modern cryptography.

\subsection{Public Key Encryption}

Public key encryption has advantages in key management compared with symmetric key encryption.
Users only need to keep their secret keys safe instead of memorizing many passwords.
The most commonly used public key encryption schemes are RSA and ElGamal (including ECC).
In RSA, we can either encrypt with a public or secret key and decrypt with the other.
In ElGamal, the public key is for encryption via Equation~(2.2), and the secret key is for decryption via Equation~(2.3).
Public key encryption allows anyone to create an encrypted message and send it to the secret key owner to establish secure communication.
We mainly focus on ElGamal here:
\begin{equation}
a, r \in \mathbb{Z}_p,
\end{equation}
\begin{equation}
pk_A = g^a,
\end{equation}
\begin{equation}
 sk_A = a.
\end{equation}

where $pk_A$ is the public key, and~$sk_A$ is the secret key.
Ciphertext $c$ is encrypted with the public key $pk_A$ and the clear message $m$ via Equation~(4):
\begin{equation}
 c = \langle c_0, c_1 \rangle = \langle g^r, m \cdot pk_A^r \rangle = \langle g^r, m \cdot g^{ar} \rangle .
\end{equation}

The secret key is required for decryption via Equation~(5):
\begin{equation}
 m = \frac{c_1}{(c_0)^{sk_A}} = \frac{m \cdot g^{ar}}{(g^r)^{a}}.
\end{equation}

The ElGamal scheme satisfies the CPA security.
Given the same input $m$, the output $c$ is different each time according to the random value $r$.
To achieve CCA security, validation is required before the decryption.
It detects if the adversary has modified the ciphertext.

\section{Chapter Summary}

In this chapter, we presented preliminary knowledge of blockchain and cryptography.
The details of the most popular blockchain, including Bitcoin and Ethereum, are explained.
They are also modeled in the state machine.
We also talked about the background of the consensus algorithms, comparing PoW with other consensus algorithms.
We introduced the basic knowledge used for blockchain in cryptography.
The hash algorithms are used in PoW, and the signatures
like ECDSA and Schnorr are used in blockchain transactions.
These algorithms and schemes will be useful to understand the following chapters in this thesis.

\chapter{Useful Proof of Work} % 这章可以更详细的讲有用的共识算法
\label{ch:consensus}

Proof of Work based blockchains require a lot of computing power and energy to run the consensus algorithm.
This process can defend against attacks in permissionless blockchains, such as Double Spending.
Therefore, the security cost of PoW networks is enormous, and miners worldwide need to join and contribute computing power.
A long-term research direction in academia effectively utilizes computing power to form a consensus with useful tasks.
But most of them changed to new consensus algorithm instead of using hash based Nakamoto Consensus.

We propose to extend the Nakamoto PoW algorithm without modifying the Bitcoin PoW algorithm.
We reinterpret the meaning of a part of the hash result and use the PoW algorithm as a file encoding algorithm for blockchain file storage.
This extension allows the PoW algorithm to obtain another meaningful new use without modifying the Bitcoin consensus.

At the same time, this scheme can transfer the security cost of blockchain consensus to the permissionless storage system.
When many miners perform meaningful encoding computing tasks, the blockchain can obtain almost free computing power as a by-product to protect the chain's security, significantly reducing the security cost that users initially needed to bear.
Ultimately, users can use blockchain in a low-cost and secure manner, sweeping away the obstacle of the high cost of users, which is a prerequisite for the large-scale application of blockchain.

\section{Overview}

In the early days, PoW-based blockchains consumed most of the computing power by Bitcoin and Ethereum.
Due to the excellent decentralization properties provided by PoW, Bitcoin is still the benchmark for decentralization.
However, the problems of the PoW algorithm are also apparent.
For example, the emergence of mining pools has reduced the degree of decentralization and the huge energy consumption of the entire network for consensus security.
The consumption of energy and computing power by blockchains is the use of huge costs to protect the security of blockchains.

To this end, people have tried to find answers from multiple angles.
On the one hand, they are looking for consensus algorithms requiring little computing power.
On the other hand, people are also looking for useful Proof of Work.
With the deepening of research, some algorithms can already replace PoW: PoS uses staking instead of computing power, and PoC uses disk space instead of computing power.
However, PoS avoids large-scale power consumption, but the security cost still exists.
To attract asset staking, PoS blockchains must provide an annualized return rate higher than bank interest rates so that users are willing to stake assets on the blockchain.
Since the current blockchain security cost is ultimately transferred to coin buyers, users share the total security cost of the blockchain through the purchase of coins and the payment of gas during the use of the blockchain.
Similarly, although PoC blockchains do not consume a lot of electricity resources, the hard drives purchased by miners do not store meaningful user data but only keep the "lottery" data required for mining, which still has security costs and causes a certain amount of resource waste.

After some early PoS trials of blockchain projects, Ethereum also switched to the PoS algorithm in the middle of 2022. From the current results, switching to PoS saves a lot of power consumption, but users still need to pay expensive gas fees to use blockchains. The essence of this gas is the security cost of the blockchain. The transition of Ethereum to PoS cannot reduce user costs.

There are two directions for useful Proof of Work.
Since the Proof of Work based on hash used in the Bitcoin consensus cannot be applied to scientific computing or general computing, the main research direction is to explore the replacement of PoW algorithms: trying to derive consensus from useful computing, also known as PoUW.
On the other hand, the progress of directly using the hash computing power required by the Bitcoin consensus for useful purpose computing is not making much progress.
Even the founder of Ethereum once posted in 2019 that useful Proof of Work may not be feasible.
However, the inability to use hash for general computing does not indicate that there is no meaningful computing.
In this context, this paper proposes to extend the PoW algorithm, reinterpret the meaning of a part of the hash result, and use the PoW algorithm as a file encoding algorithm for blockchain file storage.
Since file storage has a wide range of usage needs, it can be deployed on a large scale.
As the research on this algorithm deepens, this simple algorithm breakthrough has the opportunity to bring revolutionary changes to the entire blockchain world.
Its ultimate impact is likely to accelerate the large-scale deployment of blockchains and significantly reduce the cost for users to use blockchains.
In essence, this depends on the impact of the algorithm extension on the security cost of blockchains, making the meaningless hash calculation in the consensus process valuable.

Specifically, why the PoW algorithm can be applied to the permissionless storage system and how to use the algorithm to build a storage system will be discussed in detail in Chapter~\ref{ch:storage}.

\section{Background}

\subsection{Security and Computation}

The security of cryptography is related to computing power.
This is because cryptographic techniques rely on the mathematical problems, and the difficulty of these problems is closely related to the computing capacity.
An attacker with unlimited computing power can crack these problems and break the cryptographic system.

The situation of blockchain consensus is similar.
PoW and other blockchain consensus algorithms lack final consistency, meaning there may be multiple chain forks of different lengths in a blockchain network. Suppose an attacker has an absolute computing power advantage, he could generate arbitrary long chain to confirm his transactions, and discard others' transactions.
This is a typical Double Spending Attack, which threatens the security of the blockchain.

Computing power is an important guarantee of the security of blockchain networks. In the PoW consensus algorithm, the higher the computing power, the higher the cost for attackers to launch attacks, reducing the success rate of attacks.

The higher the total computing power, the higher the proportion of computing power that attackers need to control, increasing the cost of attacks. Therefore, maintaining a relatively high total computing power is an important measure to ensure the security of blockchain networks.

The concentration of computing power will reduce the security of blockchain networks. If attackers only need to control the computing power of a few mining pools to reach a computing power share of more than 50\%, then the cost of attacks will be greatly reduced.

Decentralization is an important foundation for the security of blockchain networks. If blockchain networks are too centralized, attackers will find it easier to control the networks, reducing the security of blockchain networks.

\subsection{Bitcoin Security Cost Growth}

\begin{figure}[!t]
    \centering
    \includegraphics[width=1\textwidth]{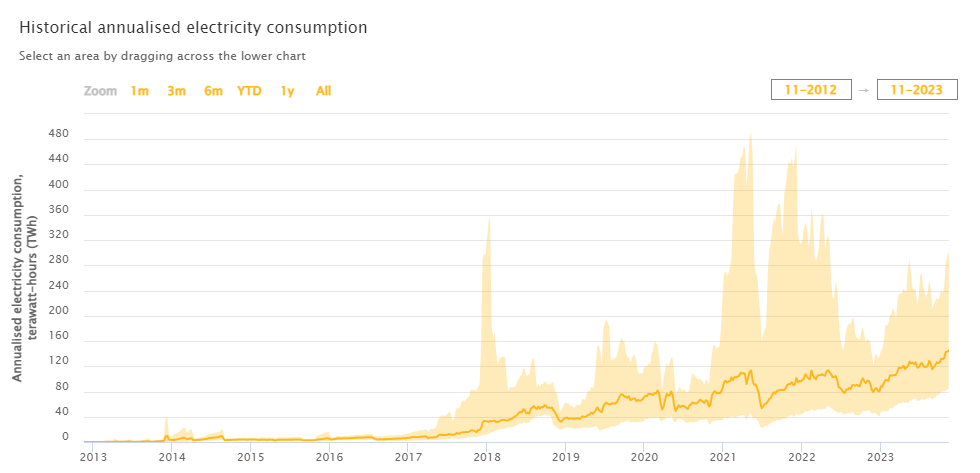}
    \caption{Bitcoin electricity cost in terawatt-hours annually by year}
    \label{fig_bitcoin_cost}
\end{figure}

Bitcoin has achieved a million-fold growth in price and market capitalization in its first decade.
Its success has exceeded that of a simple open-source software system, and even compared to commercial unicorns, Bitcoin's growth rate is beyond imagination.
Many factors are driving this miraculous growth, such as the first "bank" that does not need to trust a third party, the mysterious founder Satoshi Nakamoto, and many people who have become rich because of Bitcoin.

Here, we describe the growth flywheel brought by the consensus algorithm from a very technical perspective.
It is related to the cost of PoW.

The total coins supply limit and fixed production within a certain period make miners compete for the benefits of block production rights, resulting in internal competition: to increase the probability of mining coins, miners need to upgrade equipment and improve computing power, leading to an increase in total computing power (shown in Figure~\ref{fig_bitcoin_cost}, data source\cite{cbeci}).
After the total computing power increases, it will also make the number of coins rewarded by the miners' existing computing power decrease.
The problem with this mechanism is that the cost of producing a Bitcoin is constantly increasing.
If the selling price of Bitcoin remains the same, the profit of miners will be constantly squeezed until the price is maintained at a balance point, eliminating miners with high costs, and the price tends to be stable.

From the sales point of view, due to the continuous increase in the production cost of Bitcoin, holders generally will not sell Bitcoin below the purchase cost or production cost, and the price will be supported.

New technologies, such as more efficient computing power chips, can always break miners' balance each time.
The single-Bitcoin production computing power reaches a new high by using new devices with higher energy-to-computing power conversion rates to replace old devices.
The barrier of computing power also makes the blockchain more secure.

From the perspective of the blockchain system, from the initial small amount of computing power protection to the globally significant computing power consumption unit, the blockchain itself did not pay the cost.
It only relies on the mechanism to make a large number of miners contribute computing power to protect the security of the blockchain for their interests.

\subsection{PoS and Security Cost}

\begin{figure}[!t]
    \centering
    \includegraphics[width=1\textwidth]{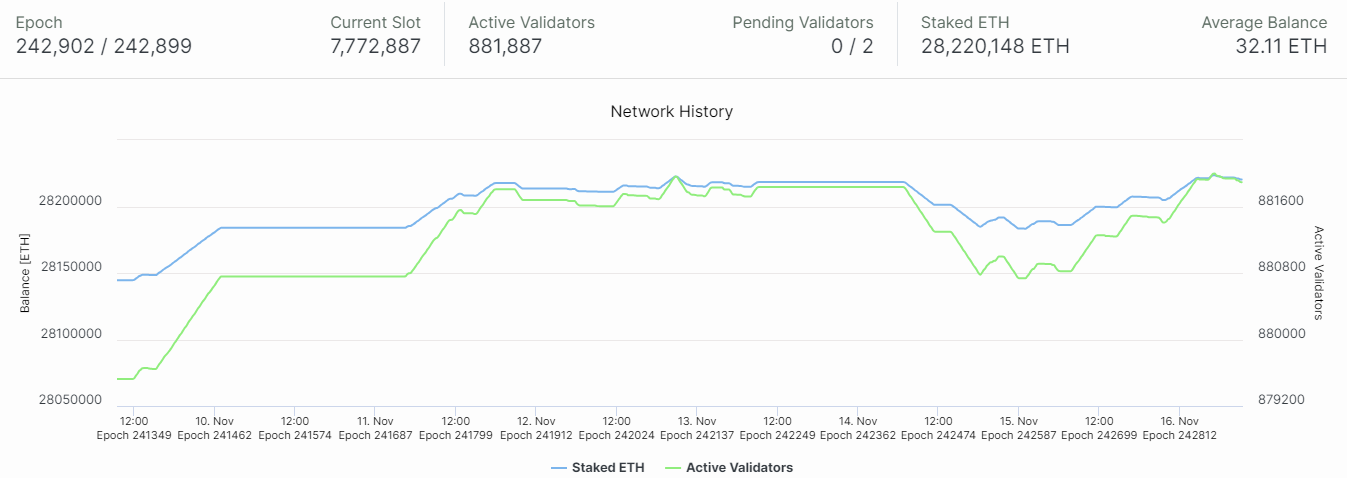}
    \caption{Ethereum staking cost}
    \label{fig_ethereum_staking_cost}
\end{figure}

The electricity cost in Bitcoin is a security cost.
In PoS, the electricity for mining is saved, but the security cost remains.
Ethereum turns to PoS in 2022.
Instead of burning electricity resources, PoS uses capital for chain security, similar to Bitcoin.

The money saved in the bank could get around 4\% interest.
To attract users staking the coins on the chain, Ethereum must provide interest with an Annual Percentage Rate (APR) higher than the bank to ensure the capital stays on the chain, shown in Figure~\ref{fig_ethereum_staking_cost} (data source \cite{beaconcha}).
As the total value of Ethereum staking keeps growing, the cost of staking interest is huge.
Table~\ref{eth_security_cost} shows the ETH Staking at the price and the security cost per block.

\begin{table}[h!]
    \centering
    \caption{PoS Ethereum security cost summary}
    \label{eth_security_cost}
    \begin{tabular}{| c | c |} 
        \hline
        % Col1 & Col2 & Col2 & Col3 \\ [0.5ex] 
        % \hline\hline
        ETH Staking & 28245172 ETH \\ 
        \hline
        ETH Price & 1985 USD \\
        \hline
        APR & 3.9\%  \\
        \hline
        Interest by Year & 2.186B USD \\
        \hline
        Interest by Day & 5.99M USD \\
        \hline
        Blocks per Day & 5760 blocks \\
        \hline
        Cost per block & 1040 USD \\
        \hline
        Date & 2023/11/17 \\
        \hline
    \end{tabular}
\end{table}

\subsection{Coin-based Security Engine}

\begin{figure}[!t]
    \centering
    \includegraphics[width=1\textwidth]{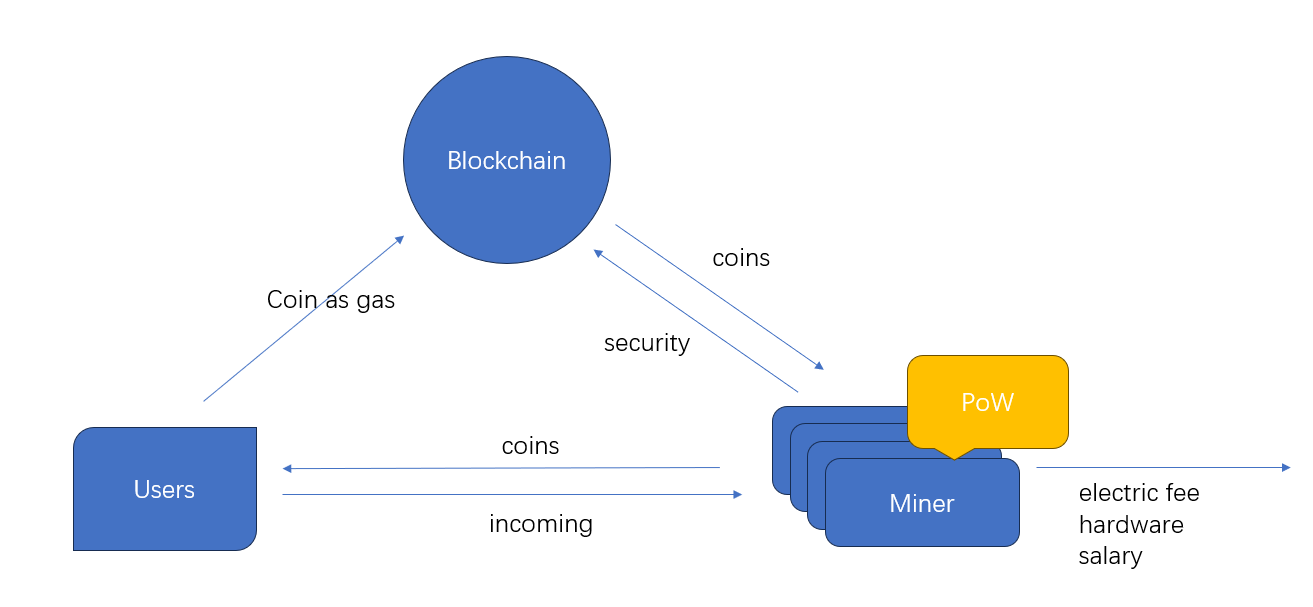}
    \caption{Coin as the Security Engine}
    \label{fig_coin_security_engine}
\end{figure}

Since the emergence of blockchain technology, mainstream public blockchains have been associated with native coins.
In addition to the wealth effect of coin price increases attracting new users, the biggest task of the coin is to ensure sustainable resource input through economic incentives to miners, and the input of computing power directly guarantees the security of the blockchain.
Therefore, coin has always been the security engine of blockchain, shown in Figure~\ref{fig_coin_security_engine}.

In the PoW consensus algorithm, miners generate new blocks by calculating hash values and receive corresponding coin rewards.
Therefore, the higher the coin price, the higher the miner's income, and the more motivated they are to invest computing power to maintain the blockchain network.

Higher coin prices mean higher overall market capitalization for coins. The wealth effect will attract more miners to join the mining competition, and the blockchain will gain more electricity and computing power, which also increases the security cost of the blockchain.

In the PoW consensus algorithm, the higher the computing power, the higher the cost for attackers to launch attacks, reducing the success rate of attacks.
Therefore, higher coin prices mean higher security costs, which will increase the attacker's costs and, in turn, reduce the success rate of attacks.

Most blockchains use coins to pay on-chain transaction fees, giving coins intrinsic use value.
This design originated from Bitcoin, considering that in the future when the Bitcoin halving policy makes it almost only a small number of new coins produced by mining, there are still relatively stable transaction fees in each block that can incentivize miners to continue to invest resources. Ethereum also adopts a similar design, changing the transaction fee to a gas fee, from paying for the space for issuing transactions to paying for the computing power and storage space for program execution.
However, in essence, applying native coins is mainly to create demand. In use, coins generate consumption and flow in addition to value storage.

Compute power protects the security of blockchains, but what drives miners to contribute compution power is the fixed total amount of coins. Therefore, native coins have always been the security engine of blockchains.

\subsection{Known Issues for Proof of Work}

From an economic perspective, the goal of creating Bitcoin is to establish a peer-to-peer electronic currency system that no third party can control. From a technical perspective, a "bank" needs to meet basic security requirements, and a blockchain-based bank should be as decentralized as possible to meet the requirements of peer-to-peer transactions.
These properties require cryptographic protection and computing resources to provide security and resist fraudulent activities such as Double Spending that may be caused by 51\% Attack.

Currently, the PoW consensus algorithm has strong decentralization properties and meets most of the needs of blockchain. 
However, people still complain about the huge energy consumption caused by the PoW mechanism.
In addition, our research shows that the security of PoW has an upper limit, and the proposed PoW upgrade algorithm can solve the security upper limit problem and efficiency problem.

\subsubsection{Energy consumption}

% PoW algorithm requires a lot of computational power, which means it requires a lot of electricity. The amount of electricity consumed by the Bitcoin network each year is equivalent to the total consumption of some countries in the world. This not only leads to energy waste but also environmental problems. Although some cryptocurrencies have tried sustainable methods to solve this problem, such as using renewable energy, this problem still exists.

Bitcoin is often criticized for its massive energy consumption, and it is also a fact that PoW-based blockchains often consume a lot of electricity.

This argument extends to a more core question: whether it is worth consuming resources on a blockchain system. People hold different views on this.

Those who support Bitcoin argue that the banking system consumes many resources beyond electricity.
The cost of opening branches nationwide, security fees, auditing, and IT system costs may far exceed Bitcoin.

Opponents support blockchains with less energy consumption. Environmentalists are more concerned about energy consumption, and scientists are mainly worried about whether computing power can be used for other purposes.
For example, artificial intelligence training and scientific computing are often mentioned.

From the demand perspective, most of the human needs are created. More needs create more services, which means human prosperity.
Using electricity to produce virtual currencies and maintain the safe operation of blockchain systems is a new demand brought about by the emergence of blockchain.
Users have raised new demands for decentralized banking and assets (contribution to decentralized security).

In some regions with abundant natural resources but high electricity transmission costs, cryptocurrency mining has also brought wealth and employment opportunities to the local area.

\subsubsection{Mining Pool caused Centralization}

\begin{figure}[!t]
    \centering
    \includegraphics[width=1\textwidth]{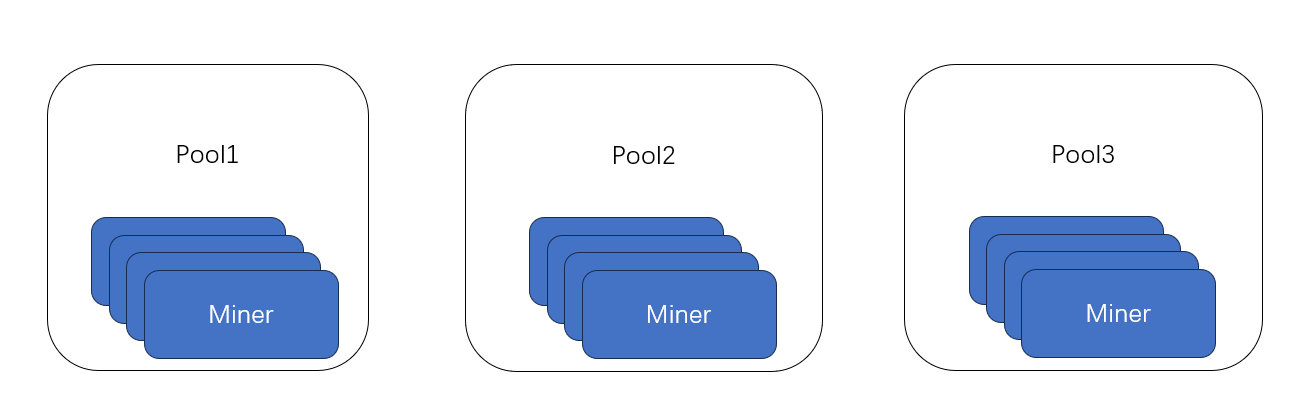}
    \caption{Mining Pools}
    \label{fig_mining_pools}
\end{figure}

PoW mining pools may lead to centralization issues.
As PoW requires significant computing resources, the miners with sufficient computing power have more chance to find new blocks and receive rewards.
This may result in smaller miners being excluded from mining, further concentrating the network's computing power and control. Additionally, the network may become vulnerable to attacks due to the concentration of computing power.

Mining pools allow multiple miners to combine their computational power to mine blocks more efficiently and share the rewards, shown in Figure~\ref{fig_mining_pools}.
However, this also leads to centralization because the largest mining pools can control the majority of the network's computational power and, therefore, the ability to determine which transactions are added to the blockchain.
This puts the network at risk of a 51\% Attack, where a group of mining pools with more than 51\% of the computational power can control the blockchain and potentially compromise its security.

\subsubsection{The Sustainable of Mining}

\begin{figure}[!t]
    \centering
    \includegraphics[width=1\textwidth]{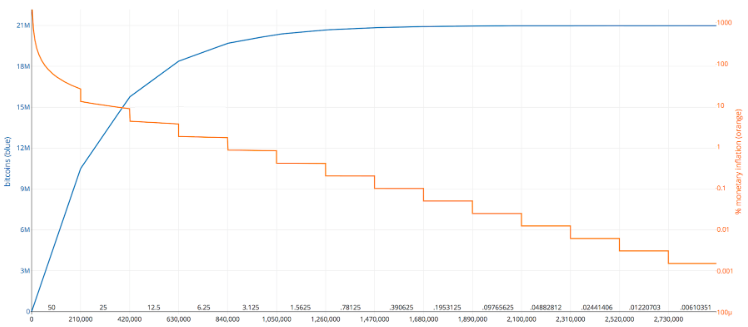}
    \caption{Bitcoin Halving every Four Years}
    \label{fig_bitcoin_halving}
\end{figure}

Bitcoin uses human nature to establish a mechanism to gain more computing resources to protect its security constantly.
This design is ingenious, but it also has some flaws and potential risks in the long run, such as Bitcoin halving shown in Figure~\ref{fig_bitcoin_halving}.

In the current PoW model, the currency price affects the security of the blockchain.
A blockchain that is recognized by a large number of miners, who will invest resources to ensure the security of the blockchain and earn profits.

In the PoW network, the miner's profit is the difference between the input of resources and the selling price of the currency.
Therefore, having a high-security and low-cost blockchain in the current model is impossible. Unless a way to subsidize miners' income is found, reducing users' investment in gas is possible.

\subsubsection{Security Limitation}

Blockchain security is limited because of its computation capacity limitation.
Its security limit is estimated based on the following assumptions:

When all 21M bitcoins are mined.
All bitcoins are sold at a market peak price of 60k USD.
The proceeds from the sale are used to purchase the electricity needed for mining.
The electricity price is at an average price of 0.05 USD/KWh.

Based on these assumptions, we can easily calculate that Bitcoin can purchase about 25200 TWh of electricity.
It is a huge number, but it is still finite.

$$ {{ 2100 0000 BTC * 60000 USD/BTC } \over { 0.05 USD/KWh }} = 25200000000000 KWh = 25200 TWh $$

Due to the fixed supply of Bitcoin, it is difficult to finance the purchase of more electricity from the market unless the price of Bitcoin reaches an all-time high.
The analysis does not include the hardware cost of purchasing mining machines, labor costs, and mining profits.

These factors could all reduce the total amount of electricity that Bitcoin can purchase.
Thus, we said Bitcoin has a finite security limitation.

\subsubsection{High Cost for the End User}

\begin{figure}[!t]
    \centering
    \includegraphics[width=1\textwidth]{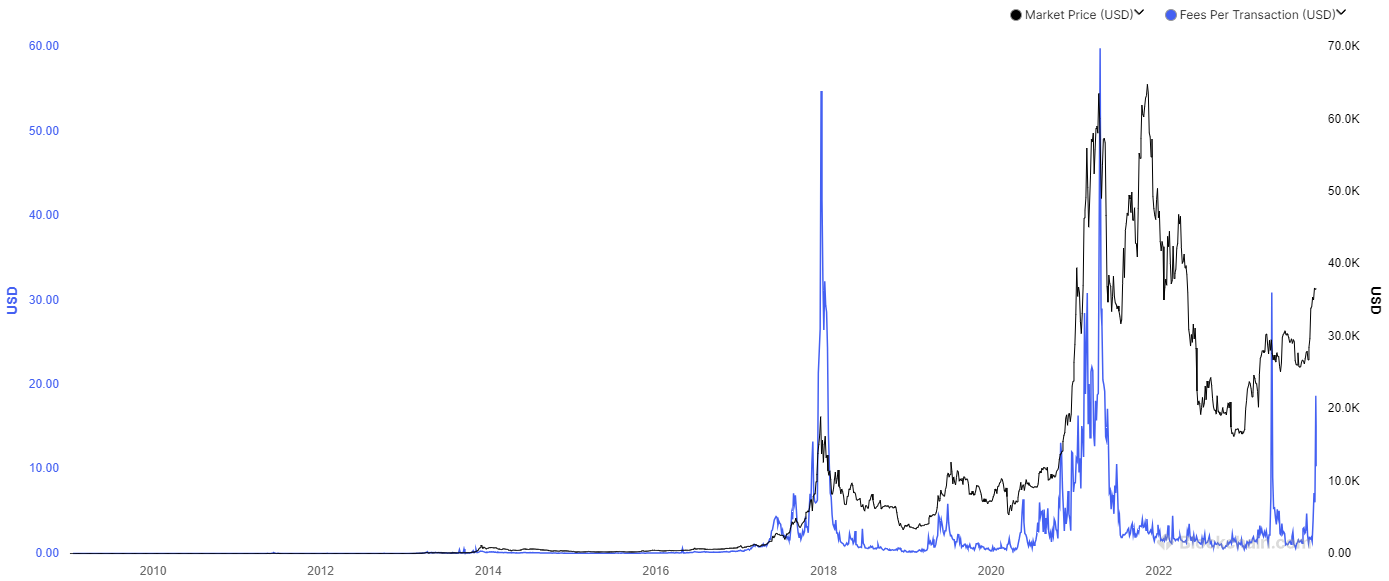}
    \caption{Bitcoin fee per Transaction}
    \label{fig_bitcoin_fee_per_transaction}
\end{figure}

The ingenuity of Bitcoin lies in transferring the security costs it needs to the final blockchain users.
This helped the rapid growth of the Bitcoin system in the early days because it incentivized miners to participate in mining, thus ensuring the security of the Bitcoin network.

However, as the price of Bitcoin continues to rise, the cost of mining is also becoming higher and higher.
It led to the increasing cost of Bitcoin transactions, shown in Figure~\ref{fig_bitcoin_fee_per_transaction}.
For ordinary users, the cost of using Bitcoin for transactions has become too high, which has become a limitation of Bitcoin's inability to become popular on a large scale.

\subsection{Comparing with Proof of Stake}

PoW and PoS are currently the most widely used blockchain consensus algorithms.
With the official transition of Ethereum to PoS in 2022, the algorithm has officially entered large-scale use.

PoS is a consensus algorithm proposed in 2013 and first implemented in the Peercoin system. PoS is currently entering a popularization stage, with more and more public chains adopting the PoS consensus algorithm.

However, PoS also has some disadvantages, the most important of which is the "Nothing at stake" problem.
The "Nothing at stake" problem is because assets are used as the condition for participating in consensus and are not consumed during the consensus process.
When a fork occurs, participants can participate in consensus on multiple forks simultaneously, making the fork unable to converge.
To address this problem, many punishment mechanisms have been added to the PoS algorithm. If malicious behavior is found, the validator will be punished.

After Ethereum switched to the PoS consensus, the consensus energy consumption was significantly reduced.
However, the security costs stayed the same after converting from mining to staking. 
The security cost is passed on to the end users by purchasing ETH to pay for gas.
The users ultimately pay the security costs of ETH.
The basic cost of interaction with a smart contract to transfer stablecoins on Ethereum is around \$10.
People expect the popularization of L2 to continue reducing usage costs.
Still, due to the centralization and security of L2, users generally take a wait-and-see attitude, only transferring small amounts of assets across chains to L2.

In PoW, the security costs are provided by miners, who recover costs through block rewards from mining.
In PoS, the security costs are provided by stakers, who recover costs through block rewards from staking interests.
However, miners and stakers could sell the native coins to recover costs; the end-users purchase coins and consume them for transaction fees or gas. Therefore, the security costs of PoS are still ultimately borne by users through gas payments.

\subsection{History for Proof of Work}

PoW is now famous as a blockchain consensus algorithm.
However, the algorithm was not created for consensus at the beginning.

\subsubsection{PoW for Email Anti-Spam}

Proof of Work (PoW)\cite{dwork1992pricing} has been proposed to prevent email spam.
The idea is to require the sender of an email to perform a certain amount of computational work before the email is sent.
This work is designed to be CPU intensity for computers to perform but easy to verify.

The email sender would include the result of this computation in the email, which the recipient's email server could verify. 
If the computation is correct, the email will be accepted; otherwise, it will be rejected. 
Making the computation difficult would be prohibitively expensive for spammers to send many emails.

However, there are several challenges with using PoW for email anti-spam.
First, it could lead to delays in email delivery, as the computation would need to be performed before the email is sent.
Second, it would require widespread adoption to be effective, as spammers could send emails that do not require PoW.

While PoW was an excellent proposal for preventing spam in some contexts, it is less popular than the machine learning approaches in the NLP field, such as the Bayes algorithm.
Many years later, people discovered its new usage in consensus.

\subsubsection{PoW for Consensus}

Since Satoshi Nakamoto designed Bitcoin in 2008, PoW has become famous for its consensus algorithm.
Miners compete to solve PoW puzzles in order to add new blocks to the blockchain.
The winner of each round is rewarded with a certain amount of cryptocurrency.
This incentivizes miners to participate in the network and helps to secure it.

PoW solves the Byzantine Generals Problem from a new perspective as a consensus algorithm. Traditional consensus algorithms are based on communication, requiring nodes to communicate multiple times to reach consensus.
This would be very difficult in a network with a large number of nodes, as there may be node failures and network congestion.

PoW uses a computational method to solve the Byzantine General's Problem. In PoW, nodes obtain the right to broadcast blocks by solving a computational problem.
This allows PoW to effectively reach consensus even in a network with a large number of nodes.

PoW's computational method also makes it have lower communication complexity. In PoW, nodes only need to broadcast blocks; no other communication is required.
This allows PoW to effectively save communication resources in a network with a large number of nodes.

\section{Related Work}

\subsection{Proof of Work and Alternatives}
In Chapter~\ref{ch:preliminaries}, we introduced blockchain consensus algorithms.
Due to the high energy cost, people are looking for better alternatives for the PoW algorithm.
This includes other consensus like PoS (Proof of Stake) and PoC (Proof of Space) and PoW improvements as useful PoW.

In the useful PoW, people also search for different directions: 1. to replace excessive usage of hash functions with tasks or 2. to make current Nakamoto consensus hash computation one more usage.
Proof of Useful Work (PoUW) is the direction to find a different algorithm to replace the Nakamoto consensus.
Useful PoW is not intended to save energy but uses the consensus computation for more meaningful tasks.

% Many useful PoW proposes that instead of simply solving complex mathematical problems, the computing power used for PoW should also be utilized for a useful purpose. This can be achieved by tying the PoW mechanism to a useful computational task, such as scientific calculations, protein folding, or machine learning.

% By performing useful computations in addition to PoW, the energy consumption of the network can be put to productive use rather than simply wasted on solving meaningless problems. This approach can also incentivize the use of renewable energy sources, as it encourages the development of energy-efficient computing systems.

\begin{figure}[!t]
    \centering
    \includegraphics[width=0.7\textwidth]{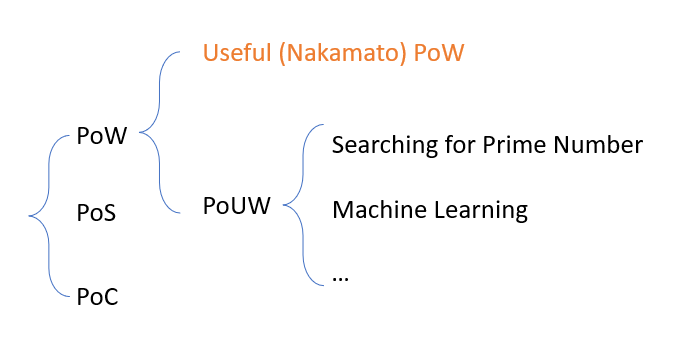}
    \caption{Consensus Methodology Approaches}
    \label{fig_consensus_methodology_approaches}
\end{figure}

% The concept of useful PoW is still in its early stages of development and has yet to be widely adopted. However, it is seen as a promising avenue for addressing the issue of energy consumption in blockchain technology while still maintaining the security and decentralization benefits of PoW.
Different from PoUW, the direction of making hash computation useful did not make a breakthrough since blockchain emerged\cite {buterin2019hard}.
The useful PoW we proposed in this work shows the new usage of hash-based computation for permissionless storage networks.
In Figure~\ref{fig_consensus_methodology_approaches} it shows different approaches to reduce the energy consumption or make the consensus algorithm useful.

\subsection{Proof of Useful Work}

Since the direction of making hash calculations themselves useful has yet to make a breakthrough, people have begun to look for alternatives to the Nakamoto algorithm in the PoUW\cite{ChallengesPoUW2022} direction.
Primecoin\cite{king2013primecoin} is the early blockchain system that first adopted useful PoW.
The mathematical twin prime numbers are used to replace the hash computation to make the hard problem solve for consensus, which is also useful for math purposes.
Permacoin\cite{miller2014permacoin} makes use of an extremely large archival data file to store portions of the public
dataset.
The Permacoin uses data storage as the scenario but changes the Nakamoto consensus hash algorithm.
Coinami\cite{ileri2016coinami} is applying the DNA sequence alignment as PoW.
Orthogonal Vectors\cite{ball2017proofs} is used as the hard problem to generate PoW for the more general computation.
Conquering Generals\cite{loe2018conquering} and \cite{haouari2022novel} use the NP-hard problem for the consensus.
Several solutions are proposed for PoUW with machine learning and artificial intelligence fields, including Coin. AI\cite{baldominos2019coin}, BlockML\cite{merlina2019blockml}, \cite{lihu2020proof} and Proof of federated learning\cite{qu2021proof}.
Proof of Useful Randomness\cite{seyitoglu2021proof} propose to generate commitments during consensus and use them in cryptographic operations later.
HDCoin\cite{ma2022hdcoin} miners use HDC to generate, train, and test machine learning models that are verified on accuracy metrics.
The highest accuracy model earns the block reward.
Proof of exercise\cite{shoker2017sustainable} and Proof of Computation\cite{dragos2022proof} work on the PoW with matrix computation.
% PoUW hopes to find algorithms to use computing power for scientific computing problems, while also being able to derive consensus. This direction of PoUW completely modifies the Nakamoto algorithm, but it still relies on computing power resources, not capital.
% The proposals range from math problems and general computation\cite{loe2018conquering, dragos2022proof, ma2022hdcoin} to artificial intelligence\cite{baldominos2019coin, merlina2019blockml, lihu2020proof, qu2021proof}.

\subsection{Limited Usage Scenarios}

The PoW consensus algorithm used by Satoshi Nakamoto is based on the hash algorithm.
Hash algorithms can hardly be used in scientific computing, thus Vitalik believed that useful PoW may not be feasible\cite{buterin2019hard}.
The PoUW changes the hash-based algorithm for various mathematical and scientific computing algorithms, including searching for prime number, machine learning, and matrix calculations.
Some of them, such as the prime number, are useful in math, but it is hard to attract users to pay for the computation.
Ideally, a scenario in which users' high demand can cover the cost for the cost of consensus computation.
It could bring down the blockchain end-users usage cost.

However, useful computing is not limited to general computing.
A special purpose computation may have large-scale of use scenarios as well.
% We can find dedicated computing and application scenarios to use PoW-based hash algorithms as a special encoding algorithm.

\section{Make Use of Nakamoto PoW}

We propose to extend Nakamoto PoW for a new usage for the permissionless storage network.
The extension does not modify Nakamoto hash-based PoW.

\subsection{Reuse the Hash for Encoding}

\begin{algorithm}[h]
    \caption{Nakamoto Proof of Work Algorithm}
    \begin{algorithmic}[1]
    \State Get $identity$ of miner
    % \While {$True$}
    \State Get $hash$ of current block 
    \State Get $data$ to write into the next block
    \State Calculate current difficulty $D$ based on history blocks
    \State Let $nonce = 0$
    % \While {$Hash(hash || data || id || nonce ) \geq D $}
    % \State Let $nonce = nonce + 1$
    % \EndWhile
    \Repeat
    \If{Receive new block from the network}
    \State Exit and restart this algorithm based on the new block hash
    \EndIf
    \State Let $nonce = nonce + 1$
    \Until{$Hash( hash || data || identity || nonce ) < D $} %or {Got other's new block}
    \State Broadcast the new block worldwide
    % \EndWhile
    \end{algorithmic}
\end{algorithm}

Nakamoto PoW algorithm used hash calculations as a hard puzzle to solve.
Since this PoW algorithm is straightforward, it is not easy to use for other useful calculations, especially artificial intelligence.

For general computing, it is difficult to utilize hash computing power.
This is because hash calculations cannot be applied for general computing.
They are designed to be one-way functions, meaning it is straightforward to compute the hash of a given input but very difficult to find an input that will produce a given hash.

However, hash calculations can be used for more basic tasks like file data encoding.
This is because file encoding can be viewed as a one-way function.
We can generate a hash of a file and then use that hash to represent the file.
This can be useful for storing files in a distributed system because it allows us to verify a file's preservation without downloading it.

Even the task of file encoding that can be implemented has been considered to be inconsistent with the design goals of general encoding and decoding algorithms.
This is because most encoding and decoding algorithms are designed for efficiency.
They are designed to minimize the amount of data that needs to be transmitted.

However, we need an algorithm with asymmetric computation costs for encoding and decoding in a permissionless file storage system.
This verifies if the storage network honestly keeps the user storage file.

In conclusion, using hash calculations for useful tasks is a relatively rare scenario. However, there are some cases where it can be useful.

\subsection{PoW as the Encoding Algorithm}

\begin{algorithm}[h]
    \caption{Our Extended Useful Proof of Work for Encoding}
    \begin{algorithmic}[1]
    \State Get $id$ of node
    \State Get the difficult $L$ for encoding
    \State Get message $m$ to encode
    \State Let $nonce = 0$
    \Repeat
    % \If{Got other's new block}
    % \State Exit and restart this algorithm based on the new block hash
    % \EndIf
    \State Let $nonce = nonce + 1$
    \Until{The last $L$ bits of $Hash(id || nonce )$ equal to $m$} %or {Got other's new block}
    \State Output $nonce$ as $r$ as the encoded value
    % \EndWhile
    \end{algorithmic}
\end{algorithm}

Unlike normal encoding algorithms, which process data efficiently, PoW encodes the data slowly and expensively by intent.
The reason for the necessity for slow and expensive encoding is explained in Chapter~\ref{ch:storage}.

While the block generated during mining and consensus broadcasts globally on-chain, the miners reuse the computation for local file encoding.
The encoding result is saved on the local disk off-chain.

\subsection{Remain PoW for Consensus}

\begin{figure}[!t]
    \centering
    \includegraphics[width=1\textwidth]{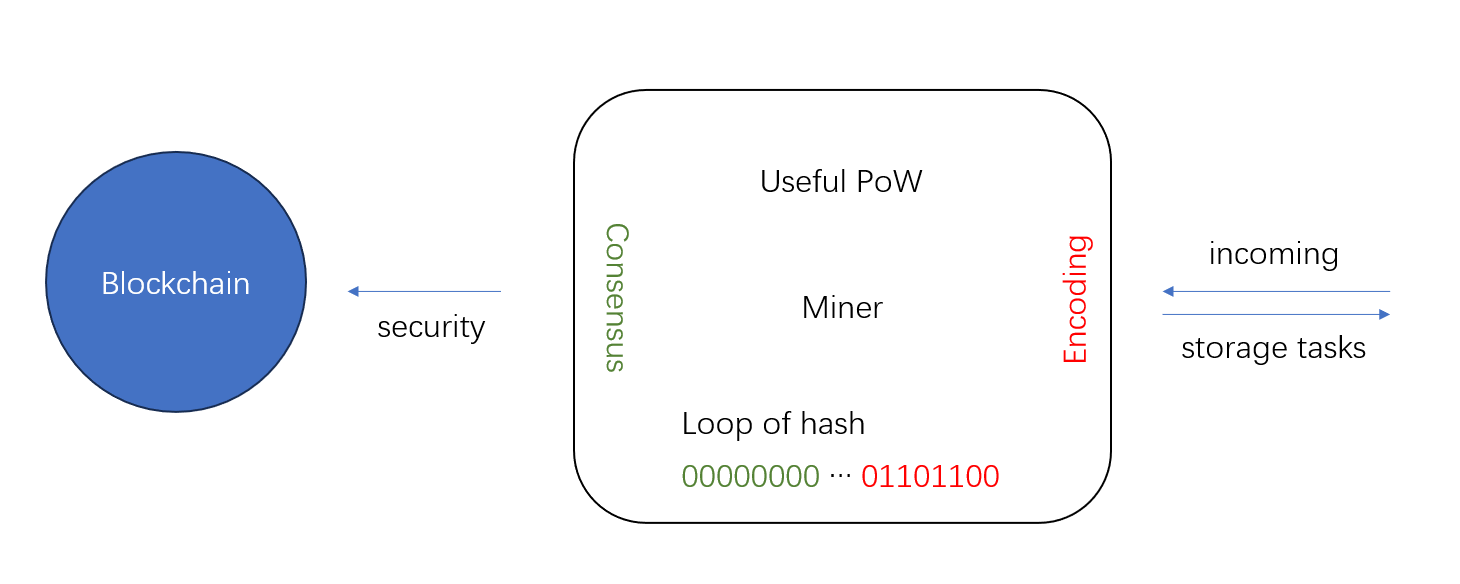}
    \caption{Remain PoW for consensus}
    \label{fig_remain_PoW_for_consensus}
\end{figure}

Satoshi Nakamoto first used the PoW algorithm for consensus, making it possible for blockchain systems to operate under large-scale decentralized conditions.
In the past few decades, the Nakamoto consensus has been well-received by the academic community and has been fully demonstrated.

Practice proves the theory.
Bitcoin has been running safely for over ten years, and its security has been verified by practice.
It is risky to modify the algorithm rashly.
Therefore, modifying the Nakamoto consensus is necessary for the extended useful consensus algorithm to ensure sufficient security.

Although Bitcoin has run safely in the past, people are still worried about the future security of PoW because coin prices cannot always rise.
Useful PoW improvements based on storage have the opportunity to compensate for the potential security risks of PoW, making the security of PoW blockchains decouple from coin prices, shown in Figure~\ref{fig_remain_PoW_for_consensus}.

\subsection{Data Encoding over Blocks}

\begin{figure}[!t]
    \centering
    \includegraphics[width=1\textwidth]{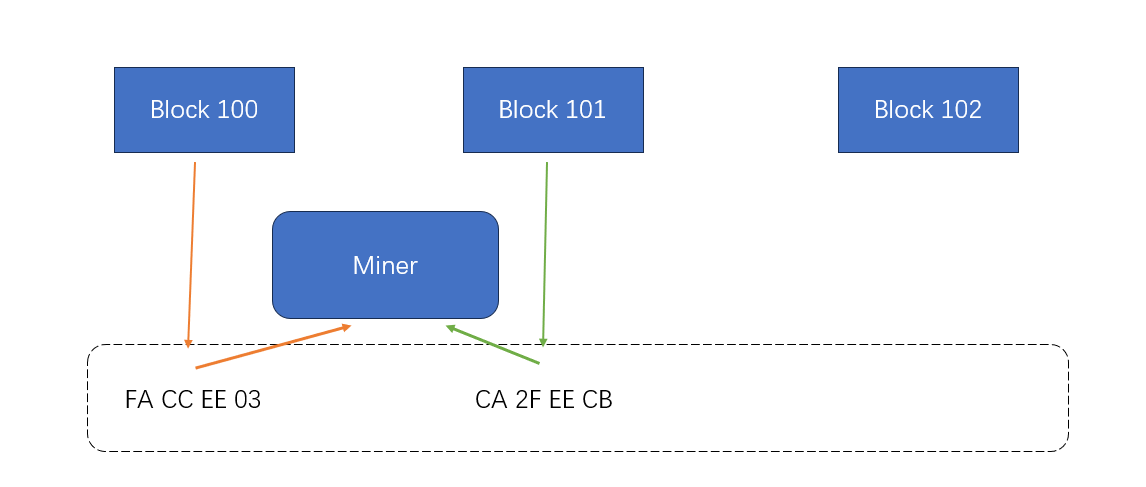}
    \caption{The file data encoding over chain blocks}
    \label{fig_encode_over_blocks}
\end{figure}

We combine the consensus algorithm and the encoding algorithm into one.
The calculation time for fixed-length encoded content is basically fixed in the encoding process.

In the consensus calculation, there is a large amount of randomness.
In most cases, the algorithm runs the consensus algorithm to find the next block based on the current block information and the transaction information to be packed into the next block. 
If other miners are lucky to find a new block, they will broadcast it to the entire network as soon as possible.
At this point, rational miners will stop the exhaustive calculation of the block of the same height to avoid wasting computing power and block forks, shown in Figure~\ref{fig_encode_over_blocks}.
Miners continue to calculate the next block based on the latest received block.

For encoding tasks, the length of a file block may be large, so the blockchain may have generated several blocks before the file block is wholly encoded.
Whenever a new block is generated, the miner should immediately mine based on the new block, and the file encoding should also be encoded based on the new block.
The block information of the encoding process should be included in the data encoding results.

\subsection{Encoding Difficulty}

The data encoding has an encoding difficulty, different from the mining difficulty.
The bit length of the encoding input value decides the encoding difficulty.
When the encoding difficulty is set, the number of hash operations for a value encoding can be estimated.
It takes around $2^n$ times of hash operations for a given $n$ bits of encoding data input.

The computation speed of CPU and GPU are evolving.
When most modern devices have reached higher computational power, we should collectively increase the difficulty of encoding.
The blockchain system will thereby receive more computational power protection, and this will also increase the cost of proof of replication, enhancing data security.

\section{Experiment}

\subsection{Encoding and Decoding Time for Different CPUs}

\begin{figure}[!t]
  \centering
  \includegraphics[width=0.8\textwidth]{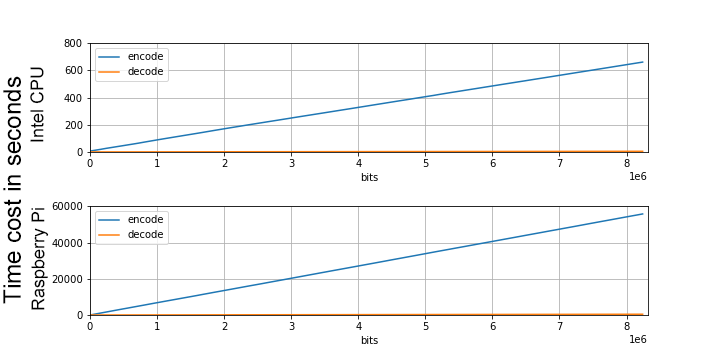}
  \caption{The time cost for encoding and decoding for 1MB data}
  \label{fig_encode_decode}
\end{figure}

In this experiment, we run the algorithm on different speed CPUs shown in Figure~\ref{fig_encode_decode}.
The encoding difficulty is set to 8 bits, encoding one-byte data for each iteration.
We choose different platforms, such as the Intel i5-6300U processor and a very old Raspberry Pi 1 Model B+.
The time cost of a modern CPU and an embedded device, Raspberry Pi, is significantly different, but both devices take a while.
It shows that even older devices can join the replication mining, and it is not cheap work for the latest CPU.

Decoding time on different CPUs is constant.
Figure~\ref{fig_encode_decode} shows that decoding is much faster than encoding.
It only takes one hash operation for each decoding loop.
It is observed that decoding is much cheaper than encoding.
Given a fixed amount of storage space, the miner will tend to keep the replicated data rather than the original content.
Converting the replication to the original is cheap and fast.
The challenge for verification requires the miner to access the replication data, which cannot be produced on the fly.

\subsection{Encoding Time in different Difficulty}

\begin{figure}[!t]
    \centering
    \includegraphics[width=0.8\textwidth]{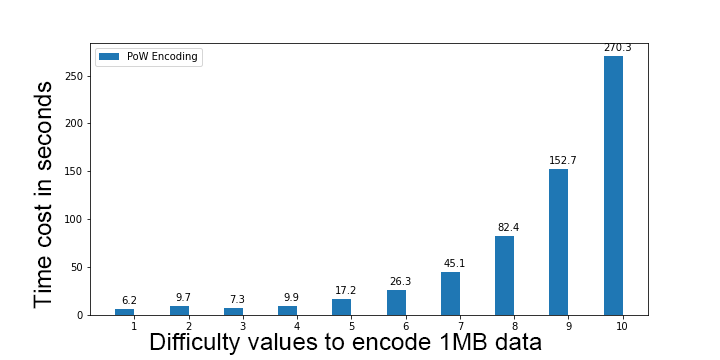}
    \caption{The file data encoding over chain blocks}
    \label{fig_encode_difficulty}
\end{figure}

In Figure~\ref{fig_encode_difficulty}, the overall trend shows the time costs growing exponentially.
The test was done on the Intel i5-6300U processor.
The wider bits were set per encoding, and the more time was taken in each encoding iteration.
For the fixed length of data, increasing the replication difficulty will take a long time for encoding.
The wider bits lead to fewer iterations of encoding for a fixed data length.
The decoding also takes fewer hash operations under the wider encoding difficulty.
It reduces the overall decoding time.
As we can observe, the encoding difficulty increased when the encoding width from 1 bit to 2 bits, but the number of decoding iterations was reduced by half.
That is the reason that 2-bit difficulty takes a longer time, even than 3-bit.

\section{Impact}

\subsection{Reduce the End-User Cost for Blockchain Usage}

In the early days of blockchain technology, the performance and scalability of public chains were the focus of attention.
Before these problems were solved, the cost of using public chains was not widely considered.
However, once blockchain technology can support a sufficient number of users, the cost of use will become a more important prerequisite for the large-scale use of blockchain.

The cost of use is the security cost that is allocated to each blockchain user. As long as it is a public chain system based on native coins, users will bear this cost in the form of purchasing native coins as fees or gas.
The higher the security cost of the PoW blockchain, the more expensive the coin price, so people have to pay more costs to use the blockchain.
Under such conditions, it is difficult for blockchain to achieve large-scale applications. 
Unlike traditional Internet applications, the cost of using blockchain cannot be modified by simple product design but is closely related to the underlying consensus algorithm and security.

Using the useful Proof of Work, we added a storage subsystem to the original system based on native coins.
Paying for storage resources is a very common business model, which has been fully verified in the popular wave of cloud computing in the past few years.
Suppose the computing power used to prove the availability of files in storage is used to protect the security of the blockchain.
In that case, blockchain users can avoid bearing the high cost of security because the computing power required for security is a byproduct of using computing power for the storage business.
Therefore, useful Proof of Work can significantly reduce the cost of using the blockchain for users.

\subsection{Remove the Mining Pools}

Decentralization is an important concept in blockchain because it can eliminate the risk of relying on a trusted third party.

In Bitcoin, the computing power is driven by the native coin. In solo mining, only one miner can win the block reward each time, regardless of the number of participants.
Under the conditions of increased computing power and full competition, the probability of most miners winning the reward is extremely small.

Mining pools were born under this premise. Because mining pools aggregate the computing power of many miners, they have a greater chance of winning the block reward. Miners who join mining pools will also have lower but more stable rewards. However, the emergence of mining pools also reduces the degree of decentralization of blockchain.

When computing power can be consumed locally for useful purposes, and the income from consuming computing power is greater than the income from mining pools, users are more likely to use their computing power locally rather than connect to mining pools to earn income. 
Useful Proof of Work allows us to bypass PoW mining pools and allow most miners to solo mine, further improving the decentralization of PoW blockchains.

\subsection{Avoid Greedy Energy Consumption}

Reducing the waste of electricity is not about not consuming electricity but about not greedily consuming electricity for the sake of the blockchain's security (to obtain coin rewards).

In the early days, mining Bitcoin only needs a computer.
As the public's understanding of Bitcoin technology deepened, people found that graphics cards could provide more powerful computing power than CPUs.
Later, with the rise of Bitcoin prices, people designed ASIC chips for dedicated mining machines, which increased the energy efficiency of computing power.
Therefore, after years of development, mining has become an industry of hardware equipment arms race.

Today, blockchain mining mainly requires two conditions: 1) professional equipment and 2) cheap electricity resources.

Due to the obsolescence of professional equipment, from the time it is produced to the time it is eliminated due to efficiency problems, it has a complete life cycle.
In this life cycle, the operation of the equipment means profit. Therefore, rational miners will try their best to make the mining machines run at full power under the condition of meeting the profit requirements, which also means that electricity resources will be greedily consumed.

Useful Proof of Work can improve this situation. Assuming that the blockchain itself does not reward, or the reward is very small, miners mainly rely on the storage part of the useful PoW algorithm to make a profit.
At this time, rational miners will only start PoW calculation when there is a coding demand.
Otherwise, they will set the mining machine to low power mode and try their best to reduce power consumption to increase their own profits. In such a mode, useful Proof of Work can inhibit the greedy consumption of electricity in mining and achieve an environmental protection function.

\subsection{Blockchain Storage as the new Security Engine}

\begin{figure}[!t]
    \centering
    \includegraphics[width=1\textwidth]{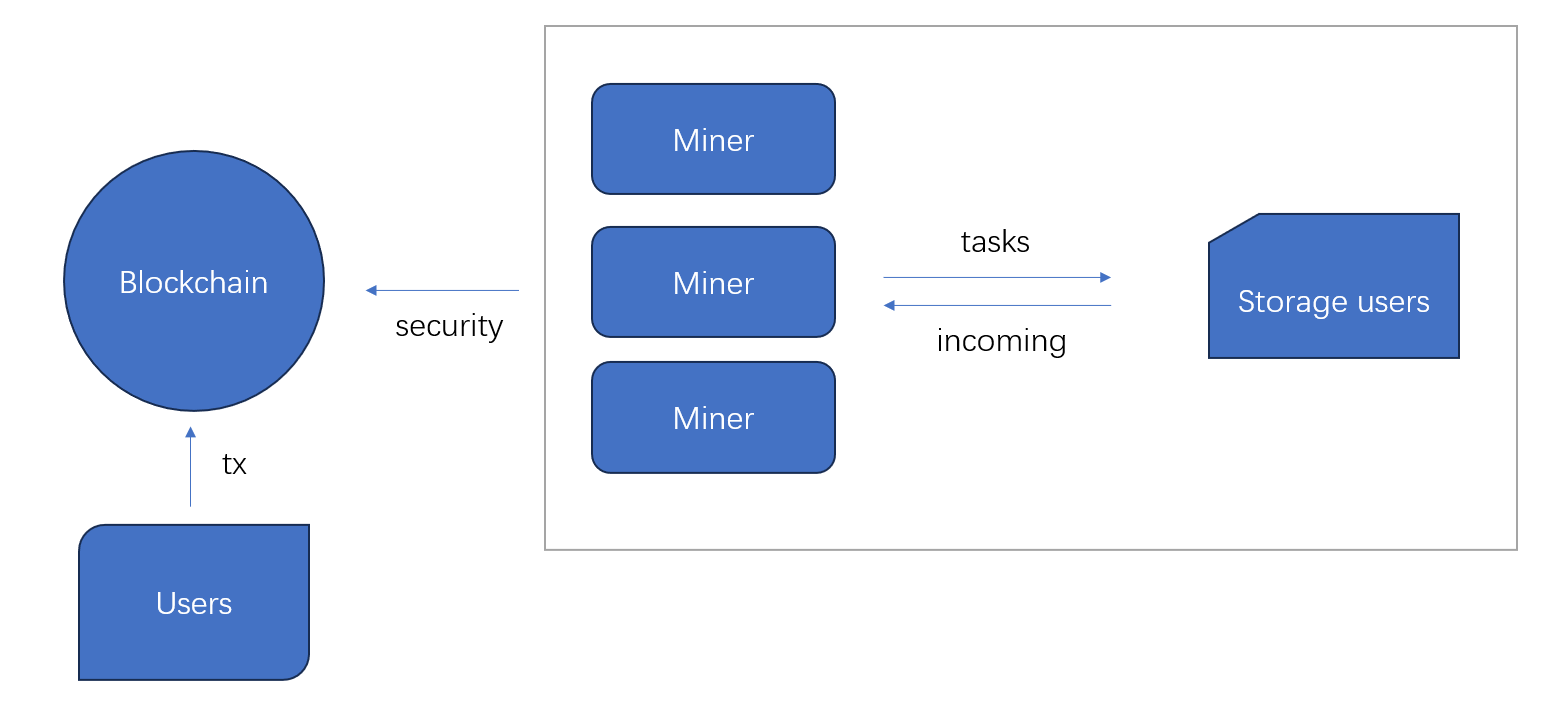}
    \caption{Storage as the new Security Engine}
    \label{fig_storage_security_engine}
\end{figure}

Previously, we analyzed the mainstream mode of current blockchain systems, which is to adopt the native currency of public chains as the security engine of blockchains.
Whether it is a blockchain using PoW or PoS, as long as the native currency is used as the transaction fee or gas of the blockchain, the security cost will eventually be shared by users.

From the user's point of view, the price of the native currency is meager in the early stage, but the successful blockchain in the later stage will mean the price of the native currency will soar, which will lead to an increase in the cost of use at the same time.
Although it can generate wealth effects in the direction of speculation, this model is unsuitable for blockchain applications.

We introduce the permissionless storage as the consensus engine, enabling blockchains to pay transaction fees without relying on native currencies for the first time while having sufficient security protection, shown in Figure~\ref{fig_storage_security_engine}.
Miners under this new consensus engine can use useful Proof of Work to earn storage profits denominated in stablecoins.
The storage consensus engine transforms the original lottery-style profit model (i.e., many miners, only one gets the block reward each block) into a fixed income for all miners participating in running useful tasks, and computing power can be fully utilized. Since storage is sustainable, miners who participate can periodically earn rewards if they prove they have honestly stored data.

The new engine reduces the cost of using blockchains for users.
Users still need to pay transaction fees or gas, but this does not include sharing chain security costs, so the fees can be as low as a few cents and can be paid in stablecoin form.
This will lower the user threshold in large-scale blockchain applications.

\section{Chapter Summary}

In this chapter, we first extend the PoW consensus algorithm for another useful usage.
To address the existing weakness of PoW, we proposed an upgraded algorithm without changing the original Nakamoto Proof of Work by reinterpreting the hash output.
The extra usage is generated during the same hash computation; there is almost no more cost.
It also means useful computation tasks can cover the consensus cost.

While performing the computation, blockchain may receive high-security protection, which will cost the chain end-users very little fee.
Thus, the extended PoW can resolve the contradictory problem that user no longer need to pay the chain security cost.

In the next chapter, we will dive into how the PoW computation can be applied in a permissionless storage network and why a permissionless one must use such an algorithm to ensure data storage security.

\chapter{Permissionless Storage as Security Engine} % 这里可以扩展成整个区块链存储系统，使用第三章的算法
\label{ch:storage}

In Chapter~\ref{ch:consensus}, we proposed a method to make the Nakamoto PoW useful.
In this chapter we would design and build a permissionless storage network with the useful PoW as the core to prevent cheating such as Outsourcing Attack.
The goal is to make the permissionless storage as the new security engine for PoW blockchain to replace current coin driven security engine like Bitcoin.
Idealy, the PoW contributed by useful task during the storage process has already been paid by the end-users, thus coin is not required to remain a high price to keep the PoW contributors profitable.
Storage as the new engine could result in a true decentralized blockchain without a high usage cost.

Blockchain technology was first created for a decentralized bank: Bitcoin.
Most of the blockchains created since Bitcoin are financial related, including Ethereum.
Ethereum aims to build more applications rather than just a simple bank with transactions.
However, non-financial blockchain applications are still away from massive usage.
Projects such as IPFS and Filecoin are attempting to solve the permissionless storage network problem.
IPFS implements a protocol similar to BitTorrent, which can be used for building a decentralized website or downloading files.
However, since IPFS was not backed by any blockchain system, it lacks an incentive mechanism.
The data owners must ensure enough nodes are hosting their content so that the users can enjoy high-speed downloads and a good user experience.
Filecoin aims to fix the weakness of IPFS by adding an incentive layer with blockchain.
It proposes a very good academic framework, which reveals the fundamental problem in the permissionless decentralized storage network is the Outsourcing Attack.
The solution to the Outsourcing Attack is the Proof of Replication, which is the key to ensure data security.

As we propose a new Proof of Replication by reusing the Proof of Work in Nakamoto Consensus, we design new permissionless storage network by studying existing distributed storage system.
Besides, we also proposed our Proxy Re-Encryption scheme for the permissionless storage network in this chapter.

\section{Overview}

Until now, the scenarios where the current consensus of hash power can be used without modifying the Nakamoto Consensus algorithm are very limited.
It was once believed that Useful Proof of Work did not exist\cite{buterin2019hard}, until we discovered that simple hash algorithms could be used for data encoding.
As it happens, an encoding-slow and decoding-fast algorithm\cite{2017PoRep} is also needed in a permissionless storage network to prove that users' data is being stored honestly.
The Proof of Work hash algorithm used in the Nakamoto consensus just meets the requirement.
Proof of Replication is a very computationally intensive algorithm, but it is also the core algorithm of a permissionless storage network used to combat Outsourcing Attacks.
This discovery has tightly linked blockchain and storage together and can potentially change the entire industry by using computational power for scenarios that are valuable to users.

This chapter aims to build a permissionless storage system with useful PoW.
The replication process will feed the blockchain extra computation, providing the PoW through the useful tasks.
As a result, the blockchain will get free computation from the storage network, reducing the end-users usage cost.

To achieve this goal, we studied the existing distributed storage systems and cryptographic schemes.
Distributed systems like GFS and Ceph show the great direction of how to build a permissionless storage network.
It is helpful to reference the design to build our protocol.
Existing Filecoin also comes with great lessons to learn.
One important design we learn is to avoid too complex logic on blockchain.
The permissionless storage only relies on the blockchain for payment.
The verification must be off-chain as it costs heavy computation.
Meanwhile, existing decentralized storage can simplely use logic access control.
But in the permissionless storage system, anyone could join the network as a service provider.
It is important to use cryptographic access control scheme.
We introduces Proxy Re-Encryption for the permissionless access control in this chapter.

\section{Related Work}

Many works and projects are aiming to build a blockchain-based permissionless storage system.
But our work has quite different goals.
By introducing and integrating the storage system, we aim to bring the storage to the blockchain as a new consensus engine.
Previously, the blockchain consensus and security is powered by coin as the engine.
The miners contribute the computations to win the coins as a reward.
Our work turns the meaningless hash calculation into a meaningful encoding task.
To implement such a new engine, a permissionless storage market is demanded.
We study the knowledge and designs related to the storage systems in this section.

\subsection{Data Security in Permissionless Storage Network}

\subsubsection{Outsourcing Attack}
\label{outsourcing_attack}

\begin{figure}[!t]
    \centering
    \includegraphics[width=1\textwidth]{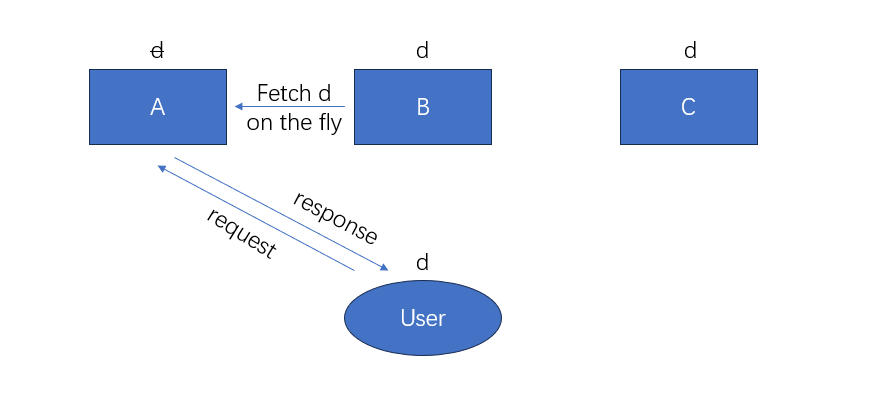}
    \caption{Outsourcing Attack}
    \label{fig_outsourcing_attack}
\end{figure}

Outsourcing Attack is a type of attack that can be launched against permissionless storage network, shown in Figure~\ref{fig_outsourcing_attack}.
Malicious storage resource providers could commit to store more data than the amount they can physically store, relying on quickly fetching data from other storage providers.

The attacker then launches an Outsourcing Attack against the network by intentionally corrupting or deleting their data, which in turn causes the nodes storing their data to return incorrect or missing data when requested by other nodes.
This can cause serious disruptions to the network and potentially compromise the integrity and availability of all data stored there.

One way to mitigate Outsourcing Attack is through replication with Proof of Work.
This involves replicating data across multiple nodes in the network and using a Proof of Work algorithm to ensure that the data is being stored honestly and accurately.
By doing so, the likelihood of an attacker successfully corrupting or deleting the data is greatly reduced, as they would need to attack a majority of the nodes storing the data to succeed.

\subsubsection{Proof of Replication}

\begin{figure}[!t]
    \centering
    \includegraphics[width=1\textwidth]{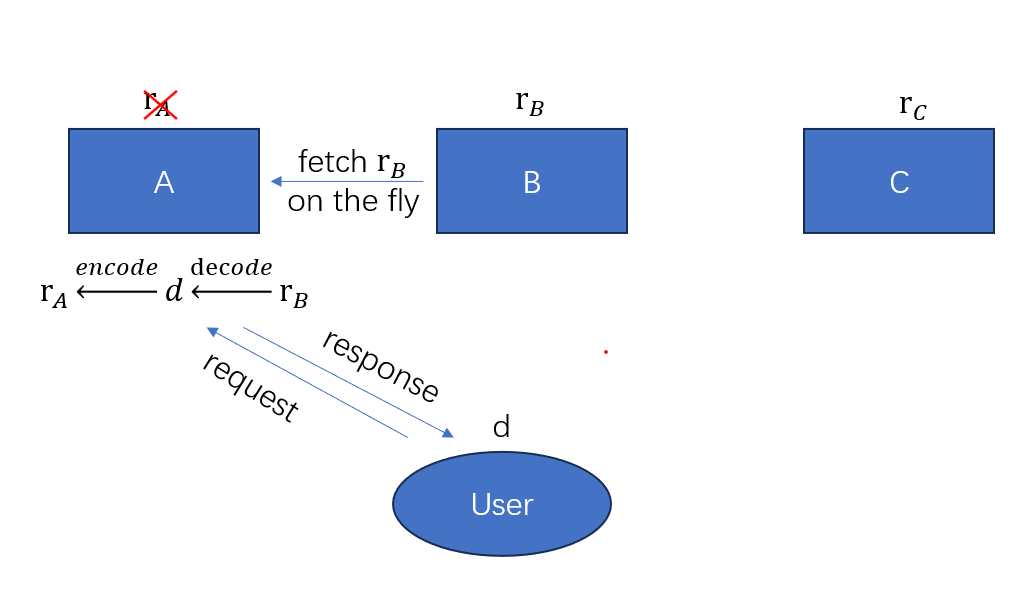}
    \caption{Proof of Replication}
    \label{fig_proof_of_replication}
\end{figure}

Proof of Replication (PoRep) is a proof algorithm in decentralized storage systems.
It is designed to verify whether a node has stored a copy of a specific piece of data or not.
PoRep ensures that the data remains safe and secure, even if some of the nodes storing the data are compromised.

The basic idea behind PoRep is to ask nodes to prove that they have made a complete and faithful copy of a piece of data by providing cryptographic proof that other nodes can verify.
This is achieved by selecting a set of unique and random challenges the node must respond to with a specific portion of the stored data.
The node then provides cryptographic proof of possession of the entire data, shown in Figure~\ref{fig_proof_of_replication}.

PoRep is a computationally expensive algorithm that requires a significant amount of disk I/O and disk space.
The goal of PoRep is to ensure that nodes can only claim to have stored a copy of the data if they actually do so.
By requiring nodes to prove their possession of the data, PoRep provides a means of detecting and rejecting nodes attempting to carry out outsourcing attacks.

Overall, PoRep is a crucial element in building secure and decentralized storage systems, and it is an active area of research in the blockchain and distributed systems communities.

\subsection{Cloud-based Storage Network}

A decentralized storage network is a network of computers and nodes that store data in a distributed manner instead of relying on a central server or service.
In a decentralized storage network, data is broken down into smaller fragments, encrypted, and stored on multiple nodes.
This ensures that the data is resilient to any single point of failure.
It also makes it difficult for one party to tamper with or access the data without authorization.

Decentralized storage networks typically use a peer-to-peer architecture, where each node communicates with other nodes to share and verify data. This approach offers several advantages over traditional centralized storage, including increased security, scalability, and reduced costs. Because there is no single point of failure, decentralized storage networks can be more resistant to cyber attacks or data breaches.

Decentralized storage networks are commonly used in blockchain-based systems, where they are used to store transaction data and other important information. They are also used by individuals and organizations who want to store data more securely and resiliently without relying on centralized cloud storage services.
Some popular decentralized storage networks include IPFS, Filecoin, and Storj.

Before diving into our work, we review the existing storage-related solutions.

\subsubsection{GFS}

\begin{figure}[!t]
    \centering
    \includegraphics[width=1\textwidth]{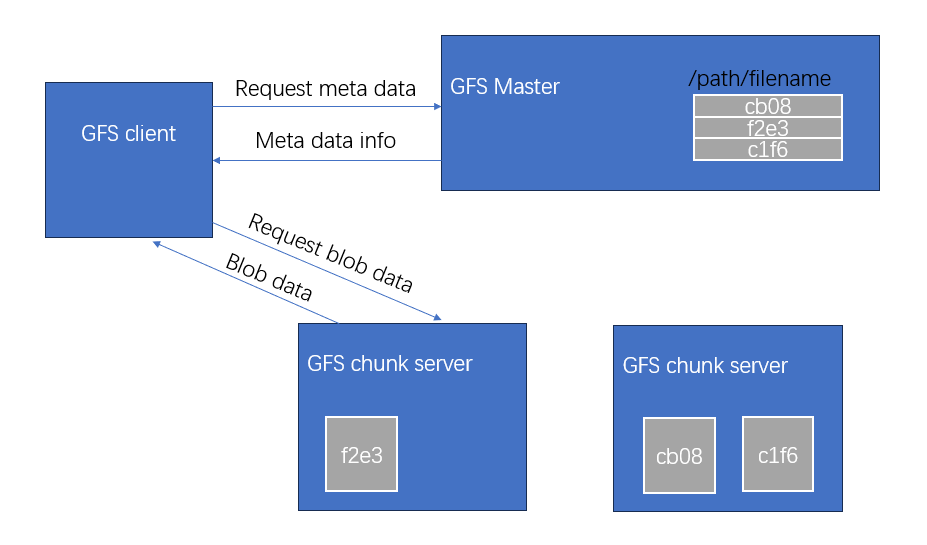}
    \caption{Google File System}
    \label{fig_GFS_overview}
\end{figure}

With the HTTP protocol\cite{fielding1999hypertext}, browser technology\cite{vetter1994mosaic,mcenery1995internet,ayers1995using}, and the popularization of the Internet, Web 1.0 allows people to create websites to share information.
With the advent of search engines and big data in Web 2.0, traditional single-machine systems cannot meet people's needs, and they need to turn to distributed systems.
GFS is the theoretical foundation of Google's big data storage.

Google File System\cite{ghemawat2003google} (GFS) is a distributed file system developed by Google to meet the needs of large-scale data processing applications.
GFS was designed to handle data-intensive applications that require the processing of large files.
Many Google applications, including Google Search and Google Maps use it.

GFS uses a master/slave architecture, with a single master node controlling multiple slave nodes, shown in Figure~\ref{fig_GFS_overview}.
The master node manages metadata, such as the location of data blocks and file access permissions, while the slave nodes store data blocks.
Each data block is replicated on multiple slave nodes to ensure high availability and reliability.

GFS is designed to be fault-tolerant, with automatic detection and recovery from node failures.
It also supports high throughput and scalability, allowing for efficient processing of large amounts of data.
GFS has influenced the development of other distributed file systems, such as the Hadoop Distributed File System (HDFS).

Overall, GFS is a key component of Google's infrastructure for processing large amounts of data and has contributed to the development of big data technologies.
It is a good reference for designing similar goal systems.

\subsubsection{Ceph}

\begin{figure}[!t]
    \centering
    \includegraphics[width=1\textwidth]{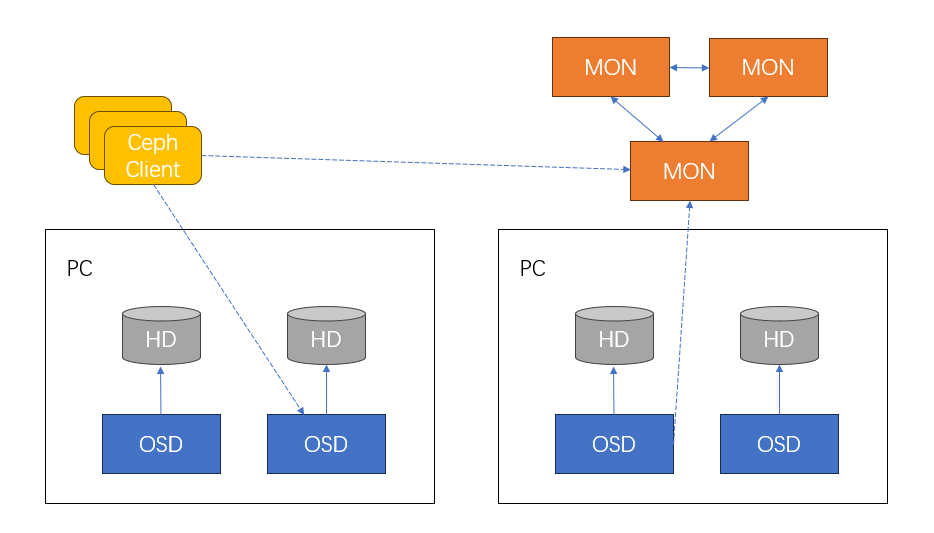}
    \caption{Ceph System}
    \label{fig_CEPH_overview}
\end{figure}

Ceph\cite{weil2007ceph,weil2006ceph} is an open-source, software-defined storage platform that provides scalable object, block, and file-system storage.
It is designed for large-scale deployments and offers high performance, reliability, and scalability.
Ceph is designed for high availability.
Compared to the simplicity and efficiency of GFS, Ceph's design has more robust scalability.
It aims to bring high-performance storage over low-price devices, such as PCs.

In Ceph, a computer process monitors a storage device named Object Storage Daemon (OSD), shown in Figure~\ref{fig_CEPH_overview}.
The monitors (MON) cluster keeps communication to ensure the high availability with Paxos consensus\cite{lamport2019byzantine,lamport2001paxos}.
There may be more than one storage device on a server; an OSD process watches each device.
% For a server, there are multiple disks, and each disk corresponds to an OSD process. 
There must be at least one monitor in the system, and each server can have at most one monitor.
The system should have at least three monitors to keep minimal high availability.
The system has an odd number of monitors ($2n+1$), which run the Paxos algorithm to discover and monitor OSDs.
The OSDs report the status to the monitors in each time fragment, called view in the epoch.
Thus, the monitors keep the knowledge of the whole storage system.
When the state of an OSD changes, the system triggers a rebalance to maintain data availability.
During the rebalance period, performance will be affected, but storage service will not be interrupted.

\begin{figure}[!t]
    \centering
    \includegraphics[width=1\textwidth]{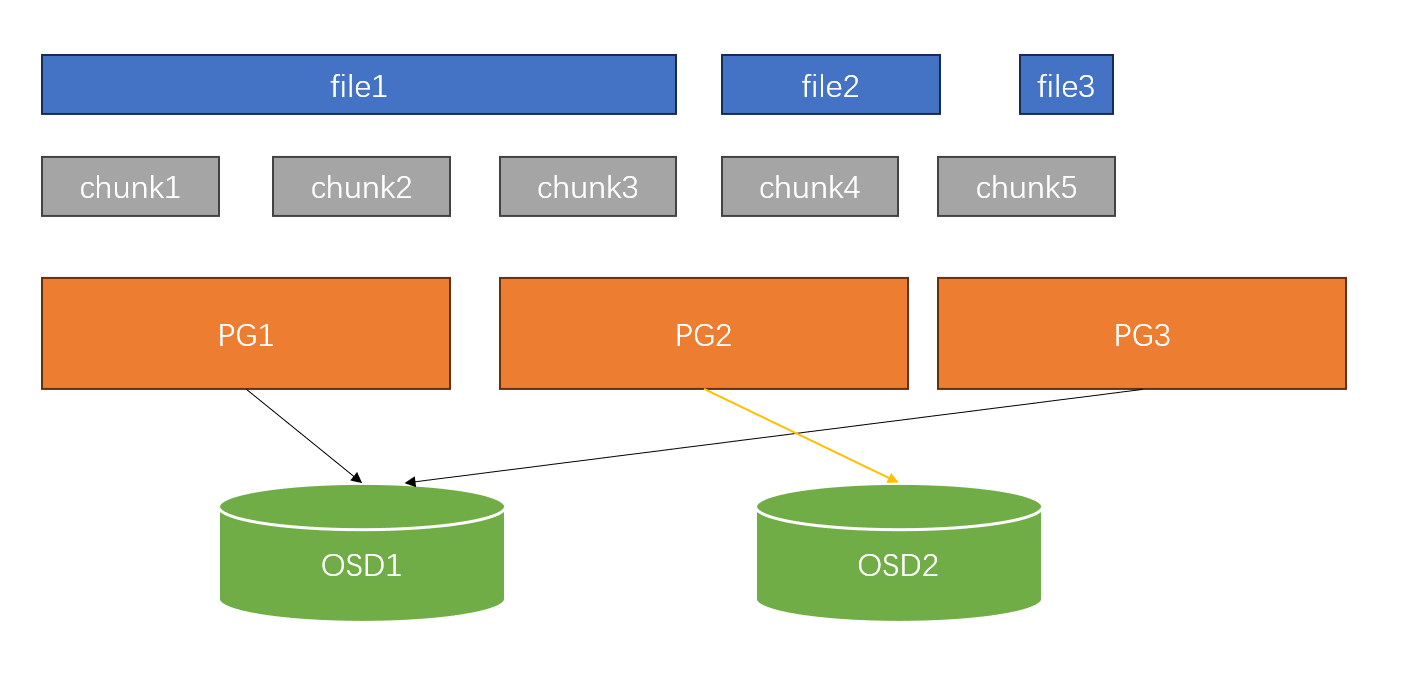}
    \caption{Ceph Placement Groups and OSDs}
    \label{fig_CEPH_PGs}
\end{figure}

Placement Groups (PG) is the logical concept of mapping data to groups.
Any data, such as user files, can be split into chunks and then kept by PGs average.
The PGs will be mapped to the different OSDs.
In the rebalance process, PGs move between OSDs, shown in Figure~\ref{fig_CEPH_PGs}.

Ceph has the key idea of controlled replication under a scalable hashing (CRUSH), pseudorandom hash algorithm.
It avoids the table look-ups like metadata query to GFS master in Figure~\ref{fig_GFS_overview}, reducing MON server's heavy load.

With the above concepts, Ceph managed to provide a Reliable Autonomic Distributed Object Store (RADOS).
Based on RADOS, the RADOS Block Device (RBD) interface serves as a hard drive in the cloud for virtual machines.
RADOS Gateway (RGW) interface serves like an Amazon S3, providing storage service like the bucket's object storage over a REST API.
Ceph File System (CephFS) interface provides a POSIX-compatible file system, which can be mounted in Linux like an NFS.

\subsection{Decentralization Storage Network}

\subsubsection{Filecoin}

Filecoin is a blockchain storage network that aims to create a more efficient and secure way of storing and accessing data.
It was created by Protocol Labs.

Filecoin allows storage providers to rent out unused hard drive space in exchange for a cryptocurrency called FIL.
This space can then be used to store files, with the blockchain network ensuring their availability and integrity through a combination of repeated Proof of Replication (PoRep) named Proof of Spacetime (PoSt).
One of the key features of Filecoin is its use of a Proof of Replication mechanism in the blockchain storage system, which ensures that data is being stored on the hard drives of storage providers.
This is done by having storage providers prove that they have replicated a piece of data to their hard drive, making it much more difficult to cheat the system.

\begin{figure}[!t]
    \centering
    \includegraphics[width=1\textwidth]{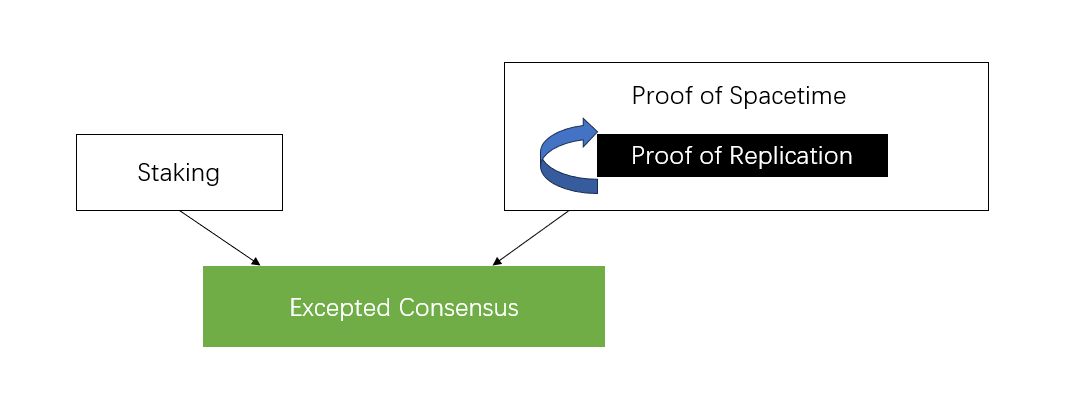}
    \caption{Filecoin Expected Consensus}
    \label{fig_Filecoin_consensus}
\end{figure}

In addition to PoRep and PoSt, Filecoin also uses the Expected Consensus (EC) as the blockchain consensus mechanism, allowing storage providers to get rewards for FIL coins by proving that storage resources are still in occupation over time, shown in Figure~\ref{fig_Filecoin_consensus}.
The EC consensus is similar to Proof of Staking, but combined with the concept of Proof of Space, the PoSt is a factor of chance to reward.
In the early times, the storage only needs to show the storage occupation with staking the FIL to win the FIL rewards.
Thus, the generated data was filled in the storage providers' hard drives at the early stage to win the FIL coins.
% This helps to ensure that files remain available and accessible over long periods of time.

Overall, Filecoin aims to create a permissionless and secure way of storing data by leveraging the power of blockchain networks and cryptography.
Its storage proof approach helps to ensure that files remain available even if individual storage providers go offline or cheat (such as Outsourcing Attack), while its use of cryptographic techniques helps to ensure the integrity and security of the data.
However, the economic model makes it hard to ensure that useful data is stored in the storage resource.
As early pathfinders, we can learn much experience from the Filecoin project.

\subsubsection{Arweave}

\begin{figure}[!t]
    \centering
    \includegraphics[width=1\textwidth]{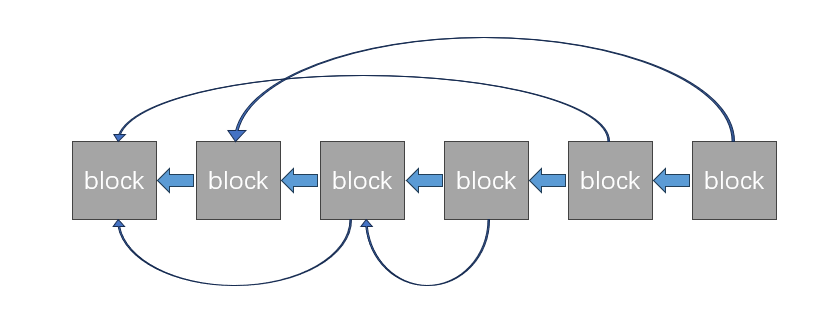}
    \caption{Arweave: Blockweave and Proof of Access}
    \label{fig_AR_blockweave_and_PoA}
\end{figure}

Arweave is a decentralized blockchain storage platform that allows users to store and retrieve data permanently and immutably.
It uses a unique Proof of Access (PoA) mechanism for both chain consensus and data integrity, shown in Figure~\ref{fig_AR_blockweave_and_PoA}.

Unlike all the blockchain storage systems, which usually only store the storage content hash information on the chain, the Arweave network applies a bold design to put content into blocks.
Users can store their data on the network by paying a one-time fee, which is used to reward miners who provide the storage space.
Once the data is stored on the network, it is replicated across multiple nodes, ensuring it is always available and cannot be lost or tampered with.

Unlike a traditional blockchain, where each entry in the ledger is linked to the one before it, Blockweave links to the entry before it and a randomly selected previous block in the system, weaving the data together. 
In addition, a consensus algorithm called Proof of Access incentivizes miners to store and provide access to the data rather than just validating transactions.

Arweave has many use cases, including storing archival data, web pages, applications, and digital assets.
It has gained popularity in the NFT space as a way to store and retrieve user data and digital assets securely and efficiently.
However, Arweave can only provide the permanent storage.
It does not allow a user to remove the content.
Thus, the payment for the content is expensive as it covers the lifetime storage cost.
This is due to its blockweave and PoA design, which means the system can only provide such a business model to the users and miners.

\subsubsection{Chia}

Chia is an open-source, decentralized blockchain platform that uses a new consensus mechanism called Proof of Space (PoSpace).
PoSpace is designed to be more environmentally friendly and accessible than other consensus mechanisms, such as Proof of Work (PoW), which requires miners to consume large amounts of electricity on computation to earn rewards.

In Chia, miners allocate unused disk space to form the consensus for the network and exchange for block rewards for Chia coins (XCH).
The more disk space a miner allocates, the higher their chances are of winning a block reward.

Compared with Filecoin and Arweave, Chia is not a storage project.
The disk space is used for consensus, but no useful data is stored within those hard disk resources.
Although the energy is saved, the chain must maintain high-cost storage resources for competition and security.
Thus, Chia's design did not address the conflict between high security and low usage cost.

\subsubsection{Problem of Blockchain Storage}

Filecoin and Arweave combine blockchain with storage, where miners or nodes provide storage space, and the blockchain acts as an independent entity that is not controlled by any third parties to provide permissionless storage services to end users.

With the popularity of blockchain, the concept of blockchain storage networks has become a trend in recent years.
But unlike asset blockchains such as Bitcoin and Ethereum, end-users have not widely accepted storage blockchains.

The concept of blockchain storage learns the use of coins in asset blockchains.
When asset blockchains use coins to force the consumption of native coins, the blockchain obtains the computing power needed for security.
However, the forced use of coins as a value exchange for storage in storage blockchains does not benefit the storage itself.
Still, the disadvantage is that users and storage providers have to use floating-value coins to settle storage business, which creates more possibility of impermanent loss.

\subsubsection{Nostr Protocol}

\begin{figure}[!t]
    \centering
    \includegraphics[width=1\textwidth]{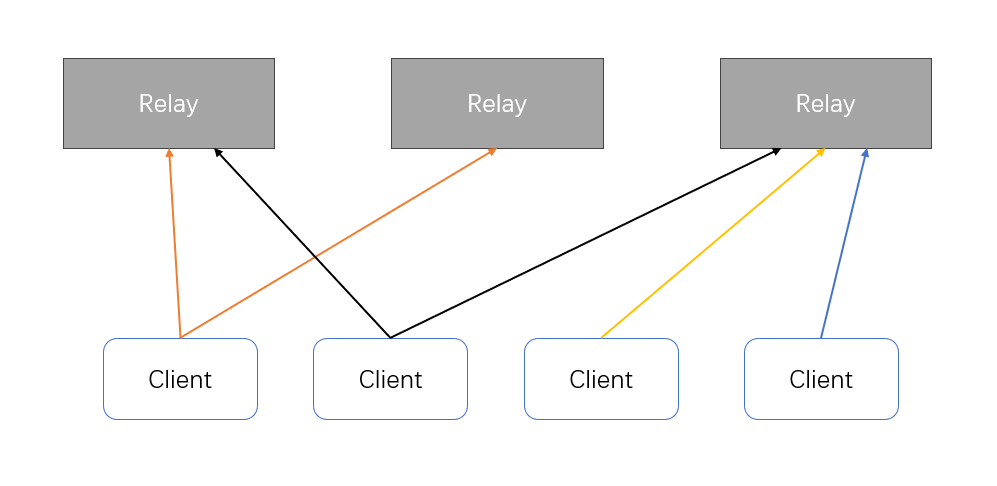}
    \caption{Nostr Architecture}
    \label{fig_nostr_arch}
\end{figure}

In recent years, the mode of Ethereum has become a popular paradigm of blockchain app design.
The term dapp is defined by Ethereum, indicating the smart contract development on Ethereum.
Due to the success of decentralized finance (DeFi), the concept was proved in the finance application field.

However, in the other fields of blockchain, such as the storage network, the dapp mode is hard to apply.
The smart contract lacks the ability to visit off-chain data to verify the user storage.

Nostr protocol was proposed to use cryptographic method for the decentralized social network.
Its architecture is simple that the relies does not communicate with each other, shown in Figure~\ref{fig_nostr_arch}.
We borrow the concept to build the permissionless storage network away from blockchain, except that the payment still use the stablecoin over blockchain.

% \section{Related Work}

\subsection{Proxy Re-Encryption and Proof of Replication}

In 1998, Blaze, Bleumer, and Strauss\cite{blaze1998divertible} proposed the first proxy re-encryption scheme based on a cyclic group\cite{elgamal1985public}.
In 2010, Weng et al.\cite{weng2010chosen} proposed a CCA and collusion-resilience PRE scheme.
After 2010\cite{nunez2017proxy}, most PRE schemes\cite{canetti2007chosen,libert2008unidirectional} were based on bilinear pairing\cite{chu2007identity}\cite{matsuo2007proxy}\cite{ateniese2009key} or lattice algebra structure\cite{kirshanova2014proxy}.
The data uploaded must be encrypted in the decentralized storage network.
However, most PRE schemes generate the new ciphertext during the re-encryption.
A ciphertext would be replicated with each miner's unique identity.
Any modification of the ciphertext would lead to more expensive PoRep operations.
A PRE scheme is ideal for not generating new ciphertext during the frequent permission-sharing actions under the decentralized storage network scenario.
Thus, we propose the new PRE scheme.

While the new PRE schemes are diving into more complex algebra structures, the use scenarios of PRE still need to be improved.
As business companies back cloud services, the cloud and mobile still need to utilize PRE schemes fully.
Encryption will prevent data analysis and incur extra storage and computation costs.
Owing to the blockchain, we foresee that blockchain-based decentralized applications will heavily rely on cryptographic schemes.
Web3 allows the user to own their data.
The decentralized storage network requires a pure cryptographic access control feature.
PRE is ideal, but PoRep is mandatory.

The first blockchain, Bitcoin, proposed the Proof of Work (PoW) as the consensus algorithm\cite{2008Bitcoin} after PoW was used earlier for anti-spam purposes\cite {dwork1992pricing}.
Computation was used as the resource for consensus, such as voting.
Later, the storage space as a resource was studied, which can be classified into two categories.
The Proof of Space intends to replace Proof of Work as the consensus algorithm.
This replacement can bring down the cost of electricity by PoW, but junk data needs to be filled in the hard disk so far.
Conversely, the Proof of Storage algorithm focuses on storing useful data.
However, this algorithm cannot agree on a consensus.
It only shows the proof of the data stored.
Proof of Replication is an extension of Proof of Storage, which convinces the owner that the unique storage resource keeps the data.
In the permissionless blockchain, the PoRep is the key algorithm.
It is nice to design schemes working with PoRep for the storage features.
Filecoin uses SDR\cite{2017SDR} as the PoRep encoding and proves with the zero-knowledge-based algorithm\cite{2018PoSpace,2018scaling}.
Filecoin lets users decide how to encrypt their data.
Therefore, cryptographic access control for decentralized storage networks has yet to be implemented.
The improved PRE scheme is worth studying.

In 1998, Blaze, Bleumer, and Strauss\cite{blaze1998divertible} proposed the first proxy re-encryption scheme.
The ciphertext can be re-encrypted into another by the proxy authorized by the owner.
Although the first PRE scheme is not collusion-resilient, it shows the possibility of changing the ciphertext's key or password without decryption.
In 2003, Ivan and Dodis\cite{ivan2003proxy} proposed the group-based proxy cryptography scheme.
The secret key is first divided into two parts in their unidirectional scheme.
This is the primary technique used for collision safety by many schemes.
This illuminates our idea.
In 2009, Shao et al.\cite{shao2009cca} proposed a CCA-secure scheme without pairing.
Their scheme uses double trapdoors with the big prime multiplication as the secret key.
One year later, Weng et al.\cite{weng2010chosen} proposed a CCA-secure scheme WDLC10 without pairing.
Two key pairs are used to avoid collusion for the secret key in their scheme, which is similar to Ivan and Dodis\cite{ivan2003proxy}.
Most CCA-secure schemes were based on bilinear pairing or lattice in the following years.
The PRE schemes such as AFGH06\cite{ateniese2006improved} and GA07\cite{green2007identity} are based on bilinear pairing. 
XT10\cite{xagawa2010proxy} and ABPW13\cite{aono2013key} are based on lattice (LWE).
NAL15a\cite{nunez2015ntrureencrypt} is based on lattice (NTRU).

Let us take a look at how WDLC10\cite{weng2010chosen} works.
To achieve collusion resilience, WDLC10 uses two key pairs.
The core idea of preventing collusion is to use the sum of two secret keys instead of one.
This results from the fact that the delegatee and proxy can only work together to obtain the sum of the keys but cannot learn the value of each secret.
The two keys are used for two different layers of ciphertext.

Our scheme achieved collusion resilience with one regular public/secret key pair.
Under the general concept of asymmetric encryption, the cleartext can be encrypted with a public key, and the ciphertext is decrypted with the corresponding secret key.
However, in the PRE scenario, we slightly changed the definition.
Assuming that anyone can create PRE ciphertext with the given public key, the malicious user could keep the crucial internal value, which should be discarded during the encryption.
The internal value can be used to generate new re-keys, where the owner should control the access permission to the ciphertext.
In this case, the ciphertext generated with Alice's public key may not be controlled by Alice.
This may lead to security issues.
The proposed scheme uses the secret key for encryption and decryption to ensure that only the owner can create the ciphertext.
Alice can generate a re-key with her secret and Bob's public keys.

To summarize, BBS98\cite{blaze1998divertible}, Ivan and Dodis\cite{ivan2003proxy}, and WDLC10\cite{weng2010chosen} use groups.
WDLC10\cite{weng2010chosen} improved many features compared to BBS98 (including the most important feature, collusion resilience).
To archive collusion resilience in "hashed" ElGamal, two key pairs are required for WDLC10\cite{weng2010chosen}.
Shao et al.\cite{shao2009cca} used double trapdoors.
Most of the other PRE schemes are based on bilinear pairing or lattice.

\section{Storage System Design}

In Chapter~\ref{ch:consensus}, we introduced useful Proof of Work. In addition to maintaining the original use of Proof of Work in blockchain consensus, we proposed to reuse the hash operation in Proof of Work to encode data files.
This chapter gives more details on the hash usage of the data encoding.
We design and implement the permissionless storage network for the goals of both storage data and supply blockchain security. 

\subsection{Replication with Proof of Work}

\begin{figure}[!t]
    \centering
    \includegraphics[width=0.8\textwidth]{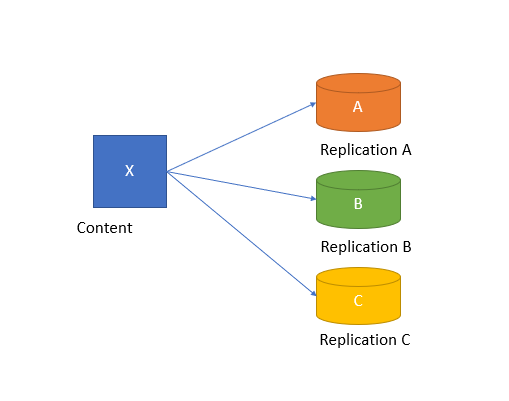}
    \caption{Encode for Proof of Replication}
    \label{fig_replication_encode}
\end{figure}

In the previous Section~\ref{outsourcing_attack}, we introduced Outsourcing Attacks and proof of replication in a permissionless file storage system.
The Proof of Replication solves the attack issue with the core idea of an encoding algorithm that is slow to encode and fast to decode.
We propose to use the Nakamoto Proof of Work algorithm, which consumes a lot of computing power in blockchain consensus, as the encoding algorithm for the Proof of Replication that is mandatory in permissionless file storage systems to ensure storage data security, shown in Figure~\ref{fig_replication_encode}.

In the traditional field of encoding algorithms, efficiency is pursued.
Using hash as a means of encoding and decoding is undoubtedly inefficient.
However, in a permissionless storage network, the hash of Proof of Work for encoding perfectly meets the asymmetric requirements of the encoding algorithm in Proof of Replication.
Moreover, we can adjust the encoding difficulty and the difficulty ratio between encoding and decoding.

\subsection{Storage Architecture}

\begin{figure}[!t]
    \centering
    \includegraphics[width=1\textwidth]{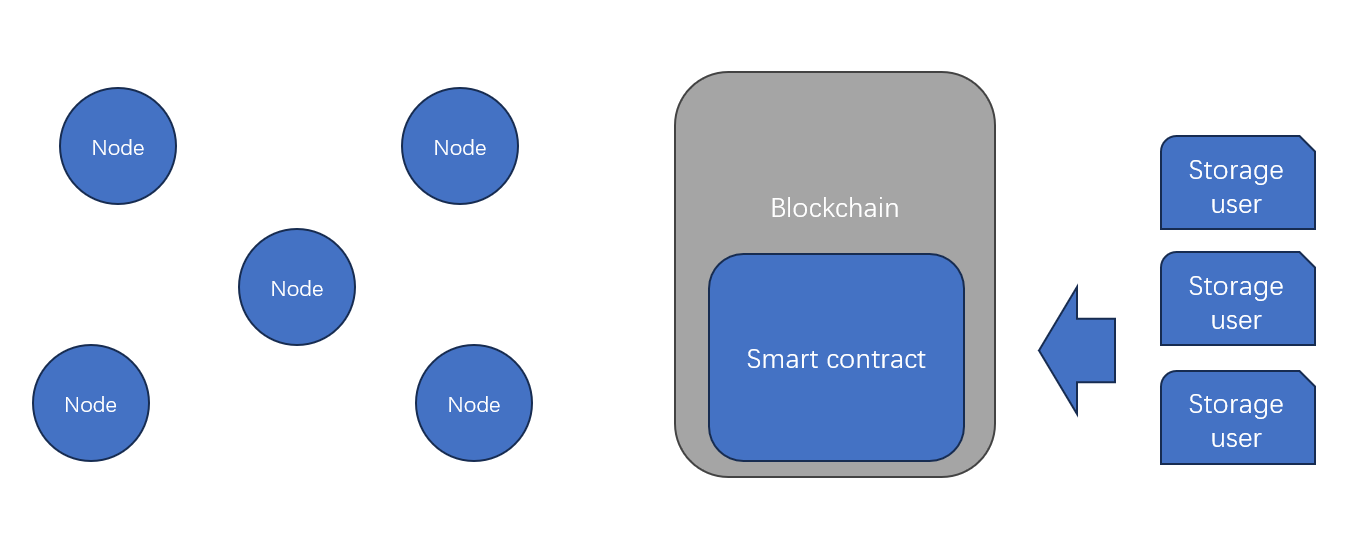}
    \caption{Storage Architecture}
    \label{fig_storage_arch}
\end{figure}

Based on the experience of previous blockchain storage and inspired by the nostr protocol, we designed a permissionless storage network that minimizes the reliance on on-chain operations of the blockchain, shown in Figure~\ref{fig_storage_arch}.
The permissionless storage network can be used as a security engine for blockchain and make payments for storage via smart contracts.
Our improvement is that the complex storage logic, such as verification, goes off-chain as much as possible.

Our storage design comprises users, nodes, resource providers, and blockchains with smart contracts.
Users submit a file storage request, upload the file data to the node, and call the smart contract to start storage.
Blockchain is mainly used for the payment.

\subsection{Storage Protocol}

In a permissionless storage system, there are the roles of the user $U$, the storage resource provider $P$, the verifier $V$, the storage node $N$, and the blockchain $B$.

\subsubsection{Setup}

In the setup phase, the nodes pre-register on the blockchain's smart contract $B$.
User $U$ initializes the on-chain request through blockchain $B$ on a smart contract.
The smart contract call returns which storage node is the primary node $N$ and how many redundancies the system will keep for files of $U$.
The smart contract pre-charges $U$ some stablecoins, such as USDT, so $N$ can request $B$ to pay $P$.

Meanwhile, the other nodes can apply to blockchain B's smart contract as the backup nodes N2 and N3, depending on the redundancy size.

\subsubsection{Process}

\begin{enumerate}
    \item The user $U$ sends a request to node $N$ for an action such as upload/new, rename/move, or remove a file with a signature.
    \item The primary node $N$ for $U$ calculates the new file system state of $U$ from the existing state and the operation, then confirms the state via blockchain $B$.
    \item Once confirmed, the resource providers $P$s would download the chunks from $N$ and replicate them with an extended PoW algorithm with their unique ID. After replication is done, the $P$s would report the replicated content hash to node $N$.
    \item The verifier $V$ download the replicated chunks.
    \item The blockchain system $B$ issues a challenge $c$. The resource provider $P$ responds to the challenge with the calculated proof $p$.
    \item The verifier $V$ checks the proof $p$. If incorrect, the $V$ would inform $N$ of the proof that $p$ is wrong. $N$ would perform a recheck. 
    \item If the recheck shows $p$ is wrong, $N$ would pause $P$'s payment and $V$ would get a reward. $N$ would transfer the duty of chunk storage to another $P$.
    \item If the $N$ is offline or not responding, the $V$ may propose to make another $N$ from the backup nodes as the new primary node.
\end{enumerate}

\subsubsection{Protocol}

\begin{table}[h!]
    \centering
    \caption{Data fields recorded in the blockchain smart contract}
    \label{storage_data_fields}
    \begin{tabular}{| c | c |} 
        \hline
        % Col1 & Col2 & Col2 & Col3 \\ [0.5ex] 
        % \hline\hline
        User data size & Used for calculate the payment rate. \\ 
        \hline
        Data replication number & It costs more with more redundancies. \\
        \hline
        User primary node & The node in charge of the user metadata of the file space. \\
        \hline
        User backup nodes & A list of backup nodes. \\
        \hline
        User prepaid balance & In stable coin and no refundable. \\
        \hline
        Last payment time & UNIX timestamp. \\
        \hline
        Storage start time & UNIX timestamp, inmutable. \\
        % \hline
        % Date & 2023/11/17 \\
        \hline
    \end{tabular}
\end{table}

The client can perform operations like uploading, renaming, and removing the files to the primary node once the blockchain assigns a primary node for a user.
The client first uploads a file via HTTP or HTTPS to the primary node.
The primary node calculates the hash of the uploaded content.
Then, the client initializes the request for the file addition.
With the signed request from the user and the verified content, the primary node changes the user file space.
Depending on the size of new uploaded content, the primary will assign the storage providers for the new content.
The same is true for the rename and remove file operations; the primary node will change the state of file space for the user.

The primary node of a user will interact with the blockchain smart contract.
This update will only change the total size and the redundancy of the user content, which will impact the storage's payment rate.
The user file space information is not submitted to the smart contract.
A storage node can be a primary or a backup node for different users.

\begin{figure}[!t]
    \centering
    \includegraphics[width=1\textwidth]{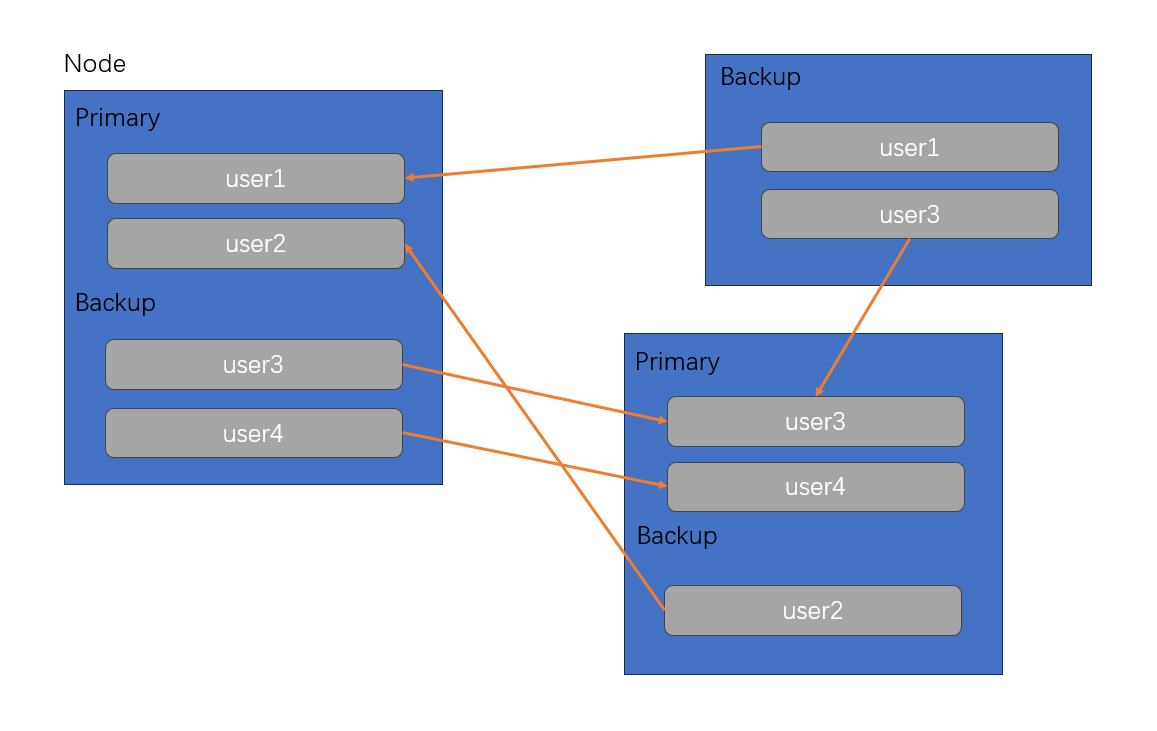}
    \caption{Storage node as primary and backup role for user}
    \label{fig_storage_primary_backup}
\end{figure}

The backup nodes will watch the change of the primary node of the particular users, shown in Figure~\ref{fig_storage_primary_backup}.
A node can play the primary node for some users and the backup node for others.
Any unexpected failures, such as system crashes or network connection loss, including the primary node cheating, will be verified by the backup nodes.
In the smart contract, the node's information is recorded in order.
The second node will take the primary role if the primary node is not working.

\subsection{Verification}

In storage verification, we adopt an optimistic mode.
The blockchain system issues a challenge.
Each service provider responsible for storing data must prove that it is honestly storing data periodically, so it is the prover.
Finally, for each chunk of replication, there can be many verifiers. If the verifier does not explicitly point out that the prover's response to the challenge is wrong, the challenge is considered to be passed.
At the same time, if the response is pointed out to be wrong, there will be a reward.

\subsubsection{Naive Verification}

The naive verification requires the verifiers to download the replicated file data from the storage provider every time.
The replicated file data hash was committed in the blockchain or off-chain.
The verifier can easily check the hash of replicated data.

\subsubsection{Interactive Verification}

Since the naive verification requires downloading each time, an improved verification method is to initiate interactive challenges and responses.
This method only requires the verifier to download and save the replicated data at begin, which saves a lot of bandwidth.

\subsection{Storage Incentive}

In Filecoin and Arweave systems, blockchains issue their coins or tokens to pay for storage.
In Bitcoin and Ethereum, the native coin is mainly used to pay for transaction fees or gas.
Existing blockchain file storage systems attempt to use the native coin to pay for storage fees, and resource providers will receive blockchain storage fees denominated in coins.

Since the price of coins is highly volatile compared to USD or stablecoins, consumers and storage providers may suffer losses when coin prices go up or down.
Although the price of storage hardware has been declining in recent years, the cost of storage is relatively stable in the short term.
Therefore, using stablecoins rather than coins to pay for storage service fees is more reasonable.

It is believed the reason for using coins for storage is to give coins more use scenarios, which may increase the coin price.
However, there is a difference between using coins for investment and value settlement.

If the native coins of Bitcoin or Ethereum lose their fundamental use case: paying transaction fees on the chain. Miners will find it challenging to sell native coins after investing resources in mining because there are almost no use cases, and only the storage value function remains.
Therefore, it is a correct design to find use cases for coins in blockchains, but using native coins to price services may not be a suitable scenario. Our analogy is that using company stock to buy breakfast is inappropriate.
In the exchange of service value, stablecoins are more suitable.

We designed to use stablecoins to pay for the permissionless storage.
For end users, it is easy to estimate the generated fees accurately.

Unlike general smart contracts, in addition to transfers between ordinary users, the blockchain will reward storage providers based on the evidence submitted periodically by storage resource providers.
As long as the storage provider correctly responds to the storage challenge, the storage provider can receive stablecoin income periodically.

\subsubsection{Analysis}

The blockchain storage was a popular topic in the recent years.
With the successful of Bitcoin and Ethereum, people look forward to expand blockchain to the field more than the pure finance.
The attempts to build blockchain based storage are still continuing but not in massive adoption yet.

There are several aspects preventing people adopting blockchain based storage.
The primary issue is about the cost.
The Proof of Replication must be adopted in permissionless storage network, including blockchain storage.
Existing systems such as Filecoin uses zero knowledge based verification, leading to an expensive jump in the blockchain storage cost.
Not only the storage is expensive, the blockchain usage cost is high enough.
Comparing with the DeFi dapp who helps users making money, people are not willing to use storage application especially when both the storage and chain cost are too high.
In our solution, we make permissionless storage system less rely on the blockchain as much as possible.
Moreover, the storage does not use ZK based verification.
The storage system also help the blockchain to reduce the end users cost.
The protocol tries to bring down the user buden as much as possible.

The existing blockchain storage also uses the coins like most of the public blockchain systems.
Beyond that, the coins in blockchain storage are not only used for transaction but also for the storage.
This makes the storage price fluctuation like the transaction fee.

\subsection{Discussion}

This design uses nostr-based architecture for a permissionless storage network.
Compared with the previous blockchain storage design for the permissionless storage network, the new design splits the verification role from the blockchain.
The storage node judges the malicious storage provider based on the cryptographic storage proof.

The smart contract on blockchain $B$ keeps the user's payment, ensuring no trusted third party takes the pre-charged storage fee.
The storage node $N$ plays the role of metadata server and sometimes as the CDN to cache user data.
The data chunks are kept by the storage provider $P$s.
It is similar to the design of GFS, as the main difference is that our protocol runs over a permissionless environment.
Thus, the verifier role $V$ should check the $P$ periodically.

According to Ceph, the storage node $N$ plays the MON role in Figure~\ref{fig_CEPH_overview}.
The difference is that MON is in charge of the whole picture of OSDs in the system, but the storage node $N$ only cares for specific users.
The MONs run the Paxos algorithm to ensure the leader is live or vote to change the leader.
The storage node $N$ and its backups perform a similar mechanism through the blockchain $B$ as the consensus is built into the blockchain system.
It is easy for the chain to get all the backup nodes of $N$.
The rule can be set that when $2/3$ backup nodes vote to mark $N$ as invalid, the blockchain $B$ would pass the leader role to the next backup node.

\section{Proxy Re-Encryption for Permissionless Storage}

The permissionless storage system provides the user with a private file space to store data.
By default, cryptography should be applied in this system to ensure the uploaded content is encrypted.
A significant difference between permission and permissionless storage is that the permission system can use an authenticated mechanism to control users' access permission.
If the system access control is secure, encrypting the user's content is unnecessary.
But in a permissionless system, the content is kept by semi-trusted storage providers worldwide.
Cryptography must be applied to the content.

Usually, standard symmetric encryption can be used in a permissionless storage system such as AES.
The user encrypts the content locally and uploads the cipher text onto the storage node.
However, the access control requires the encrypted content to be accessed by the specific user with permission granted.
This suggests more advanced cryptography rather than existing password-based symmetric encryption.

Proxy Re-Encryption is the cryptography scheme for cryptographic access control.
It enables the pure cryptographic method to grant access permission for the storage system.
We propose a Proxy Re-Encryption scheme for the permissionless storage network.

\subsection{PRE Model}

We review the model of collusion-resilience PRE.
A CCA collusion-resilience proxy re-encryption scheme is an algorithm:
\begin{equation}
 (KeyGen, RenKeyGen, Enc, ReEnc, Dec).
\end{equation}

% $Setup(k)$:

$KeyGen()$: The algorithm generates the public/secret key pair $(pk, sk)$.

$ReKeyGen(sk_A, pk_B)$: The re-encryption key generation algorithm accepts the secret key $sk_A$ of Alice and the public key $pk_B$ of Bob.
It outputs a re-encryption key $rk_{A \rightarrow B}$.

$Enc(sk, m)$: The encryption algorithm takes the public key $sk$ and the clear message returns the encrypted message $c$.

$ReEnc(rk_{A \rightarrow B}, c_A)$: The re-encryption algorithm transfers the encrypted message $c_A$ into the ciphertext $c_B$ using the re-encryption key $rk_{A \rightarrow B}$.
% Bob's secret key can decrypt The transformed ciphertext $c_B$.
Bob's secret key can decrypt the transformed ciphertext $c_B$.

$Dec(sk, c)$: The user decrypts the ciphertext with his secret key and encrypted $c$, e.g., $c_A$ or $c_B$.
It outputs the cleartext $m$.

\textbf{{Correctness.}
} Correctness is ensured for any $m \in \mathcal{M}$ and any key pair of $(pk_A, sk_A)$, $(pk_B, sk_B)$, following the conditions Equations (7) and (8):
\begin{equation}
Dec(sk_A, Enc(sk_A, m)) = m,
\end{equation}
\begin{equation}
Dec(sk_B, ReEnc(ReKeyGen(sk_A, pk_B), Enc(sk_A, m))) = m.
\end{equation}

% $$

\textbf{{Security definition.}}
Security for a CCA collusion-resilience PRE scheme is defined in the game between an adversary $\mathcal{A}$ and a challenger $\mathcal{C}$.
There are two ciphertexts from the cleartext message for the PRE scheme: encrypted cipher $m \rightarrow c_A$ and re-encrypted $m \rightarrow c_B$ are required for chosen-ciphertext security.
\vspace{2mm}

\textbf{{Phase 1.}} The adversary $\mathcal{A}$ issues queries $q_1, \dots, q_m$, of which $q_i$ is one of the following:
\begin{itemize}
  \item Uncorrupted key generation query: The challenger $\mathcal{C}$ computes $(pk_i, sk_i) \leftarrow KeyGen()$, and sends the $pk_i$ to the adversary $\mathcal{A}$.
  \item Corrupted key generation query: The challenger $\mathcal{C}$ computes $(pk_j, sk_j) \leftarrow KeyGen()$, and sends the $(pk_j, sk_j)$ to the adversary $\mathcal{A}$.
  % \item Corrupted key generation query: The challenger $\mathcal{C}$ computes $(pk, sk) \leftarrow KeyGen()$, and sends the key pair $(pk, sk)$ to the adversary $\mathcal{A}$.
  \item Re-encryption key generation query: The challenger $\mathcal{C}$ computes\\
   $rk_{1 \rightarrow 2} \leftarrow ReKeyGen(sk_1, pk_2)$, and sends the $rk_{1 \rightarrow 2}$ to the adversary $\mathcal{A}$. Here, $sk_1$ and $pk_2$ must be from different key pairs. This query allows any key pair except that $\mathcal{A}$ cannot know the value of $sk_1$.
  % including $sk_1 = sk_A$ and $pk_2 = pk_B$ 
  % \item Encryption query: The challenger $\mathcal{C}$ computes $c \leftarrow Enc(pk, m)$, and sends the ciphertext $c$ to the adversary $\mathcal{A}$. $pk$ can be any public key including $pk_A$ or $pk_B$.
  \item Re-encryption query: The challenger $\mathcal{C}$ computes $c_2 \leftarrow ReEnc(rk_{1 \rightarrow 2}, c_1)$, and sends the ciphertext $c_2$ to the adversary $\mathcal{A}$.
  \item Decryption query: The challenger $\mathcal{C}$ computes $m \leftarrow Dec(sk, c)$, and sends the cleartext $m$ to the adversary $\mathcal{A}$. Here, $sk$ cannot be $sk_1$ or $sk_2$.
\end{itemize}

\textbf{Challenge.} After the adversary $\mathcal{A}$ ends up Phase 1 , $\mathcal{A}$ chooses from two equal-length messages $m_0, m_1 \in \mathcal{M}$, and sends to the challenger $\mathcal{C}$.

% \textls[-5]{
The challenger $\mathcal{C}$ receives $m_0, m_1$. $\mathcal{C}$ flips a random coin $\delta$ Equation (9), and computes $c$,
% }
\begin{equation}
\delta \leftarrow \{0,1\},
\end{equation}
\begin{equation}
c \leftarrow Enc(sk_A, m_\delta),
\end{equation}

% $$y \leftarrow \{0,1\}$$
% $$c_B \leftarrow ReEnc(ReKeyGen(sk_A, pk_B), Enc(sk_A, m_y))$$
then sends the $c$ Equation (10) back to the adversary $\mathcal{A}$.

\vspace{2mm}

\textbf{Phase 2.} The adversary $\mathcal{A}$ continues to issue queries $q_{m+1}, \dots, q_{max}$, which $q_i$ can be one of the queries:
\begin{itemize}
  \item Uncorrupted key generation query: The challenger $\mathcal{C}$ responses are the same as in Phase 1.
  % \item Uncorrupted key generation query: The challenger $\mathcal{C}$ responses same as in Phase 1.
  \item Corrupted key generation query: The challenger $\mathcal{C}$ responses are the same as in Phase 1.
  \item Re-encryption key generation query: The challenger $\mathcal{C}$ responses are the same as in Phase 1.
  % \item Encryption query: The challenger $\mathcal{C}$ responses same as in Phase 1. Both $m_0$ and $m_1$ can be used in this query.
  \item Re-encryption query: The challenger $\mathcal{C}$ responses are the same as in Phase 1.
  \item Decryption query: The challenger $\mathcal{C}$ responses are the same as in Phase 1, except that $c \neq c_A$ and $sk \neq sk_A$, or $c \neq c_B$  and $sk \neq sk_B$.  

\end{itemize}

\textbf{Guess.} The adversary $\mathcal{A}$ outputs $\hat{\delta} \in \{0,1\}$.
\vspace{2mm}

Referring to adversary $\mathcal{A}$ as an IND-PRE-CCA adversary, we define the advantage of the adversary $\mathcal{A}$ in attacking scheme $\Pi$ as
\begin{equation}
 \mathbf{Adv}_{\Pi, \mathcal{A} }^{IND-PRE-CCA} = | \Pr[\delta=\hat{\delta}] - \frac{1}{2}|.
\end{equation}

% \begin{Definition}
A PRE scheme $\Pi$ is said to be $ (t, q_u, q_c, q_{rk}, q_{re}, q_d, \epsilon )$-IND-PRE-CCA secure,
if for any t-time, IND-PRE-CCA adversary $\mathcal{A}$ makes at most $q_u$ uncorrupted key generation queries, 
at most $q_c$ corrupted key generation queries, at most $q_{rk}$ re-encryption key generation queries, at most $q_{re}$ re-encryption queries, and at most $q_d$ decryption queries; thus we have
\begin{equation}
 \mathbf{Adv}_{\Pi, \mathcal{A} }^{IND-PRE-CCA} \leq \epsilon.
\end{equation}
% \end{Definition}

\subsection{PRE Features}
% \label{sec4}

Our PRE scheme is adopted with the PoRep in a permissionless decentralized storage network.
The ciphertext $ReEnc$ is an optional operation in the definition.
This CCA and collusion-resilience PRE scheme is based on "hashed" ElGamal.
ElGamal is one of the most important asymmetry cryptographic schemes based on CDH assumption.
Both discrete logarithm and ECC can be used for the ElGamal implementation.

\subsubsection{Features}

\textbf{{Collusion resilience.}} Collusion resilience (collusion safe) states that the proxy and the delegate (Bob) can collude to obtain the delegator's (Alice) secret key.
In BBS98\cite{blaze1998divertible}, $rk_{A \rightarrow B} = \frac{sk_B}{sk_A} $, proxy and delegate (Bob) can calculate $sk_A = rk_{A \rightarrow B} \cdot sk_B $.
Collusion resilience (collusion safe) is an important feature.
In any case, $sk_A$ should be safe because it is related to more than the current ciphertext.
All the ciphertexts generated by Alice are bound with the security of $sk_A$.
Our scheme is collusion resilience due to the novel method of re-key generation, inspired by the bidirectional scheme of Ivan and Dodis\cite{ivan2003proxy}.

\textbf{Bidirectional.} Delegation from $A \rightarrow B$ allows re-encryption from $B \rightarrow A$.
It is observed that unidirectional and bidirectional delegation can be applied in different use cases.
It is nice to distinguish between unidirectional and bidirectional proxy re-encryption.
The bidirectional PRE refers to the fact that it can generate $rk_{B \rightarrow A}$ from $rk_{A \rightarrow B}$.
% Unidirectional is usually harder to design.
WDLC10\cite{weng2010chosen} used the two layers for unidirectional encryption, where layer 2 cipher can be converted into layer 1 cipher by $rk_{A \rightarrow B} = \frac{\Delta}{sk_{A1}+ sk_{A2}}$.

The bidirectional scheme means the re-encrypted ciphertext can transfer back to the original cipher.
It depends on how the re-encryption key is designed.
In BBS98, $ rk_{A \rightarrow B} = \frac{sk_B}{sk_A} $ and the reversed key $ rk_{B \rightarrow A} = \frac{sk_A}{sk_B} $ can be easily calculated.
Obviously, this reversed encryption key can be applied to all the ciphertext generated by Bob.
In the Ivan and Dodis 2003 bidirectional ElGamal scheme, $rk = g^{r(x_2-x_1)}$ also can be reversed, but due to the random $r$, the reversed key can be only applied to Bob's current ciphertext.
Comparing the two scenarios, BBS98's bidirectional feature leads to more privacy issues than Ivan and Dodis\cite{ivan2003proxy}.
% In our scheme it sounds that Bob can fake Alice's ciphertext by using the Alice's $rk_{A \rightarrow B}$, but this requires B to obtain Alice's random number $r$ so that B can generate the similar cipher and re-encrypt back with Alice's $rk_{A \rightarrow B}$.

\textbf{Noninteractive.} The re-encryption key generation requires Alice to use Bob's public key.
Bob is not involved in the interaction of re-key generation.

\textbf{Proxy invisibility.} The user sending messages to Alice does not need to be aware of the existence of the proxy.
The same %Please ensure the intended meaning is retained
applies to Bob, the delegate.

\textbf{Key optimality.} Bob should keep a constant number of secrets, regardless of how many delegations he accepts.

\textbf{Nontransitive.} A proxy re-encryption scheme is transitive if the proxy has the right to re-delegate decryption permission.
Moreover, it combines several re-encryption keys to produce a new re-key (e.g., from $rk_{A \rightarrow B}$ and $rk_{B \rightarrow C}$ one can obtain $rk_{A \rightarrow C}$).
Our scheme is nontransitive, as generating a re-key requires Alice's authorization to prevent transitive action on a proxy.

\textbf{Transferability.} This property was first considered by Ateniese et al. in\cite{ateniese2006improved}, catches the inability of collusion of the proxy and the delegates to re-delegate decryption rights (i.e., producing new re-encryption keys).
The proxy has $rk$, and Bob knows $g^r$ and $sk_B$, which can generate a new re-key for another user. 
% This is why a semi-trusted proxy is important in PRE that proxy should not reveal the $rk$ to the public.

\subsection{Proposed Scheme}

\textbf{Setup}

In the CCA-secure and collusion-resilience PRE scheme, $g$ is the generator of a cyclic multiplicative group $\mathbb{G}$ of prime order $p$.
$sk_A$ Equation (16) is the secret key and $pk_A$ Equation (15) is the public key of Alice. $sk_B$ Equation (18) is the secret key and $pk_B$ Equation (17) is the public key of Bob.
% $$ k_{Enc} = pk_{A} = g^a $$
% $$ k_{Dec} = sk_{B} = b $$
% $$ rk_{A \rightarrow B} = (\frac{pk_B}{pk_A})^d = g^{bd-ad} $$

$m$ is the clear message of $l_0$ bits length in the binary message space denoted by $\mathcal{M}$.
$w$ is the random bits of $l_1$ length.
$H$ is the hash function, where $H_1:\mathbb{Z}_p \cdot \mathbb{Z}_p \rightarrow \mathbb{Z}_p $, $H_2:\{0,1\}^{l_0} \cdot \{0,1\}^{l_1} \rightarrow \mathbb{Z}_p $, $H_3:\mathbb{G}^2 \rightarrow \{0,1\}^{l_0+l_1} $, $H_4:\mathbb{G} \cdot \{0,1\}^{l_0+l_1} \rightarrow \mathbb{Z}_p $.

$KeyGen()$: The key generation algorithm generates the public/secret key pair $(pk,sk)$ for the user:
\begin{equation}
 a,b\in \mathbb{Z}_p,
\end{equation}
\begin{equation}
 pk_A = g^a,
\end{equation}
\begin{equation}
 sk_A = a,
\end{equation}
\begin{equation}
 pk_B = g^b,
\end{equation}
\begin{equation}
 sk_B = b.
\end{equation}

$ReKeyGen(sk_A, pk_B)$: The re-encryption-key-generating algorithm accepts the secret key $sk_A$ from Alice and the public key $pk_B$ from Bob.
The algorithm returns the re-encryption key $rk_{A \rightarrow B}$.

Re-key $rk_{A \rightarrow B} = (\frac{pk_B}{pk_A})^d = (\frac{g^b}{g^a})^d = g^{bd-ad}$, where the $pk_A$ can be derived from $sk_A$.
When re-encrypting $D_B = D_A \cdot rk_{A \rightarrow B} = (g^a)^d \cdot g^{bd-ad} = g^{bd} $, the re-key can only be issued by Alice, who knows $d = H_1(sk_A, r)$.

$Enc(sk_A, m)$: The encryption algorithm takes the secret key $sk_A$, and the clear message $m$ returns the encrypted message $c_A$ via Equation (19):
% $$ c_A = \langle D, r, E, F, V, S \rangle $$
\begin{equation}
  \begin{split}
    (m || w) \xrightarrow{Enc} c_A &= \langle D_A, r, E, F, V, S \rangle.
    % &= (g^{r}, g^{ar} \cdot m)
  \end{split}
\end{equation}
where $D, r, E, F, V, and S$ are defined as Equations (20)--(24):
\begin{equation}
  \begin{split}
    D_A = (pk_A)^d, d = H_1(sk_A, r), r \leftarrow \mathbb{Z}_p,
  \end{split}
\end{equation}
\begin{equation}
  \begin{split}
    E = g^e, e = H_2(m, w), w \leftarrow \{0, 1\}^{l_1},
  \end{split}
\end{equation}
\begin{equation}
  \begin{split}
    F = H_3(g^d, E) \oplus (m || w),
  \end{split}
\end{equation}
\begin{equation}
  \begin{split}
    V = g^v, v \leftarrow \mathbb{Z}_p,
  \end{split}
\end{equation}
\begin{equation}
  \begin{split}
    S = g^s, s = v + sk_A \cdot r.
  \end{split}
\end{equation}

$ReEnc(rk_{A \rightarrow B}, c_A)$: The re-encryption algorithm transfers the encrypted message $c_A$ into the ciphertext $c_B$ using the generated re-encryption key $rk_{A \rightarrow B}$ via Equation (25).
Bob's secret key can decrypt the transformed ciphertext $c_B$.
Here, $d$ is used to generate the permission of delegation.
Only the content owner can create new $D$ or $rk$ with $sk_A$ and $r$.

Before the transferring, the validation $S \overset{?}{=} V \cdot pk_A^r $ of ciphertext should be checked to ensure Alice generates the ciphertext. Otherwise, the algorithm outputs $\bot$:
\begin{equation}
  \begin{split}
    c_A \xrightarrow{ReEnc} c_B &= \langle D_B, r, E, F, V, S \rangle\\
    &= \langle D_A \cdot rk_{A \rightarrow B}, r, E, F, V, S \rangle.
  \end{split}
\end{equation}

% \textls[-15]{
$Dec(sk, c)$: The user decrypts the ciphertext with his secret key and encrypted $c$, e.g., $c_A$ or $c_B$.
It outputs the cleartext $m$ and random bits $w$ via Equation (26).
After decryption, the validation of ciphertext should be checked $E \overset{?}{=} g^{H_2(m, w)} $. If not, the algorithm outputs $\bot$:
% }
\begin{equation}
  \begin{split}
    c_B \xrightarrow{Dec} (m || w) &= F \oplus H_3(g^d, E), g^d = D_B^{\frac{1}{sk_B}} \\
    &= F \oplus H_3(D_B^{\frac{1}{sk_B}}, E).
    % &= m
  \end{split}
\end{equation}

%\vspace{4mm}

In WDLC10\cite{weng2010chosen}, CCA-secure "hashed" ElGamal and modified version is used.
The textbook ElGamal is CPA-secure and risky in the rounded attack.
To enhance the security, "hashed" ElGamal is applied with a message authenticated mechanism.

\subsection{Security Analysis}

Our collusion-resilience PRE scheme is CCA-secure in a random oracle model, under the modified-computation Diffie--Hellman (mCDH) assumption\cite{libert2008multi}.

In this section, we prove the scheme under mCDH assumption\cite{libert2008multi} that any efficient algorithm's mCDH advantage is negligible.
%\vspace{4mm}

% \begin{Theorem}
Our PRE scheme $\prod$ is IND-PRE-CCA-secure under the assumption of the mCDH\cite{libert2008multi} in group $\mathbb{G}$, and the Schnorr signature\cite{schnorr1989efficient} is EUF-CMA-secure in the random oracle model.
An adversary $\mathcal{A}$, who asks at most $q_{H_i}$ random oracle queries to $H_i$ with $i \in \{1, \dots, 4\}$, can effectively break the $(t, q_u, q_c, q_{re}, q_d, \epsilon)$-IND-PRE-CCA of our scheme $\prod$, for any $0 < \nu < \epsilon$. Thus we have:

\begin{itemize}
  \item The $(t', \epsilon')$-mCDH problem\cite{libert2008multi} in group $\mathbb{G}$ can be solved by an algorithm $\mathcal{B}$ with \linebreak\mbox{Equations (27) and (28):}
\begin{equation}
    \begin{split}
      t' \leq t&+(q_{H_1}+q_{H_2}+q_{H_3}+q_{H_4}\\
      &+q_u+q_c+q_{rk}+q_{re}+q_d)\mathcal{O}(1)\\
      &+(q_u+q_c+4q_{re}+3q_d+(2q_d+q_{re})q_{H_2})t_e,
    \end{split}
  \end{equation}
\begin{equation}
    \begin{split}
    % \epsilon' \geq \frac{1}{q_{H_2}}(&2(\epsilon - \nu) - \frac{q_{H_1}+(q_{H_1}+q_{H_2})q_d}{2^{l_0+l_1}} - \frac{2q_d+q_{re}}{q})
      \epsilon' \geq \frac{1}{q_{H_3}}(2(\epsilon - \nu) - \frac{q_{H_2}+(q_{H_2}+q_{H_3})q_d}{2^{l_0+l_1}} - \frac{2q_d+q_{re}}{q}).
    \end{split}
  \end{equation}

  where $t_e$ is the exponential running time in the group $\mathbb{G}$.
  \item The EUF-CMA security of the Schnorr signature\cite{schnorr1989efficient} can be broken by an attacker with advantage $\nu$ within time $t'$.
\end{itemize}
% \end{Theorem}

\begin{proof}
It is assumed that the Schnorr signature\cite{schnorr1989efficient} is $(t', \epsilon')$-EUF-CMA-secure for the probability $0 < \nu < \epsilon$.
While the CDH problem (given $g, g^x, g^y$ output $g^{xy}$) is as hard as the mCDH problem\cite{libert2008multi} (given $g, g^\frac{1}{x}, g^x, g^y$ outputs $g^{xy}$), this theorem is proved under the mCDH problem\cite{libert2008multi}.
A $t$-time adversary $\mathcal{A}$ can break the IND-PRE-CCA security of scheme $\prod$ with advantage $\epsilon - \nu$.
We show how an algorithm $\mathcal{B}$ solves the $(t', \epsilon')$-mCDH problem\cite{libert2008multi} in group $\mathbb{G}$.\end{proof}

Suppose algorithm $\mathcal{B}$ accepts the input of mCDH challenge tuple $(g, g^x, g^\frac{1}{x}, g^y) \in \mathbb{G}^4$, and $x, y \leftarrow \mathbb{Z}_p$ is unknown.
Algorithm $\mathcal{B}$ plays the role of challenger playing the IND-PRE-CCA game with adversary $\mathcal{A}$. Algorithm $\mathcal{B}$'s goal is to output $g^{xy}$.
%\vspace{4mm}

\textbf{Setup.} Algorithm $\mathcal{B}$ passes parameters $(p, \mathbb{G}, g, H_1, H_2, H_3, H_4, l_0, l_1)$ to adversary $\mathcal{A}$. $H_1, H_2, H_3, H_4$ are random hash oracles controlled by the algorithm $\mathcal{B}$.
%\vspace{4mm}

\textbf{Hash Oracle Queries.} Adversary $\mathcal{A}$ may send the queries to random oracle $H_1$, $H_2$, $H_3$, and $H_4$ at any time.
Algorithm $\mathcal{B}$ has four empty lists $H_1^{list}$, $H_2^{list}$, $H_3^{list}$, and $H_4^{list}$ initially, used for storing the query parameters and result value tuples.
\begin{itemize}
  \item $H_1$ queries. With the parameters $(a, r)$, if this query exists in the $H_1^{list}$ as a tuple $(a, r, d)$, output the value $d$ as the result to adversary $\mathcal{A}$. Otherwise, choose $d \leftarrow \mathbb{Z}_p$ and add the tuple $(a, r, d)$ to the hash list $H_1^{list}$, and respond with $H_1(a, r) = d$ to adversary $\mathcal{A}$.
  \item $H_2$ queries. With the parameters $(m, w)$, if this query exists in the $H_2^{list}$ as a tuple $(m, w, v)$, output the value $v$ as the result to adversary $\mathcal{A}$. Otherwise, choose $v \leftarrow \mathbb{Z}_p$ and add the tuple $(m, w, v)$ to the hash list $H_2^{list}$, and respond with $H_2(m, w) = v$ to adversary $\mathcal{A}$.
  \item $H_3$ queries. With the parameters $(g^d, E)$, if this query exists in the $H_3^{list}$ as a tuple $(g^d, E, \alpha)$, output the value $\alpha$ as the result to adversary $\mathcal{A}$. Otherwise, choose $\alpha \leftarrow \{0,1\}^l$ and add the tuple $(g^d, E, \alpha)$ to the hash list $H_3^{list}$, and respond with $H_3(g^d, E) = \alpha$ to adversary $\mathcal{A}$.
  \item $H_4$ queries. With the parameters $(E, F)$, if this query exists in the $H_4^{list}$ as a tuple $(E, F, \beta)$, output the value $\beta$ as the result to adversary $\mathcal{A}$. Otherwise, choose $\beta \leftarrow \mathbb{Z}_p$ and add the tuple $(E, F, \beta)$ to the hash list $H_4^{list}$, and respond with $H_4(E, F) = \beta$ to adversary $\mathcal{A}$.
\end{itemize}
%\vspace{4mm}

\textbf{Phase 1.} The adversary $\mathcal{A}$ sends a series of queries as in the definition of the IND-PRE-CCA game.
The algorithm $\mathcal{B}$ holds three hash lists $K_{Uncorrupted}^{list}$, $K_{Corrupted}^{list}$, and $R^{list}$, answering the adversary $\mathcal{A}$ as follows:

\begin{itemize}
  \item Uncorrupted key generation query $q_u$.
  If the tuple $(a, g^a)$ is not in the hash list $K_{Uncorrupted}^{list}$, the algorithm $\mathcal{B}$ chooses $a \leftarrow \mathbb{Z}_p$; add the tuple $(a, g^a)$ to the hash list $K_{Uncorrupted}^{list}$.
  Respond with $pk = g^a$ to adversary $\mathcal{A}$.

  \item Corrupted key generation query $q_c$. 
  If the tuple $(a, g^a)$ is not in the hash list $K_{Corrupted}^{list}$, the algorithm $\mathcal{B}$ chooses $a \leftarrow \mathbb{Z}_p$; add the tuple $(a, g^a)$ to the hash list $K_{Corrupted}^{list}$.
  Respond with $(sk, pk) = (a, g^a)$ to adversary $\mathcal{A}$.

  \item Re-encryption key generation query $q_{rk}$. The re-key generation is from Alice's secret key and Bob's public key; both key pairs can be uncorrupted or corrupted. It is because in the re-encryption from $c_A$ to $c_B$, ciphertexts can be decrypted by either $sk_A$ or $sk_B$.

  In the case algorithm, $\mathcal{B}$ recovers $(sk_A, pk_A), (sk_B, pk_B)$ from $K_{Uncorrupted}^{list}$ or $K_{Corrupted}^{list}$.
  Then, algorithm $\mathcal{B}$ generates re-key $rk_{A \rightarrow B} = (\frac{pk_B}{pk_A})^{H_1(sk_A, r)} = (\frac{g^b}{g^a})^{H_1(a, r)}$.
  The tuple $(sk_A, pk_A, sk_B, pk_B, rk_{A \rightarrow B})$ is added to the $R^{list}$.
  Then, the $rk_{A \rightarrow B}$ is returned to adversary $\mathcal{A}$.

  For the challenge purpose, both $sk_A$ and $sk_B$ should be uncorrupted.

  \item Re-Encryption query $q_{re}$. Given $rk_{A \rightarrow B}$ and $c_A = \langle D_A, r, E, F, V, S \rangle$:
  If $S \neq V \cdot pk_A^r$, it outputs $\bot$.
  Otherwise, the algorithm returns the re-encrypted ciphertext $c_B = \langle D_A \cdot rk_{A \rightarrow B}, r, E, F, V, S \rangle$ to adversary $\mathcal{A}$.

  \item Decryption query $q_d$. The algorithm recovers $sk$ from $K_{Uncorrupted}^{list}$ or $K_{Corrupted}^{list}$.
  Run $(m, w) = Dec(sk, c)$.
  If $E = g^{H_2(m, w)} $, give $m$ back to the adversary, otherwise it outputs $\bot$.

\end{itemize}
%\vspace{4mm}

\textbf{Challenge.} When adversary $\mathcal{A}$ ends %Please ensure intended meaning is retained
Phase 1, the adversary outputs a target public key $pk^*$ and two equal-length messages $m_0, m_1 \in \{0, 1\}^{l_0}$, queries to algorithm $\mathcal{B}$. Algorithm $\mathcal{B}$ responds as follows:
\begin{enumerate}
  \item Recovers $(sk^*, pk^*)$ from $K_{Uncorrupted}^{list}$ and let $pk^* = g^a := g^\frac{1}{x}$.
  \item Let $D^* = (pk^*)^d = (g^a)^d := g^y$, so that $(g^a)^d = (g^\frac{1}{x})^{xy}$.
  We can obtain $d = xy$ as $g^a = g^\frac{1}{x}$. Then, $g^d = g^{xy}$.
  \item As $F = H_3(g^d, E) \oplus (m || w)$ defined, choose $\delta \leftarrow \{0, 1\}, w^* \leftarrow \{0, 1\}^{l_1}$ and $F^* = H_3(g^d, E^*) \oplus (m_\delta || w^*) $.
  \item Return $c^* = \langle D^*, r^*, E^*, F^*, V^*, S^* \rangle$ as the challenged ciphertext to adversary $\mathcal{A}$.
\end{enumerate}

%%\vspace{4mm}

\textbf{Phase 2.} The adversary $\mathcal{A}$ issues the queries as in Phase 1.
Algorithm $\mathcal{B}$ responds to those queries to $\mathcal{A}$ as in Phase 1.
%\vspace{4mm}

\textbf{Guess.} The adversary $\mathcal{A}$ responds a guess $\hat{\delta} \in \{0, 1\}$ to $\mathcal{B}$.
Algorithm $\mathcal{B}$ calculates $H_3(g^d, E^*) = H_3(g^{xy}, E^*) = F^* \oplus m_{\hat{\delta}} = \hat{\alpha}$.
$\mathcal{B}$ looks up the hash list $H_3^{list}$ for the tuple $(g^d, E^*, \alpha)$ where $\alpha = \hat{\alpha}$, then returns the $g^d$ as the solution $g^{xy}$ to the given mCDH instance.
%%\vspace{4mm}

\textbf{Analysis.} First, let us evaluate the simulation of random oracles.
$H_1$, $H_4$ are perfect %Please ensure intended meaning is retained.
As long as $\mathcal{A}$ does not query $(m_\delta, w)$ to $H_2$ or $(g^{xy}, E)$ to $H_3$, so $H_2$ and $H_3$ are perfect %Please ensure intended meaning is retained
.
We denote $AskH_2^*$ the event $(m_\delta, w)$ has been queried to $H_2$, and $AskH_3^*$ the event that $(g^{xy}, E)$ has been queried to $H_3$.

The challenged ciphertext is identically distributed.

Second, the simulation for the re-encryption oracle. 
The re-encryption query is perfect unless the adversary $\mathcal{A}$ can transfer the ciphertext into the new one without querying hash $H_1$ to obtain the $rk$.
We denote this event as $ReEncErr$.
Since $H_1$ plays the role of the random oracle, which is queried by adversary $\mathcal{A}$ at most $q_{re}$ times, we have
\begin{equation}
  \begin{split}
    \Pr[ReEncErr] \leq \frac{q_{re}}{q}.
  \end{split}
\end{equation}

% \textls[-5]{
Third, the simulation for the decryption oracle.
Suppose that $(pk, c)$, $c = (D, r, E, F, V, S)$ is a valid ciphertext, as the validation $S \overset{?}{=} V \cdot pk_A^r $ of ciphertext can be checked.
There is still a chance that $c$ can be generated by querying other random values to $H_3$ instead of $g^d$, where $d = H_1(sk_A, r)$.
Denote $Valid$ to be an event that $c$ is valid.
Let $AskH_3$ be the event that $(g^d, E)$ has been queried to $H_3$ and $AskH_2$ be the event that $(m, w)$ has been queried to $H_2$. Then, we have
% }
\vspace{-6pt}
\begin{equation}
  \begin{split}
    \Pr &[Valid| \lnot AskH_2]\\
    = &\Pr [Valid \wedge AskH_3| \lnot AskH_2]\\
      &+ \Pr [Valid \wedge \lnot AskH_3| \lnot AskH_2]\\
    \leq &\Pr [AskH_3| \lnot AskH_2] + \Pr [Valid| \lnot AskH_3 \wedge \lnot AskH_2]\\
    \leq &\frac{q_{H_3}}{2^{l_0+l_1}} + \frac{1}{q},
  \end{split}
\end{equation}
similarly,
\begin{equation}
  \begin{split}
    \Pr &[Valid| \lnot AskH_3]\\
    = &\Pr [Valid \wedge AskH_2| \lnot AskH_3]\\
      &+ \Pr [Valid \wedge \lnot AskH_2| \lnot AskH_3]\\
    \leq &\Pr [AskH_2| \lnot AskH_3] + \Pr [Valid| \lnot AskH_2 \wedge \lnot AskH_3]\\
    \leq &\frac{q_{H_2}}{2^{l_0+l_1}} + \frac{1}{q}.
  \end{split}
\end{equation}
{Thus,} %please confirm if indentation should be added
 we have
\begin{equation}
  \begin{split}
    \Pr [Valid| \lnot &AskH_2 \vee \lnot AskH_3]\\
    \leq &\Pr [Valid| \lnot AskH_2] + \Pr [Valid| \lnot AskH_3]\\
    \leq &\frac{q_{H_2}+q_{H_3}}{2^{l_0+l_1}} + \frac{2}{q}.
  \end{split}
\end{equation}

% \textls[-25]{
Denote $DecErr$ as the event that $Valid|(\lnot AskH_2 \vee \lnot AskH_3)$ happens during the entire simulation.
The $q_d$ times of decryption queries have been issued to a decryption oracle, and we have
% }
\begin{equation}
  \begin{split}
    \Pr [DecErr] \leq \frac{(q_{H_2}+q_{H_3})q_d}{2^{l_0+l_1}} + \frac{2q_d}{q}.
  \end{split}
\end{equation}

Denote $Good$ as the event $AskH_3^* \vee (AskH_2^*| \lnot AskH_3^*) \vee ReEncErr \vee DecErr$.
If $Good$ has not happened, the adversary $\mathcal{A}$ cannot gain any advantage in guessing $\delta$ from $m_0, m_1$, due to the random $E$ as one of the input of $H_3(g^d, E)$ and $E = g^e = g^{H_2(m, w)}$ is generated with the random bits $w \leftarrow \{0,1\}^{l_1}$.
We have $\Pr[\delta = \delta'| \lnot Good] = \frac{1}{2}$
\begin{equation}
  \begin{split}
    &\Pr[\delta = \delta'] \\
    &= \Pr[\delta = \delta'| \lnot Good]\Pr[\lnot Good]+\Pr[\delta = \delta'|Good]\Pr[Good]\\
    &\leq \frac{1}{2}\Pr[\lnot Good] + \Pr[Good]\\
    &= \frac{1}{2}(1-\Pr[Good]) + \Pr[Good]\\
    &= \frac{1}{2} + \frac{1}{2}\Pr[Good],
  \end{split}
\end{equation}
and
\begin{equation}
  \begin{split}
    \Pr[\delta &= \delta'] \\
    &\geq \Pr[\delta = \delta'| \lnot Good]\Pr[\lnot Good]\\
    &= \frac{1}{2}(1 - \Pr[Good])\\
    &= \frac{1}{2} - \frac{1}{2}\Pr[Good],
  \end{split}
\end{equation}
we have
\begin{equation}
  \begin{split}
    | \Pr[\delta &= \delta'] - \frac{1}{2} | \leq \frac{1}{2}\Pr[Good].
  \end{split}
\end{equation}

By the definition, the advantage $(\epsilon - \nu)$ for IND-PRE-CCA adversary:
\begin{equation}
  \begin{split}
    \epsilon - &\nu\\
    = &| \Pr[\delta = \delta'] - \frac{1}{2} |\\
    \leq &\frac{1}{2}\Pr[Good]\\
    = & \frac{1}{2}(\Pr[AskH_3^* \vee (AskH_2^*| \lnot AskH_3^*) \vee ReEncErr \vee DecErr])\\
    = &\frac{1}{2}(\Pr[AskH_3^*] + \Pr[AskH_2^*| \lnot AskH_3^*]\\
    &+ \Pr[ReEncErr] + \Pr[DecErr]).
  \end{split}
\end{equation}

Since $ \Pr[ReEncErr] \leq \frac{q_{re}}{q} $, $\Pr [DecErr] \leq \frac{(q_{H_2}+q_{H_3})q_d}{2^{l_0+l_1}} + \frac{2q_d}{q}$ and
$\Pr [AskH_2^*| \lnot AskH_3^*] \leq \frac{q_{H_2}}{2^{l_0+l_1}} $, we obtain
\begin{equation}
  \begin{split}
    \Pr [Ask&H_3^*]\\
    \geq & 2(\epsilon - \nu) - [AskH_2^*|\lnot AskH_3^*]\\
    &- \Pr [DecErr] - \Pr[ReEncErr]\\
    \geq & 2(\epsilon - \nu) - \frac{q_{H_2}}{2^{l_0+l_1}}\\
    &- \frac{(q_{H_2}+q_{H_3})q_d}{2^{l_0+l_1}} - \frac{2q_d}{q} - \frac{q_{re}}{q}\\
    = & 2(\epsilon - \nu) - \frac{q_{H_2}+(q_{H_2}+q_{H_3})q_d}{2^{l_0+l_1}} - \frac{2q_d+q_{re}}{q}.
  \end{split}
\end{equation}

In event $AskH_3^*$, algorithm $\mathcal{B}$ will be able to solve the mCDH instance, and consequentially, %Please ensure intended meaning is retained
the following is obtained:
\begin{equation}
  \begin{split}
    \epsilon' \geq \frac{1}{q_{H_3}}(2(\epsilon - \nu) - \frac{q_{H_2}+(q_{H_2}+q_{H_3})q_d}{2^{l_0+l_1}} - \frac{2q_d+q_{re}}{q}).
  \end{split}
\end{equation}

From the description of the simulation, the running time of algorithm $\mathcal{B}$ can be bounded by
\begin{equation}
  \begin{split}
    t' \leq t&+(q_{H_1}+q_{H_2}+q_{H_3}+q_{H_4}\\
    &+q_u+q_c+q_{rk}+q_{re}+q_d)\mathcal{O}(1)\\
    &+(q_u+q_c+4q_{re}+3q_d+(2q_d+q_{re})q_{H_2})t_e.
  \end{split}
\end{equation}

This completes the proof of Theorem 1.

\section{Chapter Summary}

In this chapter, we designed a permissionless storage network as the computational security engine of blockchains.
As long as more data is stored in this network, a blockchain can obtain security protection almost for free.
Compared with existing public chain networks, the cost of blockchain usage for users is extremely low, because each user will not cover the security cost of the blockchain.
Based on this theory, we design a storage system that, unlike previous blockchain storage systems, does not put all storage logic on the chain but calculates off-chain as much as possible to reduce on-chain operations and state space occupation.
In addition, incentives are also based on blockchains.

\chapter{Sharding for Scalability Blockchain}
\label{ch:sharding}

In the previous chapters, the useful PoW is proposed, and the permissionless storage network is designed to provide the computation required by blockchain security.
Under such a setting, the end users can enjoy low usage costs.
The end users would not pay for the high-security cost of blockchain, which made an important step towards massive blockchain adoption.

Meanwhile, the increasingly high transactions per second (TPS) of blockchain keep enlarging the capacity requirement of blockchain full nodes.
Without a sharding solution, the high performance can not be scalable.
The optimization of blockchain data structure can improve the performance of blockchains, such as the DAG.
It has successfully increased the TPS of blockchains to around 3000-6000. However, improving the DAG data structure cannot reduce the requirements for single nodes. We still need sharding technology to reduce the computational and storage space requirements for nodes to truly realize blockchains' scalability.
Everyone can build sharding nodes at a lower cost, an important prerequisite for decentralization.

\section{Overview}

In the previous chapters, the extended PoW consensus algorithm can utilize the computing power consumed during the consensus to do meaningful work to transfer the security cost to the permissionless storage.
It eliminates mining pools to improve the degree of decentralization and even upgrades the incentive mechanism.
However, blockchain consensus improvement can hardly increase the transactions per second.
To improve performance, we need to upgrade the blockchain data structure.

\section{Background}

\subsection{Layer 1 Sharding}

Layer 1 sharding is a technique used in blockchain technology to increase the scalability of a blockchain network.
Sharding refers to dividing a large database into smaller, more manageable parts called shards.
In a blockchain network, sharding breaks the network into smaller groups of nodes called shards.

Layer 1 sharding is implemented at the protocol level, which is built into the blockchain's architecture.
It is designed to allow the network to process more transactions per second by dividing the workload among multiple shards. Each shard processes a subset of the transactions in parallel rather than having all nodes in the network process every transaction.

The benefits of layer 1 sharding include improved scalability, faster transaction processing times, and lower fees. It also allows for better resource utilization by reducing the amount of computational resources required to process transactions.

However, layer 1 sharding can also introduce new challenges, such as the need for secure inter-shard communication, the difficulty of maintaining consensus across multiple shards, and the potential for reduced network security.
To overcome these challenges, various approaches to sharding have been proposed, such as using cross-shard communication protocols and hybrid sharding and consensus mechanisms.

\subsection{Layer 2}

Layer 2 is a scaling solution implemented on top of an existing blockchain network.
It is a method of dividing blockchain into smaller subsets and processing them separately from the main chain, allowing for increased transaction throughput and reduced congestion on the main chain.

One early implementation of Layer 2 is called the payment channel network.
In the off-chain, users can open an one-way or two-way payment channel with another party, which allows them to conduct multiple transactions between each other without having to broadcast them to the main chain.
This reduces the number of transactions on the main chain and improves transaction speed and cost.
Lightning Network is the example of the payment channel direction.

Another example of Layer 2 is the sidechain, which are separate blockchains that are interoperable with the main chain.
Transactions can be conducted on the sidechain without affecting the main chain, allowing for increased transaction throughput and scalability.
The rollup technique is used to secure the sidechain with main blockchain network, preventing the tamper of the history data.

Layer 2 offers several benefits, including faster transaction times, lower transaction fees, and increased scalability.
However, it also introduces new challenges, such as maintaining security and ensuring interoperability between off-chain and on-chain transactions.

\subsubsection{Optimistic Rollup}\label{op_rollup}

Optimistic Rollup is a type of Layer 2 scaling solution for blockchain that uses optimistic rollups to achieve high scalability and throughput.

Optimistic rollups work by batching transactions and submitting them to the Layer 1 blockchain in a way that is equivalent to processing them individually on the Layer 1 blockchain.
This allows for a significant increase in transaction throughput without compromising security.

Layer 2 OP is a relatively new technology, but it can potentially revolutionize how blockchains are used.
It is a promising solution for scaling blockchains to meet the demands of a growing user base.

\subsubsection{ZK Rollup}\label{zk_rollup}

Zero-knowledge rollup is a type of Layer 2 scaling solution for blockchains that uses zero-knowledge proofs to achieve high scalability and throughput without compromising security.

Zero-knowledge proofs are a cryptographic technique that allows one party to prove to another party that they know something without revealing the information itself.
This makes them ideal for use in blockchains, where it is important to protect user privacy.

L2 ZK rollups work by batching transactions and submitting them to the Layer 1 blockchain, along with zero-knowledge proof that the transactions are valid.
The Layer 1 blockchain then verifies the proof and processes the transactions.
This allows for a significant increase in transaction throughput without compromising security.

L2 ZK is a relatively new technology, but it can potentially revolutionize how blockchains are used.
It is a solution for scaling blockchains to meet the demands of a growing user base.

\section{Related Work}

Blockchain sharding is a technique used to improve blockchain scalability by breaking up the network into smaller subsets called shards.
Each shard contains a subset of the total blockchain data and can process transactions independently of other shards.

This approach allows for parallel processing of transactions across multiple shards, resulting in significantly higher transaction throughput than traditional blockchain networks.
Sharding can also reduce the computational and storage requirements for nodes in the network, as they only need to maintain data relevant to their assigned shard.

However, sharding also introduces challenges to ensure consistency and security across the entire network.
One major challenge is achieving cross-shard transactions, which require coordination and communication between multiple shards.
Other challenges include shard management, data availability, and ensuring the network's security despite the increased attack surface created by multiple shards.

There are several types of sharding:
\begin{enumerate}
    \item Sharding by the transactions sends irrelevant transactions to different miners or consensus groups.
    \item Sharding by the users, which split the network by users. Each shard keeps less transaction history data.
    \item Sharding the state (not the global state in Ethereum), as the user state (such as user balance) is not kept by all the nodes.
    \item Sharding the computation avoids all the nodes applying the same computation, reducing the overhead.
\end{enumerate}

\subsubsection{Elastico}

\begin{figure}[!t]
    \centering
    \includegraphics[width=1\textwidth]{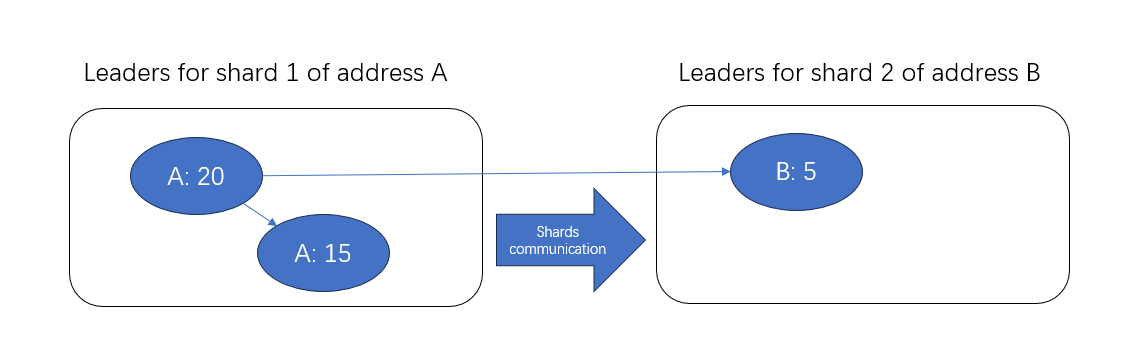}
    \caption{Elastico sharding with leader groups communication}
    \label{fig_elastico}
\end{figure}

Elastico\cite{luu2016secure} solution is sharding by the transactions, sending miners to different shards.
It uses PoW to choose leaders for each shard, shown in Figure~\ref{fig_elastico}.
Within shard, PBFT is used for consensus; thus, it uses hybrid consensus.
Elastico uses UTXO.

The key idea is to automatically parallelize the available computation, dividing it into several smaller groups, each processing a disjoint set of transactions (or shards).
All groups, each of which runs a classical byzantine consensus protocol internally to, agree on one value.
A final consensus committee is responsible for combining the shards selected by other committees, computing a cryptographic digest, and broadcasting it to the whole network.

\subsubsection{Omniledger}

\begin{figure}[!t]
    \centering
    \includegraphics[width=1\textwidth]{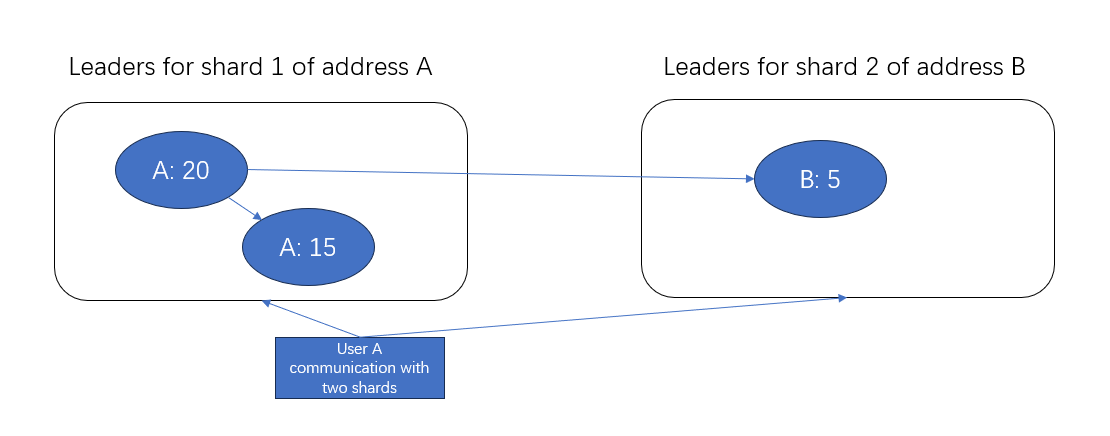}
    \caption{OmniLedger sharding with user relay cross-shard transaction}
    \label{fig_omniLedger}
\end{figure}

OmniLedger\cite{kokoris2018omniledger} uses a similar approach to Elastico with several changes.
It is sharding by the users, shown in Figure~\ref{fig_omniLedger}.
For energy efficiency, OmniLedger uses the PoS algorithm to choose leaders for shards.
Omniledger introduces the Atomix Protocol for cross-shard transactions.
There is no communication between shards; the users must pass the information by themselves for the cross-shard transactions.
OmniLedger uses UTXO as well.

\subsubsection{Monoxide}

\begin{figure}[!t]
    \centering
    \includegraphics[width=1\textwidth]{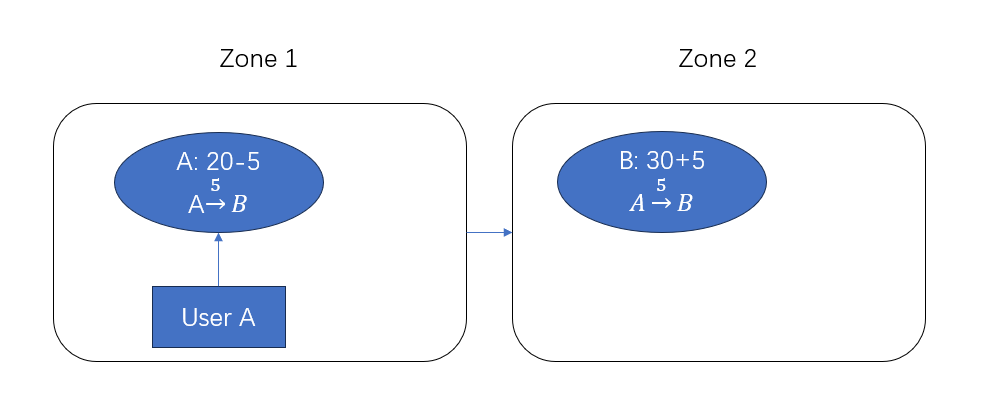}
    \caption{Monoxide sharding with Zones}
    \label{fig_monoxide}
\end{figure}

Monoxide\cite{wang2019monoxide} is sharding by the users, shown in Figure~\ref{fig_monoxide}.
Each shard is called Asynchronous Consensus Zones.
The consensus algorithm is applied in each zone (shard).
Chu ko-nu mining algorithm avoids the computation being dispersed to affect security.
In Monoxide, there are intra-shard and cross-shard transactions.
Monoxide uses an account model.
A PoS version of Monoxide is named Dioxide\cite{qu2022dioxide}.

\section{Preliminaries}

\subsection{Input and State}

\begin{figure}[!t]
    \centering
    \includegraphics[width=1\textwidth]{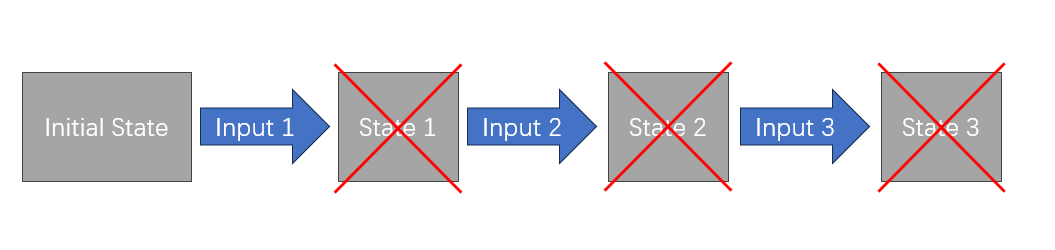}
    \caption{Transaction as input changes the state}
    \label{fig_input_and_state}
\end{figure}

A state machine can be used to model the blockchain as shown in Figure~\ref{fig_input_and_state}.
The user transactions is the input of the state machine.
The states can be removed if all the history transactions are saved.
Based on the user inputs from the earliest block, the state can be constructed.
We can look up the user balance from the current state.

In the UTXO model, the user checks the balance from the latest related transaction.
The account model is convenient for checking the latest global state.
The global state can always be reconstructed with all the user inputs in the blocks, even if we remove all states.
So, keeping all the historical transactions rather than the historical states is more important.
As time goes on, the size of history transactions grows and becomes huge.
The approach of sharding transactions becomes necessary for the blockchain's massive adoption. 

\subsection{Trie and MPT}

Trie is a high-level data structure, and we can find its detailed definition in classic textbooks.
A Trie is a tree-based data structure used for efficient prefix matching.
It is commonly used in search engines, auto-completion systems, and other applications that require fast data retrieval based on prefixes.

In the actual blockchain implementation, we cannot directly build the product based on memory, and the data needs to be persisted to disk. Therefore, we usually implement blockchain on rocksdb\cite{dong2017optimizing,dong2021rocksdb}, which is used like most key-value databases, supporting basic get, put, and prefix queries.

Rocksdb is based on LSTM\cite{o1996log,wang2014efficient} at the bottom, which is a high-performance KV storage technology. We can abstractly understand that rocksdb is a trie tree.

\begin{figure}[!t]
    \centering
    \includegraphics[width=1\textwidth]{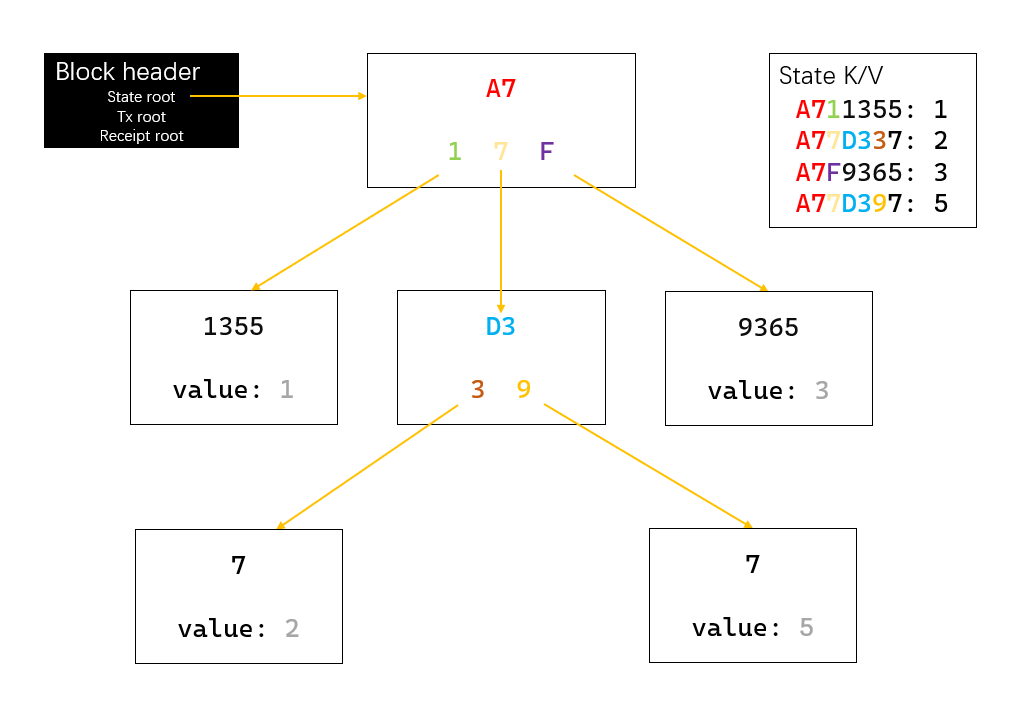}
    \caption{Ethereum uses Merkle Patricia Tree for global state}
    \label{fig_ETH_MPT}
\end{figure}

MPT is a data structure built on trie, shown in Figure~\ref{fig_ETH_MPT}.
Like we make a Merkle tree on a list and get a root hash, an MPT tree can also calculate many KV values and generate a root hash.

When fewer changes exist in a large KV, the cost of generating a new version of the old MPT tree is low, and most of the duplicate data can be reused.
If we do not consider sharding, MPT is a high-performance data structure solution.

\subsection{Transaction Tree and State Tree}

In Ethereum, MPT is a fundamental data structure for transactions and states.
Transactions are kept in the transactions tree, and history states are kept in the state tree.
Unlike Bitcoin, Ethereum does not put the transactions in the block body.
Ethereum puts the Merkle root (MPT) of the transaction tree, state tree, and receipt tree in the block header.

In the non-sharding blockchain, MPT is optimized for the on-chain storage.
However, it is not suited for sharding.
Global state is difficult to split and host distributed as the smart contract execution may require visiting any K/V pairs in the global state.
For sharding, we need to search for a different approach.

\subsection{Multi Chains Data Structure}

In the field of high-performance blockchain, many solutions choose DAG data structure.
For simplicity, we use a multi-chain data structure to implement the high-performance and scalability blockchain.
The main reason to continue the direction is because it is easier to implement sharding over the multi-chain data structure.
This concept has been used in many works, such as the high-performance solution OHIE\cite{2020OHIE} and \cite{2018Parallel}

\section{Wider Sharding Scheme}

We learned the design concept of Monoxide in blockchain sharding and developed Wider sharding.

In the Monoxide sharding, transactions are classified into two types: intra-shard and cross-shard transactions, which must be processed separately.
Each shard needs to run its consensus algorithm independently.
Since running the PoW algorithm consumes power resources, Monoxide has been working hard to improve with Chu ko-nu mining, similar to the concept of merged mining.
% In the current mainstream view, the PoS algorithm would also be very suitable for Monoxide sharding.

As our design adopts an upgraded PoW algorithm, we are still motivated to improve based on Monoxide.
Wider sharding is splitting the network by users.
But each node can handle many shards.

\subsection{How to Avoid Consensus on Shards}

\begin{figure}[!t]
    \centering
    \includegraphics[width=1\textwidth]{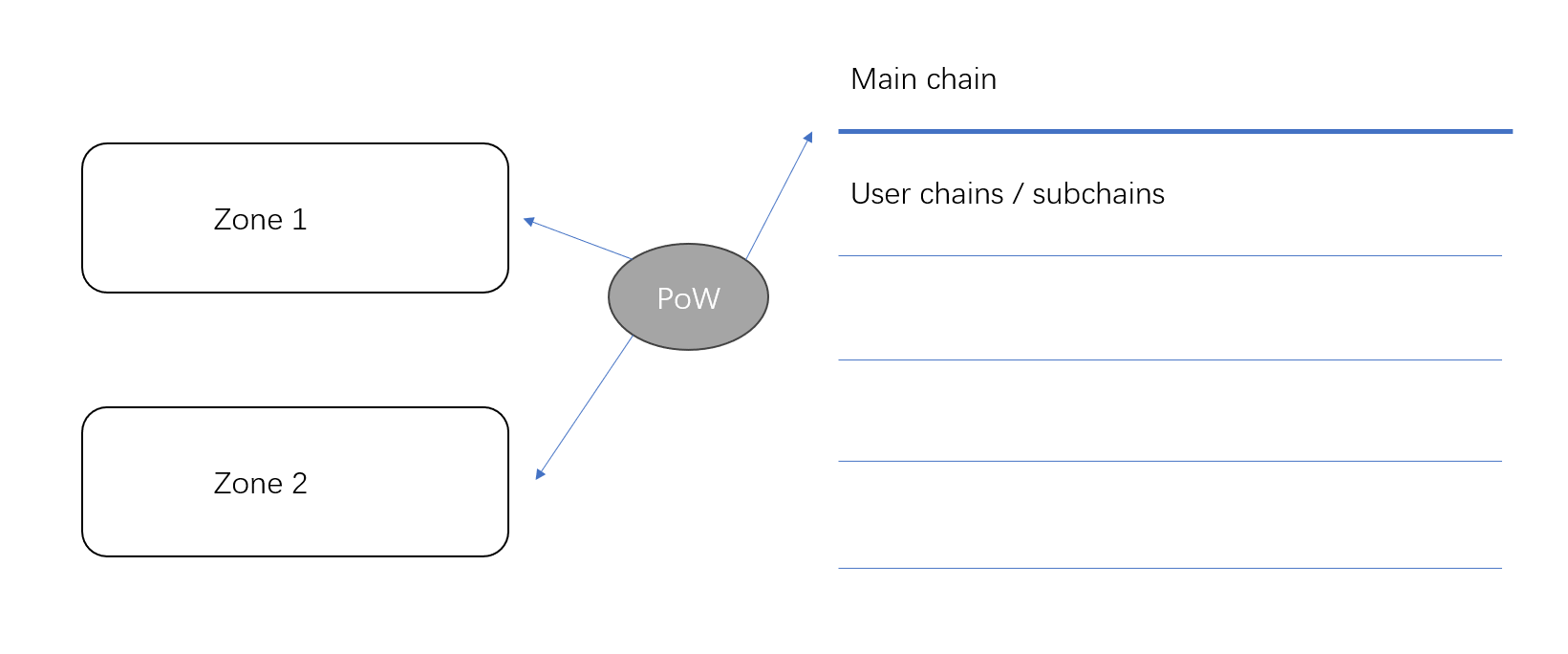}
    \caption{PoW secure each shards in Monoxide or the main chain in Wider}
    \label{fig_Wider_pow_on_main_chain}
\end{figure}

Based on Monoxide, we raised the question of under what conditions a shard does not need to run a consensus algorithm.
As we all know, consensus needs to be reached by more than one person, such as rock-paper-scissors and voting.

As long as the number of users in a shard exceeds one, the shard needs to run a consensus algorithm. The consensus algorithm determines the order of transactions, and the ordered transactions can avoid double spending and other attacks.

Following this line of thought, if we design a shard with only one account, then the consensus algorithm is unnecessary. 
In this design, the number of shards will equal the number of users, the number of shards is almost unlimited, and all transactions will be cross-shared, shown in Figure~\ref{fig_Wider_pow_on_main_chain}.

The sharding is categorized into no sharding (one shard), fixed shards, and unlimited shards.
We define $A$ as the total accounts number of the blockchain, $S$ as the total number of shards in a blockchain, and $a_{s}$ as the accounts number for the $s^{th}$ shard.
So we have:

$$ A = \sum_{s=1}^{S}{a_s} $$

Alternatively, we can define $\overline{a}$ as the average number of accounts per shard.
So we have $ A = \overline{a} \cdot {S} $.
\begin{itemize}
    \item The Bitcoin and Ethereum are not sharding, which equates to one shard as $S = 1$. Then we have $ A = \overline{a} $.
    \item Blockchain uses a fixed number of shards, then we have $S > 1$ as an integer.
    \item Our solution forces $\overline{a} = 1$, so we got $A = S$.
\end{itemize}

\subsection{Main Chain Confirms Subchains}

\begin{figure}[!t]
    \centering
    \includegraphics[width=1\textwidth]{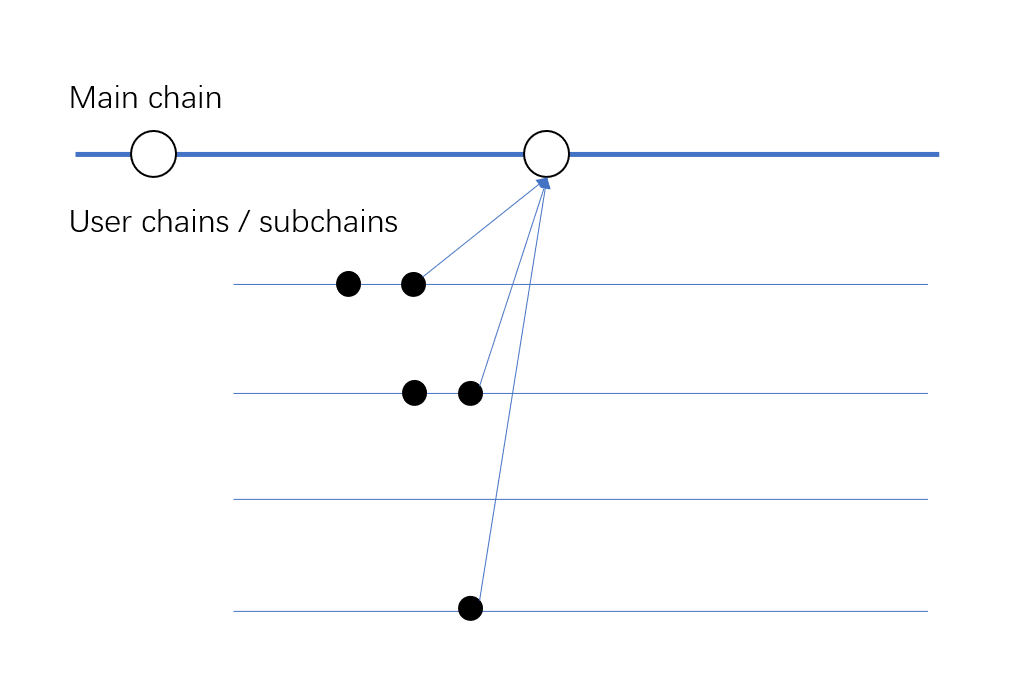}
    \caption{Main chain confirms the subchains to consensus the transactions}
    \label{fig_Wider_main_chain_confirms_subchains}
\end{figure}

At this point, we add the main chain back to the multi-chain structure and run the PoW consensus algorithm on the main chain.
Unlike the single-chain structure, transactions are no longer placed on the chain block but on the subchains corresponding to user account.
The main chain is used to confirm the latest state of multiple subchains, shown in Figure~\ref{fig_Wider_main_chain_confirms_subchains}.
Once the main chain confirms the latest transaction of a subchain, the user can no longer alter the corresponding subchain.

The main chain can confirm the updated subchains only.
In many cases, the main chain can batch-confirm many transactions.
This will increase the TPS blockchain, as one confirms in the main chain.

Assuming that we add a 1M block limit to the main chain block like Bitcoin, it can confirm about 2000 subchains in one main chain block.
Assuming each subchain has an average of 10 transactions, we can easily achieve a high TPS under such conditions.

At this point, the consensus algorithm runs on the main chain, which also solves the security threat of the system caused by the dispersion of PoW computing power in multiple shards.

\subsection{Sharding by Accounts}

\begin{figure}[!t]
    \centering
    \includegraphics[width=1\textwidth]{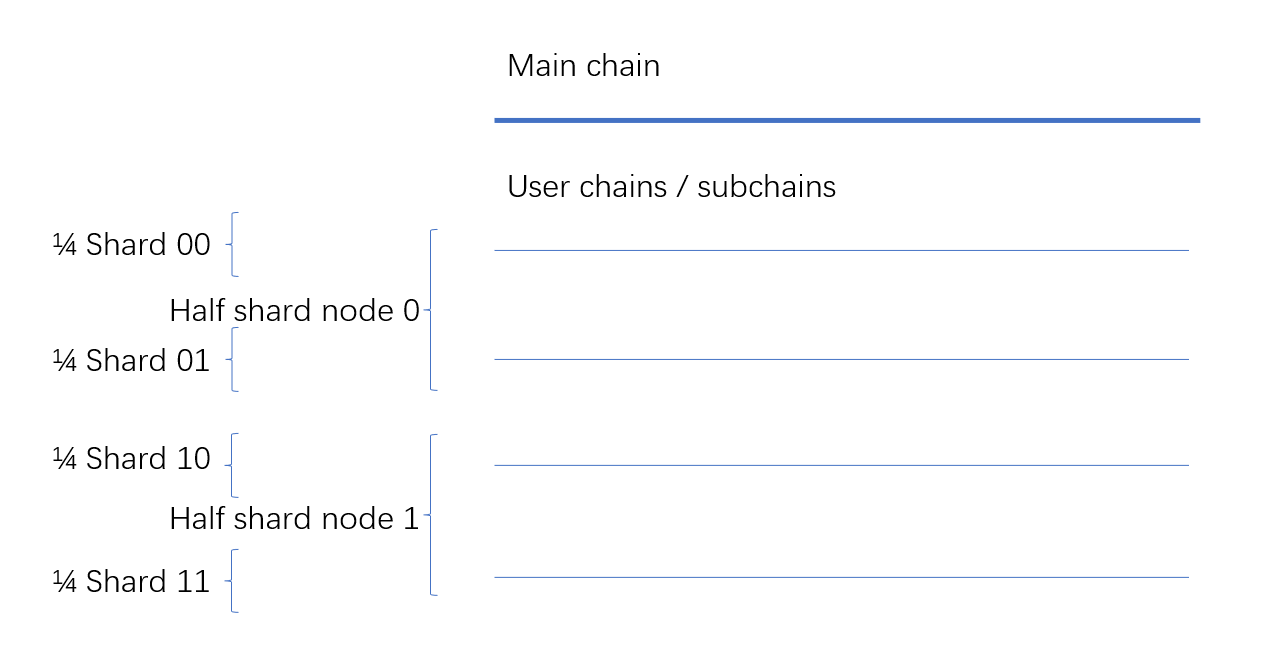}
    \caption{The subchains hosted by different nodes}
    \label{fig_Wider_sharding}
\end{figure}

As we proposed the one account per shard in the previous section, the sharding can be easily applied.
The nodes that used to act as the full nodes in the non-sharding blockchain can now turn into sharding nodes.
Each node has a unique group label indicating which subchain they should keep, shown in Figure~\ref{fig_Wider_sharding}.

In Wider, sharding by accounts is dividing the network by users.
Different from other sharding solutions, all the transactions are cross-shard transactions.
Sharding by accounts makes the users' transactions irrelevant.
A transaction sent by a user can only deduct the user's balance.

\subsection{Async Transaction Confirmation}

\begin{figure}[!t]
    \centering
    \includegraphics[width=0.8\textwidth]{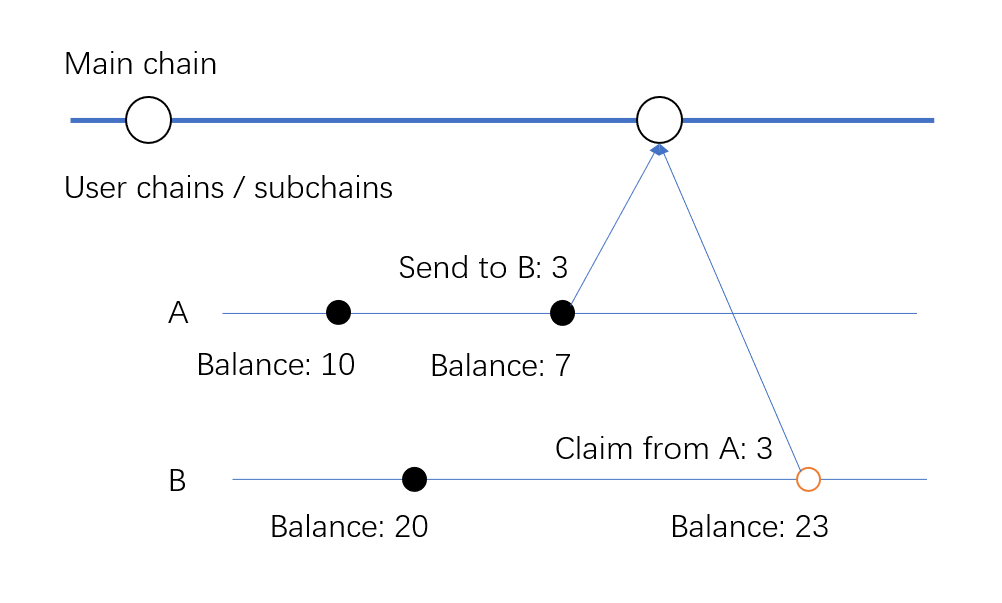}
    \caption{Async Transaction Confirmation}
    \label{fig_Wider_async_confirm}
\end{figure}

In Wider Sharding, the user transactions are recorded at the subchains.
There is no shared storage for the global blockchain state of the user balance.
Thus, the asset transfer must be operated in asynchronous mode, shown in Figure~\ref{fig_Wider_async_confirm}.

A user is assumed to send coins or tokens to another user in a non-negate value.
The sender creates a transaction with a value less than or equal to the sender's current balance.
The transaction is signed with the sender's private key.
Then, the transaction is appended to the sender's shard.
The miner who works on the main chain sees an updated shard with new transactions that are not confirmed in the main chain.
The updated subchains will be confirmed in the next main block to mine.

The balance is deduced once the subchain is confirmed, so the sender cannot send the paid balance again.
However, the receiver does not need to increase the balance at once; it is flexible to claim the fund at any convenient time, even right before the next time to transfer the fund.

\subsection{Benifit for Permissionless Storage}

In the previous chapter, we propose a permissionless storage network.
The storage network tried to reduce the reliance on blockchain as much as possible, expecting the transactions to go with blockchain.
In the storage network, the primary node accepts the off-chain operation, often requiring a batch on-chain operation.
Wider sharding works fine with such batch operations' demands.

\section{Smart Contract over Sharding}

\subsection{Transactions and Global State}

Ethereum smart contracts are designed to be callable by EOA and other smart contracts and can also call other smart contracts; this means that during the execution of the smart contract, it must be able to access the latest global state.
Suppose the currently running node or miner does not save the state data.
In that case, the system needs to obtain the data from other nodes during the running process, affecting the execution efficiency of the smart contract's virtual machine.
Or, the smart contract can only output an error result or cannot complete the execution.
Therefore, according to the design of Ethereum smart contracts, the global state is indivisible unless we modify the running rules of smart contracts, limit the data boundaries it can freely access, and design new rules based on this.

To maintain compatibility with Ethereum smart contract rules, we can choose to shard transactions but not shard the latest global state.
As mentioned before, we can treat transactions as inputs to a state machine and the global state of smart contracts as the state of the state machine. The state machine calculates the next state from the current state through input, and the calculation rule is the state transition function. In the Ethereum system, we can treat the EVM and the set of all smart contracts as the state transition function. Assuming that we have retained all of the state machine's inputs, we can reconstruct all of the states through the state transition function and the inputs. Therefore, if we maintain all the inputs, we can delete some states and only keep the latest global state.

\begin{figure}[!t]
    \centering
    \includegraphics[width=0.8\textwidth]{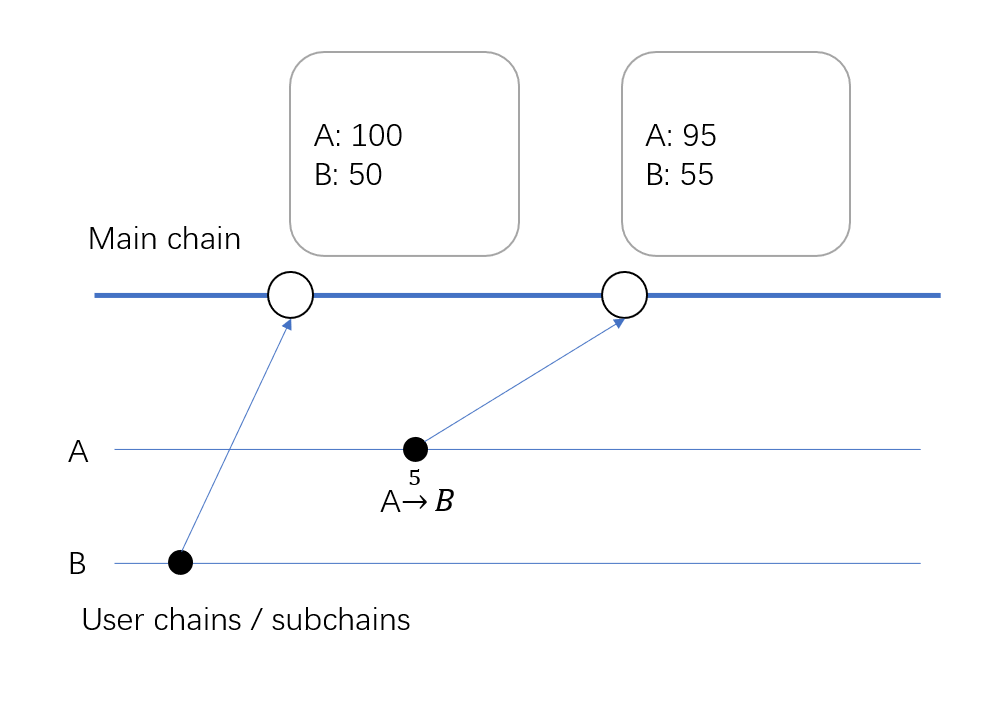}
    \caption{Wider transactions with global state}
    \label{fig_Wider_transaction_and_state}
\end{figure}

In the Wider blockchain, asynchronous confirmation is adopted because it needs to implement the transfer function of the native coin.
If the system retains the global state and implements smart contracts, then all token functions can be implemented by smart contracts, and the design of the native coin is no longer needed (use smart contract token), shown in Figure~\ref{fig_Wider_transaction_and_state}.
This works with the previous design of the useful PoW to reduce the transaction fee.
With permissionless storage providing security, the transaction fee can be stablecoin in the smart contracts.
In Wider, all transactions are placed on subchains, so it is very easy to implement transaction sharding.
Each node must save the full main chain and the latest global state.
The size of the main chain is similar to that of the Bitcoin main chain, which is linear growth, so it will not expand rapidly.
% In the next section, we will consider how to limit the size of the chain state.

\subsection{Smart Contract Implementation}

\begin{figure}[!t]
    \centering
    \includegraphics[width=1\textwidth]{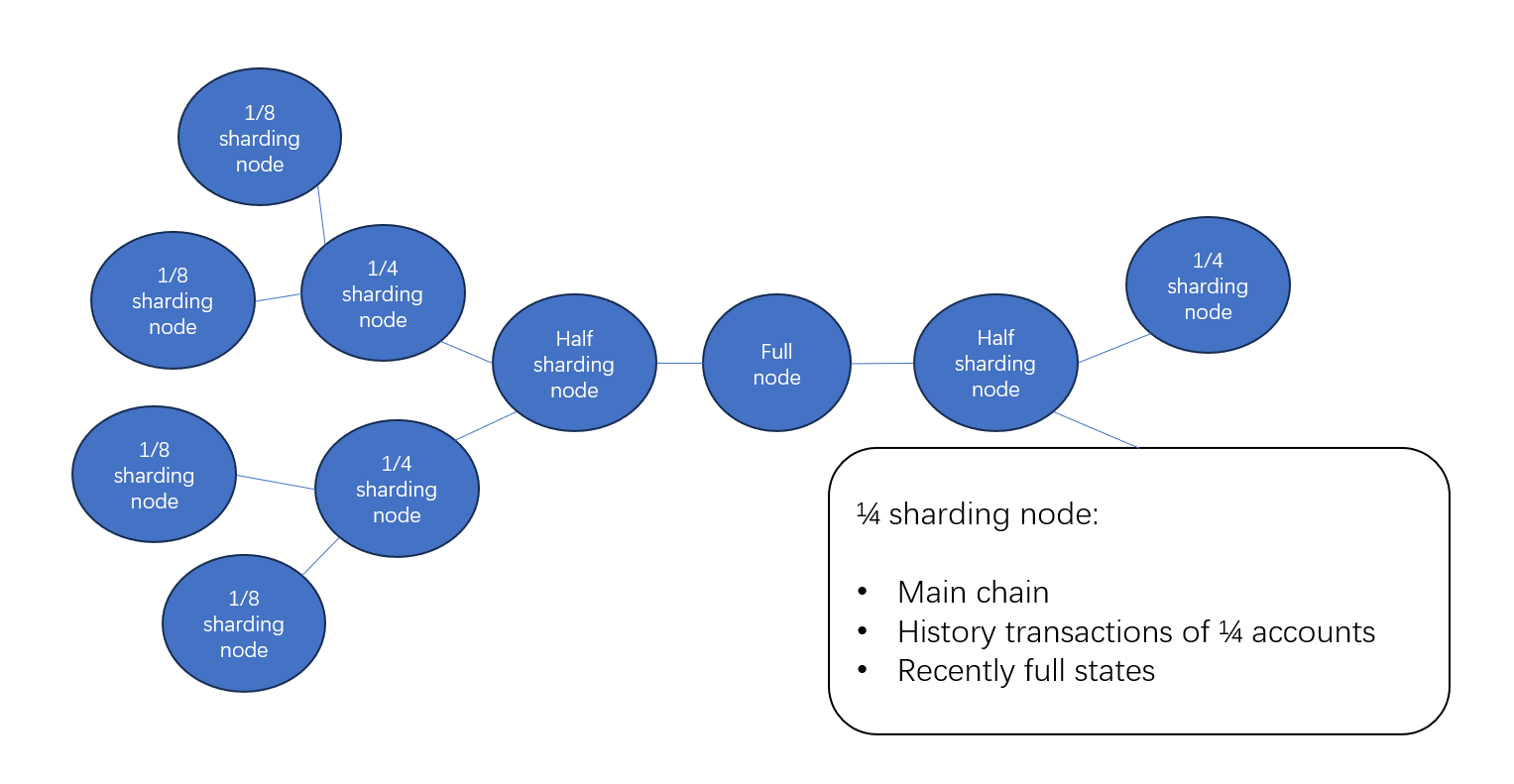}
    \caption{Nodes structure and the data kept in a node}
    \label{fig_Wider_nodes_and_data}
\end{figure}

Users can find the node corresponding to their current account address in a shard blockchain system.
The node will save all transactions sent by the corresponding account.
The new transaction sent by the user will be broadcast to all nodes and miners.
The miner will select the transactions with higher gas to run the consensus algorithm.
After the new block is generated, the new state will also be generated.
The node will only save the historical transactions of the corresponding users, as well as the main chain and the latest global state.
Figure~\ref{fig_Wider_nodes_and_data} shows the network structure, and the data should be kept in a sharding node.

Users can find the node corresponding to their current account address in a shard blockchain system.
The node will save all transactions sent by the corresponding account.
The new transaction sent by the user will be broadcast to all nodes and miners.
The miner will select the transactions with higher gas to run the consensus algorithm.
After the new block is generated, the new state will also be updated.
The node will only save the historical transactions of the corresponding users, as well as the main chain and the latest global state.

We have implemented the smart contract systems for Solidity and Python languages using EVM and Python VM.
Both systems are compatible with Ethereum smart contract standards, such as ERC20.

\subsection{Improve Performance by removing MPT}

\begin{figure}[!t]
    \centering
    \includegraphics[width=0.8\textwidth]{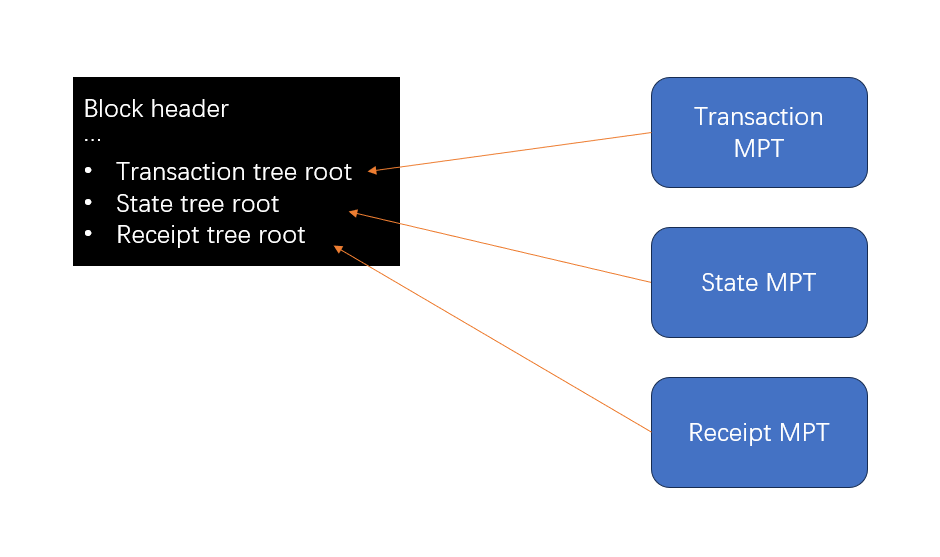}
    \caption{Ethereum block header contains the MPT roots info}
    \label{fig_ETH_block_header}
\end{figure}

Ethereum smart contracts are implemented using the MPT data structure.
There are transaction, state, and receipt MPT in each block header of Ethereum shown in Figure~\ref{fig_ETH_block_header}.
We did a simple test to insert 1 billion KV data into MPT before we tried to implement smart contracts.
The results show MPT is much slower than trie, and the overhead of the MPT data structure was very large.

For this reason, we optimized the execution of smart contracts based on the first principles.
MPT is built on top of trie, and the overhead of MPT is large; then, can we build the global state storage of smart contracts without using MPT?
We can improve the performance of smart contracts by removing the MPT.

\begin{table}[htbp]
    \centering
    \caption{The key of trie to replace state MPT}
    \label{wider_key_of_trie}
    \begin{tabular}{|c|c|c|c|}
    \hline
    \textbf{\textit{Key element}}& \textbf{\textit{Description}}& \textbf{\textit{Example}} \\
    \hline
    Prefix & Any & global\_state \\
    \hline
    Contract address & Address & 0x0000000000...0000000000001 \\
    \hline
    Variables & Data in contract & balances: 0x5bd3b58a23bc11...c126eef2d \\
    \hline
    Data owner & Owner of the value data & 0x5bd3b58a23bc11...c126eef2d \\
    \hline
    Block height & Hight in reversed order & 0999999999999997 \\
    \hline
    Block hash & Hash for block height & c22935a5d07c4...9c4a0f9f291d9c1 \\
    \hline
    \end{tabular}
\end{table}

Based on this design, we successfully removed the MPT data structure by carefully designing the trie key elements shown in Table~\ref{wider_key_of_trie} and removed the overhead of the smart contract execution.

\subsection{Reduce State Calculation Overhead}

\begin{figure}[!t]
    \centering
    \includegraphics[width=1\textwidth]{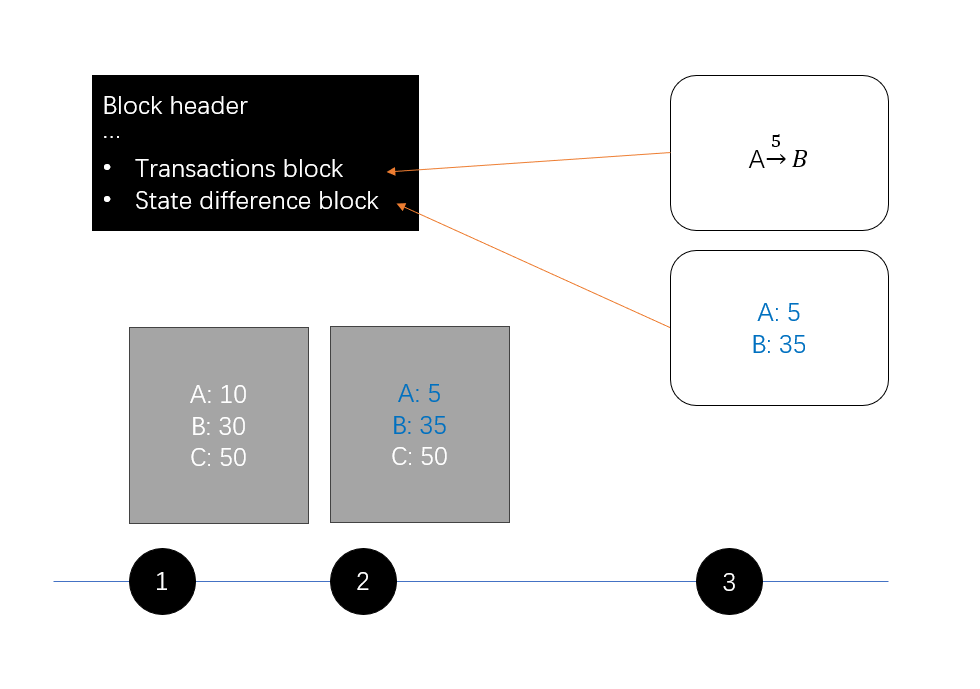}
    \caption{Wider block header contains the transaction and state difference block hashes}
    \label{fig_Wider_tx_and_state_block}
\end{figure}

When a miner finishes executing the contract according to the transaction input, the miner will start running the PoW algorithm.
The consensus algorithm allows one of the miners to generate the next block in the blockchain.
The calculated state after the miner runs the transactions will become the new global state of the blockchain.

In theory, as long as the new block is broadcasted, other nodes and miners can calculate the same new global state based on the transactions contained in the blockchain.
However, this step is full of redundancy.
Every node validates the same data.
When the number of such nodes is huge, only a portion of nodes or miners need to execute the contract calculation to verify that the new global state is equal to the calculation result of the input.

Therefore, our improvement is that the blockchain header contains two bodies, shown in Figure~\ref{fig_Wider_tx_and_state_block}.
One body contains the transaction input, and the other contains the difference between the new and old states.
We can set a ratio to randomly validate the execution of transactions on a portion of nodes, and the other portion of nodes directly applies the diff of the state body.
This can save a lot of repeated smart contract calculations.

\subsection{Limit Global State Size}

In the previous, we implemented smart contracts and optimized performance by avoiding the performance overhead of the MPT data structure.
Such a smart contract blockchain system implements the sharding of transactions (state machine inputs), but the global state is not sharded.
Although all the history of the global state can be reconstructed by replaying the inputs, which means that we can only save the latest global state, this state must be kept intact and indivisible.
Technologies similar to distributed hash tables can achieve state sharding. Still, since smart contracts need to read any content in the latest state efficiently, each node must save all the data in the latest state, which means that we need a mechanism to limit the size of the global state.

% \section{Security under Sharding}

\section{Experiment}

There are several metrics to reflect the Wider blockchain performance:

\begin{itemize}
    \item Transaction width: the number of accounts participating during the block interval.
    \item Average transactions: the average number of transactions on a subchain.
    \item Block size: the size of the main chain block, which indicates the number of subchains to confirm in a block.
    \item Block interval: the interval of main chain blocks.
\end{itemize}

We implement our sharding blockchain with Python 3.10.
The overall performance with Python is not as fast as with C++ or Rust.
Therefore, Wider may achieve better performance once we turn the solution industry-ready.
We run the experiments on the Ubuntu 22.04 LTS operation system.
The test CPU is AMD Ryzen 7 5800 @3.4GHz with 8 Cores.
The blockchain uses the RocksDB database on an SSD hard drive.
For the network, we choose the tree network to achieve better performance in broadcast communication.

\subsection{Blockchain TPS}

We measure the blockchain by adjusting the blockchain width and block size parameters.
As the roll-up technique is applied, we know all the subchain updates will be confirmed in the following main chain if the block size is unlimited.
Our experiment shows Wider can achieve 6000+ TPS by removing the block size limit.
However, there is always a fixed bandwidth in practice, so the block size can only increase for a while.
We also measure different tiers of block size by different block intervals in our experiments.

Following Bitcoin and Ethereum, we fixed the interval of blocks to 10 minutes and 15 seconds and used only one CPU core.
So, in the first experiment, we set the transaction width from 5,000 to 15,000 accounts for the 10-minute block interval and 1MB block size and set the transaction width from 100 to 1,000 accounts for the 15-second block interval and 40KB block size.

Table~\ref{wider_tps_btc} shows the result:
Under the Bitcoin-like setting (10-minute block interval and 1MB in block size), the sharding blockchain node can achieve 2000-5000 TPS.
Table~\ref{wider_tps_eth} shows the result:
Under the Ethereum-like setting (15-second block interval and 40KB in block size), the sharding blockchain node can achieve 1000-3000 TPS.

\begin{table}[htbp]
    \centering
    \caption{Transactions width and TPS in Bitcoin setting}
    \label{wider_tps_btc}
    % \begin{center}
    \begin{tabular}{|c|c|c|c|}
    \hline
    \textbf{\textit{TX Width}}& \textbf{\textit{Block size}}& \textbf{\textit{Block Interval}}& \textbf{\textit{TPS}} \\
    \hline
    5000 & 1MB & 600s & 4937 \\
    \hline
    6000 & 1MB & 600s & 5407 \\
    \hline
    7000 & 1MB & 600s & 2293 \\
    \hline
    8000 & 1MB & 600s & 3692 \\
    \hline
    9000 & 1MB & 600s & 2906 \\
    \hline
    10000 & 1MB & 600s & 2596 \\
    \hline
    11000 & 1MB & 600s & 2502 \\
    \hline
    12000 & 1MB & 600s & 2497 \\
    \hline
    13000 & 1MB & 600s & 2313 \\
    \hline
    14000 & 1MB & 600s & 2233 \\
    \hline
    15000 & 1MB & 600s & 2159 \\
    \hline
    \end{tabular}
% \end{center}
\end{table}

\begin{table}[htbp]
    \centering
    \caption{Transactions width TPS in Ethereum setting}
    \label{wider_tps_eth}
    % \begin{center}
    \begin{tabular}{|c|c|c|c|}
    \hline
    \textbf{\textit{TX Width}}& \textbf{\textit{Block size}}& \textbf{\textit{Block Interval}}& \textbf{\textit{TPS}} \\
    \hline
    100 & 40KB & 15s & 2362 \\
    \hline
    200 & 40KB & 15s & 2409 \\
    \hline
    300 & 40KB & 15s & 2026 \\
    \hline
    400 & 40KB & 15s & 2423 \\
    \hline
    500 & 40KB & 15s & 1903 \\
    \hline
    600 & 40KB & 15s & 3276 \\
    \hline
    700 & 40KB & 15s & 2491 \\
    \hline
    800 & 40KB & 15s & 2282 \\
    \hline
    900 & 40KB & 15s & 1656 \\
    \hline
    1000 & 40KB & 15s & 1461 \\
    \hline
    \end{tabular}
    % \end{center}
\end{table}

According to the high performance blockchain solution Conflux\cite{conflux}, the chain could achieve 3000 TPS with Tree-Graph ledger structure\cite{li2018scaling} and GHAST algorithm.
Our solution can achieve similar TPS.
However, the DAG-based is well known that it is hard to shard, the node to keep all the history transactions requires a large storage capacity.
Our solution makes the transactions easy.

\subsection{Transactions Verification}

\begin{figure}[htbp]
\centerline{
    \includegraphics[width=0.6\linewidth]{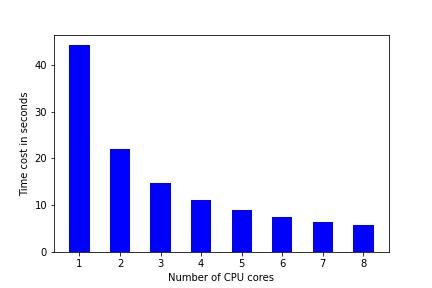}
}
\caption{The time cost for transaction verification with more CPU cores.}
\label{fig_timecost_verify}
\end{figure}

As the TPS increased, the system bottleneck is now in verification, which requires much more computation.
A single CPU core can perform limited times of the signature (ECDSA) verification operations per second.

After receiving the block or the transactions, the node and the miner verify the data.
The miner must verify all the broadcast network transactions to maximize his incoming.
After confirming the main chain block, the node verifies the corresponding subchains.
Multicore CPU can reduce the time for transaction verification.
Moreover, the verification can be outsourced to a local trusted cluster.

We experimented on the verification of 10,000 transactions with different numbers of cores.
Figure~\ref{fig_timecost_verify} shows the time cost in verification.
Thus, the transaction processing speed can rise with the help of other cores.
The computation can even be outsourced to a local cluster.

\subsection{Sharding and Full Node}

\begin{figure}[htbp]
\centerline{
    \includegraphics[width=0.6\linewidth]{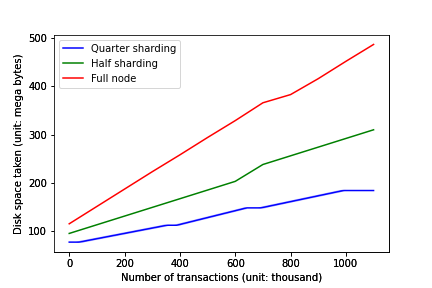}
}
\caption{The storage space cost for the full node and the sharding nodes}
\label{fig_Wider_storage_space_cost}
\end{figure}

The sharding node shall suffer less pressure than a full node in Wider, including CPU and disk usage.
With more and more nodes joining the blockchain backbone network, the new node is appended to the tree network as the outer leaf.
As there are two leaves of a parent, the leaf node only keeps half of the parent shards and the main chain.
If the early joined node does not have enough resources to host the corresponding shards, it may quit and rejoin as an outer leaf.
Since Bitcoin did not bring the tradition of incentivizing the nodes, there might be a chance to design a mechanism to reward the inner node that hosts more shards.
However, the incentive mechanism will not be discussed in this paper.

We measure the storage space taken by the RocksDB database for full node, half sharding, and quarter sharding.
The result is shown in Figure~\ref{fig_Wider_storage_space_cost}.
The sharding node uses less storage space than the full node.
Unlike the Layer 2 or off-chain solutions, all subchain transactions are publicly verifiable.

\section{Chapter Summary}

This chapter proposes the idea of sharding based on a multi-chain structure and its derivation and implementation.
Based on the multi-chain structure, we build a native coin shard blockchain system similar to Bitcoin, and we can modify it to create a blockchain system that includes smart contracts and make further optimizations.
Experiments show that the blockchain system we built can achieve 3000+ TPS and effectively reduce the chain data growth rate of blockchain nodes.

The sharding design meets the requirement for the permissionless storage network we designed in the previous chapter.
A single user can perform batch operations, and the blockchain can respond with main chain confirmation.
Combined with the useful PoW consensus we proposed, the user can enjoy cheap on-chain transactions.
This helps real-world blockchain applications land for massive adoption.

\chapter{Conclusion and Future Work}
\label{ch:conclusion}

In this thesis, we studied the issues preventing blockchain systems from massive adoption.
The major problems are related to the key conflict between high security and low usage costs, and the high TPS requirement with shading.
For the bottom line, blockchain for assets must remain decentralized, a significant difference from the web2 system.
It is because the centralized system has already crashed or been forced to shut down many times in history, from the big events like the financial crisis in 2008 to the products shutdown within Google and many others.
The decision to shut down an application service is centralized by the management, regardless many users are still using the service.

The information system will likely be as important as the fundamental infrastructure even like public goods, such as electricity, water and the road.
Imagine that an IT service heavily relied on by the public, such as Google Maps or Uber failure, would impact society negatively.
Blockchain and permissionless systems can be a better answer to future social information infrastructure and service.
However, this new technology was just born in 2008 and has yet to step into massive adoption in the past decades.

The problems in this thesis target to the most critical issues of existing the blockchain systems.
The proposed solutions significantly improve the blockchain usability without reducing security or decentralization, including the user cost and the chain performance.
The new engine may drive the blockchain to a more decentralized level.

\section{Conclusion}

In this thesis, we propose two solutions to address the massive blockchain adoption issues.
The first useful PoW in Chapter~\ref{ch:consensus} is applied to solve the blockchain usage cost issue: the conflict that blockchain requires a high-security cost and the users want the cheap transaction fee.
To reduce the blockchain usage cost, a permissionless storage system is proposed as the new security engine in Chapter~\ref{ch:storage}.
Furtherly, we designed a complete storage system, including a PRE scheme for the cryptographic permission control.

The second solution, Wider sharding, aims to solve the blockchain scalability issue of the high TPS blockchain in Chapter~\ref{ch:sharding}.
We choose the multi chains data structure for sharding rather than the popular DAG solutions.
Based on the sharding, we implemented the Ethereum-compatible smart contract system.

The sharding did not sacrifice decentralization.
Meanwhile, the useful PoW may increase decentralization by removing the mining pools in theory.
We solved the most critical issues in the blockchain infrastructure area that prevent massive blockchain adoption.
Permissionless storage also brings the blockchain more applications than finance scenarios.

\section{Limitations}

The useful PoW is facing the cold boot issue.
A PoW blockchain must have enough computation to secure the network.
Thus, the storage market may need to reuse existing payment system to cold boot the PoW for the future PoW blockchain.

In the PoRep, the prover and verifier ensure the integration of the user data in the permissionless storage network.
The verifier must keep a copy of data to check the prover's response to the system challenge quickly.
It is ideal for letting the verifier keep minimal information that could verify the honesty of data preservation.
A better cryptographic protocol is demanded.

The wider sharding solution shards the transactions, but the chain state is maintained by each node.
State explosion issue is not solved with wider sharding.
However, the token mechanism can be used to prevent the state growth unlimited.
The occupation of state space is determined by the amount of tokens.
It can be a better method to fix the state explosion issue from non-technical perspective.

\section{Future Work}

We aim to build the blockchain with permissionless storage as the new security engine.
The best way to prove our theory is to make it practical.
However, many details can be improved during the implementation.

The Proof of Replication ensures the storage is safely kept, but in permissionless storage, the speed of content retrieval is another key factor to measure the storage service quality.
It is ideal to have a cryptographic method to measure the speed of content transfer.

For the sharding solution, the smart contract has been implemented in Chapter~\ref{ch:sharding}.
It is possible to analyze the relies of different smart contracts.
The fundamental smart contracts such as ERC20s have no extra dependency, but many other applications require it.
It is possible to parallel the execution of those smart contracts without common dependencies or even the intersection of shared data.

\appendix
  \renewcommand{\appendixname}{Appendix~\Alph{section}}
\chapter{List of Acronyms}

\begin{description}
  \item[APR] Annual Percentage Rate
  \item[BFT] Byzantine Fault Tolerance
  \item[BTC] Bitcoin
  \item[CCA] Chosen-Ciphertext Attack
  \item[CPA] Chosen-Plaintext Attack
  \item[CRUSH] Controlled Replication Under a Scalable Hashing
  \item[CDN] Content Delivery Network
  \item[DAG] Directed Acyclic Graph
  \item[DH] Diffie-Hellman
  \item[DL] Discrete Logarithm problem
  \item[ECC] Elliptic Curve Cryptography
  \item[ECDSA] Elliptic Curve Digital Signature Algorithm
  \item[EVM] Ethereum Virtual Machine
  \item[ETH] Ethereum
  \item[GFS] Google File System
  \item[HD] Hard Disk
  \item[HDFS] Hadoop Distributed File System
  \item[IPFS] Inter Planetary File System
  \item[KV] Key Value, usually refer to database
  \item[LN] Lightning Network
  \item[LSTM] Log-Structured Merge-Tree
  \item[MPT] Merkle Patricia Tree
  \item[MON] Monitor
  \item[NC] Nakamoto Consensus
  \item[NFS] Network File System protocol
  \item[NFT] Non-fungible token
  \item[OSD] Object Storage Daemon
  \item[OP] Optimism Rollup Layer 2
  \item[P2P] Peer-to-Peer network
  \item[PK] Public Key
  \item[PKI] Public Key Infrastructure
  \item[PW] Password
  \item[PRE] Proxy Re-Encryption scheme
  \item[PoW] Proof of Work
  \item[PoS] Proof of Stake
  \item[PoC] Proof of Capacity, an equivalent term to Proof of Space
  \item[PoA] Proof of Access
  \item[PoSpace] Proof of Space
  \item[PoRep] Proof of Replication
  \item[PBFT] Practical Byzantine Fault Tolerance
  \item[PG] Placement Group
  \item[RADOS] Reliable Autonomic Distributed Object Store
  \item[RGW] RADOS Gateway
  \item[RDB] RADOS Block Device
  \item[SHA] Secure Hash Algorithms
  \item[SDR] Stacked DRG Proof of Replication
  \item[STF] State Transfer Function
  \item[TPS] Transaction Per Second
  \item[UTXO] Unspent Transaction Outputs
  \item[ZK] Zero Knowledge
\end{description}

%% For references, I recommend using BiBTeX. The references are stored in references.bib.
\bibliographystyle{plain}
\addcontentsline{toc}{chapter}{References} % Add the references to the table of contents.
\bibliography{references}
\end{doublespace}
%\end{doublespace}

%% It is also possible to use
%\begin{thebibliography}{99}
%\bibitem{}
%\end{thebibliography}
%% if desired.

\end{document}